\documentclass[preprint,prd,tightenlines,floatfix,
nofootinbib,eqsecnum,superscriptaddress]{revtex4-1}

\usepackage[T1]{fontenc}		% fnot encoding with polish letters !or OT4

\usepackage{amsmath,amsfonts,amssymb,amstext,mathrsfs}
\usepackage{mathpazo}

\usepackage[dvips]{graphicx}
\usepackage{epsf,float}
\usepackage{revsymb}

\usepackage{dcolumn}% Align table columns on decimal point
\usepackage{braket}
\usepackage{color,xcolor}
\usepackage{graphicx}
\usepackage{subfigure}
\usepackage{multirow}
\usepackage{tabularx}
\usepackage{pstricks}
\usepackage[section]{placeins}
\usepackage{booktabs}
\usepackage{array}

\usepackage{commath}

\usepackage{hyperref}
\newcommand{\Pom}{\mathbb{P}}

\newcommand{\Reg}{\mathbb{R}}

\newcommand{\bpta}{\mbox{\boldmath $p_{t,1}$}}
\newcommand{\bptb}{\mbox{\boldmath $p_{t,2}$}}

\newcommand{\bk}{\mbox{\boldmath $k$}}
\newcommand{\bq}{\mbox{\boldmath $q$}}
\newcommand{\bpa}{\mbox{\boldmath $p_{a}$}}

\newcommand{\p}{\partial}

\newcommand{\twosidep}[1]{\stackrel{\leftrightarrow}{\p}_{\! #1}}

\usepackage[normalem]{ulem}  % \sout{old text} for strikeout

\begin{document}

%------------------------
%\nolinenumbers
%\linenumbers
%------------------------

\title{\boldmath 
Exclusive $f_{1}(1285)$ meson production for energy ranges\\ available at the GSI-FAIR with HADES and PANDA}

\vspace{0.6cm}

\author{Piotr Lebiedowicz}
%\orcid{0000-0003-1963-6263}
\email{Piotr.Lebiedowicz@ifj.edu.pl}
\affiliation{Institute of Nuclear Physics Polish Academy of Sciences, Radzikowskiego 152, PL-31342 Krak{\'o}w, Poland}

\author{Otto Nachtmann}
\email{O.Nachtmann@thphys.uni-heidelberg.de}
\affiliation{Institut f\"ur Theoretische Physik,
Universit\"at Heidelberg, Philosophenweg 16, D-69120 Heidelberg, Germany}

\author{Piotr Salabura}
%\orcid{0000-0002-4727-3087}
\email{Piotr.Salabura@uj.edu.pl}
\affiliation{M. Smoluchowski Institute of Physics, 
Jagiellonian University, {\L}ojasiewicza 11, PL-30348 Krak{\'o}w, Poland}

\author{Antoni Szczurek
%\orcid{0000-0001-5247-8442}
\footnote{Also at \textit{College of Natural Sciences, 
Institute of Physics, University of Rzesz{\'o}w, 
ul. Pigonia 1, PL-35310 Rzesz{\'o}w, Poland}.}}
\email{Antoni.Szczurek@ifj.edu.pl}
\affiliation{Institute of Nuclear Physics Polish Academy of Sciences,
Radzikowskiego 152, PL-31342 Krak{\'o}w, Poland}

\begin{abstract}
We evaluate the cross section for the $p p \to p p f_{1}(1285)$ and $p
\bar{p} \to p \bar{p} f_{1}(1285)$ reactions at near threshold energies
relevant for the HADES and PANDA experiments at GSI-FAIR. We assume that
at energies close to the threshold the $\omega \omega \to f_{1}(1285)$
and $\rho^{0} \rho^{0} \to f_{1}(1285)$ fusion processes are the
dominant production mechanisms. The vertex for the $VV \to f_{1}$
coupling is derived from an effective coupling Lagrangian. 
The $g_{\rho \rho f_1}$ coupling constant is extracted 
from the decay rate of $f_{1}(1285) \to \rho^{0} \gamma$ 
using the vector-meson-dominance ansatz. 
We assume $g_{\omega \omega f_1} = g_{\rho \rho f_1}$, equality
of these two coupling constants, based on arguments from the naive quark
model and vector-meson dominance. The amplitude for the $VV \to f_{1}$
fusion, supplemented by phenomenological vertex form factors for the
process, is given. The differential cross sections at energies close to
the threshold are calculated. In order to determine the parameters of
the model the $\gamma p \to f_{1}(1285) p$ reaction is discussed in
addition and results are compared with the CLAS data. The possibility of
a measurement by HADES@GSI is presented and discussed.
We performed a Monte Carlo feasibility simulations of 
the $p p \to p p f_1$ reaction for $\sqrt{s}$ = 3.46 GeV
in the $\pi^+ \pi^- \pi^+ \pi^-$ (not shown explicitly) and 
$\pi^+ \pi^- \pi^+ \pi^- \pi^0$ final states using the {\textsc PLUTO} generator.
The latter one is especially promising as a peak in the
$\pi^+ \pi^- \eta$ should be observable by HADES.
\end{abstract}

%\pacs{}

\maketitle

%----------------------------
\section{Introduction}
\label{sec:intro}
%----------------------------

The production of light axial-vector mesons 
with quantum numbers $I^{G}J^{PC} = 0^{+}1^{++}$ 
is very interesting and was discussed in a number of experimental 
and theoretical papers.
For example, the $f_{1}(1285)$ meson was measured 
in two-photon interactions in the reaction 
$e^+ e^- \to e^+ e^- \eta \pi^{+}\pi^{-} (\eta \to \gamma \gamma)$
by the Mark~II \cite{Gidal:1987bn}, 
the TPC/Two-Gamma \cite{Aihara:1988uh,Aihara:1988bw},
%for the single tag case
and, more recently, 
by the L3 \cite{Achard:2001uu} collaborations.
%in untagged two-photon collisions
%where the outgoing electron and positron 
%were not detected
In such a process the 
$\gamma^* \gamma^* \to f_{1}(1285)$ vertex, associated with corresponding
transition form factors, is the building block in calculating the
amplitude. Different vector-vector-$f_{1}$ vertices and 
corresponding transition form factors were suggested
in the literature
\cite{Close:1997usa,Pascalutsa:2012pr,Pauk:2014rta,Osipov:2017ray,Milstein:2019yvz,Dorokhov:2019tjc,Roig:2019reh,Leutgeb:2019gbz,Cappiello:2019hwh,1852275}.
It was suggested in \cite{Szczurek:2020hpc} 
that a measurement of 
the $e^+ e^- \to e^+ e^- f_{1}(1285)$ reaction 
with double tagging 
at Belle~II at KEK could shed new light on 
the $\gamma^* \gamma^* f_{1}$ coupling with two virtual photons.

The $f_{1}(1285)$ meson was also measured in the photoproduction process
$\gamma p \to f_{1}(1285) p$ by the CLAS Collaboration at JLAB \cite{Dickson:2016gwc}.
The differential cross sections were measured 
from threshold up to a center-of-mass energy of $W_{\gamma p} = 2.8$~GeV 
in a wide range of production angles.
The $f_{1}(1285)$ photoproduction was studied extensively 
from the theoretical point of view; see
\cite{Kochelev:2009xz,Domokos:2009cq,Wang:2017plf,Wang:2017hug,Yu:2019wly}. 
There, the $t$-channel $\rho$ and $\omega$ exchange
(either Regge trajectories or meson exchanges)
is the dominant reaction mechanism 
for the small-$t$ behaviour of the cross section, that is,
in the forward scattering region.
The contribution of the $u$-channel proton-exchange term 
with the coupling of $f_{1}(1285)$ to the nucleon
is dominant at the backward angles
\cite{Oh:2000mq,Domokos:2009cq,Wang:2017plf,Wang:2017hug}.
In \cite{Wang:2017hug} the authors showed that 
also the $s$-channel nucleon resonance
$N(2300)$ with $J^{P} = 1/2^{+}$ may play an important role 
in the reaction of $\gamma p \to f_{1}(1285) p$ 
around $\sqrt{s} = 2.3$~GeV.
As was shown in \cite{Wang:2017hug} other contributions, 
the $s$-channel proton-exchange term,
the $u$-channel $N(2300)$-exchange term, and the contact term,
are very small and can be neglected in the analysis of the CLAS data.
The Primakoff effect by the virtual photon exchange 
in the $t$-channel was discussed in~\cite{Yu:2019wly}.
This mechanism is especially important in the forward region 
and at higher $W_{\gamma p}$ energies.

%For the $f_{1}$ $t$-channel exchange see \cite{Kochelev:1999zf,Oh:2000mq}.

The $p p \to p p f_{1}(1285)$ reaction was already measured 
by the WA102 Collaboration
for center-of-mass energies $\sqrt{s} = 12.7$ and $29.1$~GeV 
\cite{Barberis:1997ve,Barberis:1997vf,Barberis:1998by,Barberis:1999wn}.
There the dominant contribution at $\sqrt{s} = 29.1$~GeV
is most probably related to the double-pomeron-exchange
(\mbox{$\Pom \Pom$-fusion}) mechanism; 
see \cite{Lebiedowicz:2020yre}.
In \cite{Lebiedowicz:2020yre} the $pp \to pp f_{1}(1285)$ and 
$pp \to ppf_{1}(1420)$ reactions
were considered in the tensor-pomeron approach \cite{Ewerz:2013kda}.
A good description of the WA102 data \cite{Barberis:1998by}
at $\sqrt{s} = 29.1$~GeV was achieved.
A study of central exclusive production (CEP) of the axial vector mesons
$f_{1}$ at high energies (RHIC, LHC) could shed more light 
on the coupling of two pomerons to the $f_{1}$ meson \cite{Lebiedowicz:2020yre}.
As discussed in Appendix~D of \cite{Lebiedowicz:2020yre} 
at the lower energy $\sqrt{s} = 12.7$~GeV
the reggeized-vector-meson-exchange or reggeon-reggeon-exchange contributions
should be taken into account.
 
The $\omega \omega \to f_1$ and $\rho^0 \rho^0 \to f_1$ fusion
are the most probable low energy production processes.
We know how the $\omega$ and $\rho^0$ couple to nucleons.
However, the couplings of $\omega \omega \to f_1$ and
$\rho^0 \rho^0 \to f_1$ are less known. 
We note that future experiments at HADES and PANDA
will provide new information there. 
The $\rho^0 \rho^0 \to f_{1}(1285)$ coupling constant 
can be obtained from the decays:
$f_{1} \to \rho^0 \gamma$ and/or $f_{1} \to \pi^+ \pi^- \pi^+ \pi^-$.

In the present analysis we obtain the $g_{\rho \rho f_1}$ coupling
constant from the radiative decay process 
$f_{1}(1285) \to \gamma \rho^{0} \to \gamma \pi^{+} \pi^{-}$
using the vector-meson-dominance (VMD) ansatz;
see Appendices~\ref{sec:appendixA} and \ref{sec:appendixB}.
We discuss briefly our results for the $\gamma p \to f_{1}(1285) p$ reaction
and compare with the CLAS data in Appendix~\ref{sec:appendixC}.
From this comparison we estimate the form-factor cutoff parameters.

The PANDA experiment (antiProton ANnihilations at DArmstadt)
\cite{Lutz:2009ff} will be one of the key experiments
at the Facility for Antiproton and Ion Research (FAIR)
which is currently being constructed.
At FAIR, a system of accelerators and storage rings 
will be used to generate a beam of antiprotons 
with a momentum between 1.5 and 15~GeV/$c$. 
The design maximum energy in the center-of-mass (c.m.) 
system for antiproton-proton collisions is $\sqrt{s} \simeq 5.5$~GeV.

The exclusive production of the $f_{0}(1500)$ meson 
in antiproton-proton collisions 
via the pion-pion fusion mechanism was discussed
for the PANDA experiment in \cite{Szczurek:2009yk};
see also Fig.~3 of \cite{Lebiedowicz:2013ika}.
The pion-pion fusion contribution grows quickly from the threshold,
has a maximum at $\sqrt{s} \simeq 6$~GeV 
and then drops slowly with increasing energy.
The predicted cross section for the $p \bar{p} f_{0}(1500)$ final state
is $\sigma_{f_{0}} = 0.3-0.8$~$\mu$b for $\sqrt{s} = 5.5$~GeV; 
see Sec.~III~C of \cite{Szczurek:2009yk}.
At intermediate energies 
(e.g. for the WA102 and COMPASS experiments)
other exchange processes such as 
the reggeon-reggeon, reggeon-pomeron and pomeron-pomeron
exchanges are very probable; see e.g. \cite{Lebiedowicz:2013ika}.

A measurement at low energies, such as HADES@GSI would be interesting
to impose constraints on the $VV \to f_{1}(1285)$ vertices.
In this paper we wish to make first estimates of the total
and differential cross sections for 
the $p p \to p p f_{1}(1285)$ 
and $p \bar{p} \to p \bar{p} f_{1}(1285)$ reactions
at energies relevant for the HADES and PANDA experiments.
We shall present some differential distributions
for the HADES energy at $\sqrt{s} = 3.46$~GeV
and for the future experiments
with the PANDA detector at $\sqrt{s} = 5.0$~GeV.
The experimental possibilities of such measurements 
will be discussed in addition.

%----------------------------------------
\section{Theoretical framework}
\label{sec:theoretical_fr}
%----------------------------------------
%----------------------------------------
\subsection{$f_{1}(1285)$ meson production via $\omega\omega$ and $\rho\rho$ fusion mechanisms}
\label{sec:signal}
%----------------------------------------
We study central exclusive production of the $f_{1}(1285)$
in proton-proton collisions
\begin{eqnarray}
p(p_{a},\lambda_{a}) + p(p_{b},\lambda_{b}) \to
p(p_{1},\lambda_{1}) + f_{1}(k ,\lambda_{f_{1}}) + p(p_{2},\lambda_{2}) \,,
\label{2to3_reaction}
\end{eqnarray}
where $p_{a,b}$, $p_{1,2}$ and $\lambda_{a,b}$, 
$\lambda_{1,2} = \pm \frac{1}{2}$
denote the four-momenta and helicities of the protons, 
and $k$ and $\lambda_{f_{1}} = 0, \pm 1$ 
denote the four-momentum and helicity 
of the $f_{1}$ meson, respectively.

\begin{figure}[!ht]
\includegraphics[width=5.2cm]{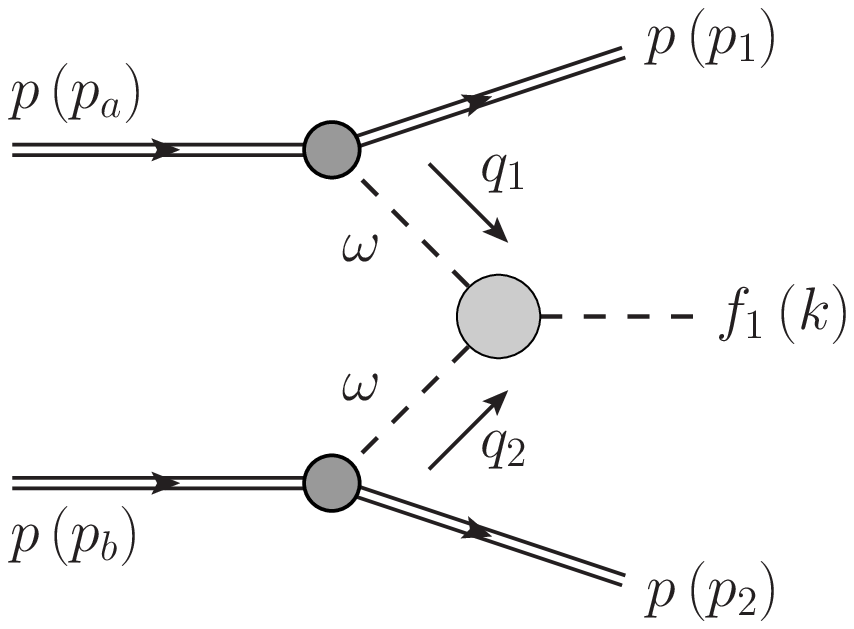}
\includegraphics[width=5.2cm]{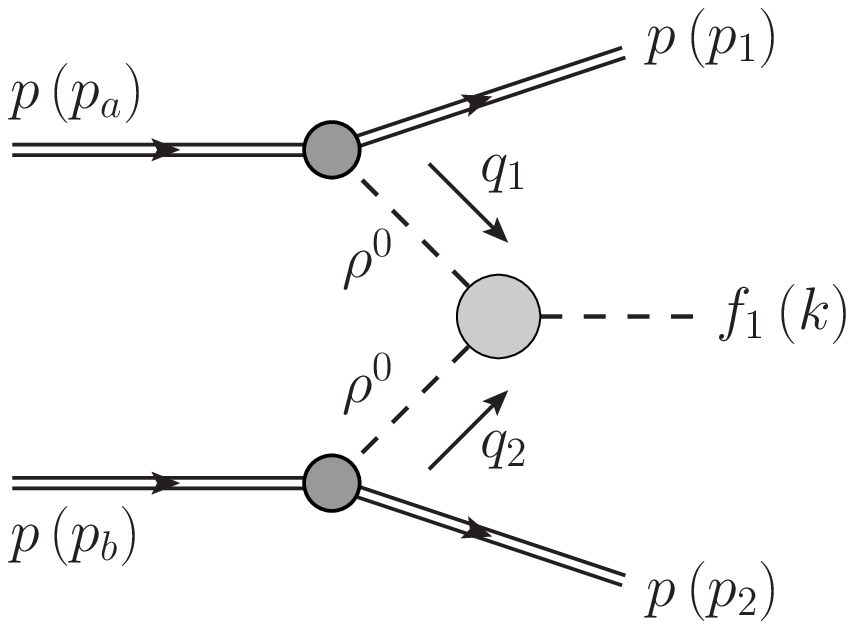}
\caption{The $VV$-fusion mechanisms 
($VV$ stands for $\omega\omega$ or $\rho^{0}\rho^{0}$)
for $f_{1}$ production in proton-proton collisions.}
\label{fig:diagrams}
\end{figure}

In \cite{Lebiedowicz:2020yre} for the reaction (\ref{2to3_reaction}) 
the pomeron-pomeron-fusion mechanism was considered
which seems to dominate at the WA102 energy of $\sqrt{s} = 29.1$~GeV.
As discussed in Appendix~D of \cite{Lebiedowicz:2020yre} 
at lower energies other fusion mechanisms may be important.
We shall take into account only the main processes 
at energies close to the threshold, 
the $VV$-fusion mechanism, 
shown by the diagrams in Fig.~\ref{fig:diagrams}.
There can be also the $a_{1}^{0}(1260) \pi^{0}$-fusion mechanism
not discussed in the present paper.
Note that due to the large width of the $a_{1}(1260)$
the decay $f_{1}(1285) \to \pi^{\pm} a_{1}^{\mp}$ can easily 
occur for off-shell $a_{1}(1260)$ and this is 
an important decay mode
in the $f_{1} \to 2 \pi^{+} 2 \pi^{-}$ channel
as will be discussed in~\cite{LNS:2020}.

%-------------------------------------------------------------
%-------------------------------------------------------------

The kinematic variables for the reaction (\ref{2to3_reaction}) are
\begin{eqnarray}
&&q_1 = p_{a} - p_{1}, \quad q_2 = p_{b} - p_{2}, \quad k = q_{1} + q_{2}, \nonumber \\
&&t_1 = q_{1}^{2}, \quad t_2 = q_{2}^{2}, \quad m_{f_{1}}^{2} = k^{2}, \nonumber \\
&&s = (p_{a} + p_{b})^{2} = (p_{1} + p_{2} + k)^{2}, \nonumber \\
&&    s_{1} = (p_{a} + q_{2})^{2} = (p_{1} + k)^{2}, \nonumber \\
&&    s_{2} = (p_{b} + q_{1})^{2} = (p_{2} + k)^{2}\,.
\label{2to3_kinematic}
\end{eqnarray}
For the kinematics see e.g. Appendix~D of \cite{Lebiedowicz:2013ika}.

The amplitude for the reaction (\ref{2to3_reaction})
includes two terms
\begin{eqnarray}
{\cal M}_{pp \to pp f_{1}(1285)} = 
{\cal M}_{pp \to pp f_{1}(1285)}^{(\omega \omega \; {\rm fusion})}
+
{\cal M}_{pp \to pp f_{1}(1285)}^{(\rho \rho \; {\rm fusion})}
\,.
\label{sum_amplitudes}
\end{eqnarray}
The $VV$-fusion ($VV = \rho^{0}\rho^{0}$ or $\omega\omega$) amplitude 
can be written as
\begin{eqnarray}
{\cal M}^{(V V \; {\rm fusion})}_{\lambda_{a} \lambda_{b} \to \lambda_{1} \lambda_{2} \lambda_{f_{1}}}
&=& (-i)\,(\epsilon^{\alpha}(\lambda_{f_{1}}))^{*}\,
\bar{u}(p_{1}, \lambda_{1}) 
i\Gamma^{(V pp)}_{\mu_{1}}(p_{1},p_{a}) 
u(p_{a}, \lambda_{a}) \nonumber \\
&& \times 
i\tilde{\Delta}^{(V)\, \mu_{1} \nu_{1}}(s_{1},t_{1}) \,
i\Gamma^{(VV f_{1})}_{\nu_{1} \nu_{2} \alpha}(q_{1},q_{2}) \,
i\tilde{\Delta}^{(V)\, \nu_{2}\mu_{2} }(s_{2},t_{2}) \nonumber \\
&& \times 
\bar{u}(p_{2}, \lambda_{2}) 
i\Gamma^{(V pp)}_{\mu_{2}}(p_{2},p_{b}) 
u(p_{b}, \lambda_{b}) \,.
\label{amplitude_f1_pompom}
\end{eqnarray}
Here $\epsilon_{\alpha}(\lambda)$ is the polarisation vector 
of the $f_{1}$ meson,
$\Gamma^{(V pp)}_{\mu}$ and $\Gamma^{(VV f_{1})}_{\nu_{1} \nu_{2} \alpha}$
are the $V pp$ and $VV f_{1}$ vertex functions, respectively,
and $\tilde{\Delta}^{(V)\, \mu \nu}$ is the propagator
for the reggeized vector meson $V$.
At very low energies the latter must be replaced 
by $\Delta^{(V)\, \mu \nu}$,
the standard propagator for the vector meson $V$.
We shall now discuss all these quantities in turn.

First we discuss the $V V f_{1}$ coupling.
We start by considering the on shell process
of two real vector particles $V$ fusing to give an $f_{1}$ meson:
\begin{equation}
V + V \to f_{1}\,.
\label{VV_to_f1}
\end{equation}
The angular momentum analysis of such reactions 
was made in \cite{Lebiedowicz:2013ika}.
The spins of the two vectors can be combined to a total spin
$S = 0, 1, 2$.
Then $S$ has to be combined with the orbital
angular momentum $l$ to give the spin $J = 1$
and parity $+1$ of the $f_{1}$ state.
From Table~8 of \cite{Lebiedowicz:2013ika}
we find that there is here only one possible coupling,
namely $(l,S) = (2,2)$.
A convenient corresponding coupling Lagrangian,
given in (D9) of \cite{Lebiedowicz:2020yre}, reads
\begin{eqnarray}
{\cal L}'_{VV f_{1}}(x) = 
\frac{1}{M_{0}^{4}}
g_{VV f_{1}}
\big( V_{\kappa \lambda}(x)
\twosidep{\mu} \twosidep{\nu} 
V_{\rho \sigma}(x) \big)
\big( \p_{\alpha} U_{\beta}(x) - \p_{\beta} U_{\alpha}(x) \big)
g^{\kappa \rho} g^{\mu \sigma} \varepsilon^{\lambda \nu \alpha \beta},\qquad
\label{VVA}
\end{eqnarray}
where 
\begin{equation}
V_{\kappa \lambda}(x) = 
\p_{\kappa} V_{\lambda}(x) - \p_{\lambda} V_{\kappa}(x)\,,
\label{C10}
\end{equation}
with $M_{0} \equiv 1$~GeV and $g_{VV f_{1}}$ a dimensionless coupling constant.
$U_{\alpha}(x)$ and $V_{\kappa}(x)$ are the fields of 
the $f_{1}$ meson and
the vector meson $V$, respectively.
For the Levi-Civita symbol we use the normalisation 
$\varepsilon_{0123} = +1$.

The expression for the $VV f_{1}$ vertex 
obtained from (\ref{VVA}) is as follows 
\newline
\hspace*{0.5cm}\includegraphics[width=120pt]{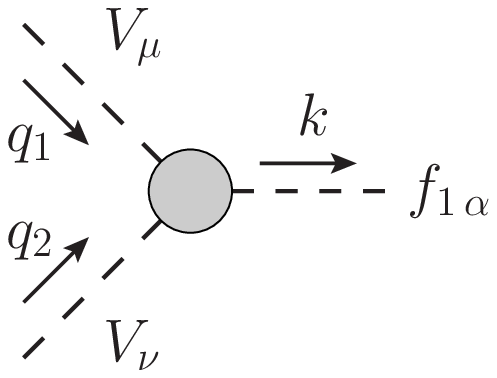}
\begin{eqnarray}
i\Gamma_{\mu \nu \alpha}^{(VV f_{1})}(q_{1},q_{2})\mid_{\rm{bare}} &=&
\frac{2 g_{VV f_{1}}}{M_{0}^{4}}
\bigl[ (q_{1}-q_{2})^{\rho}
(q_{1}-q_{2})^{\sigma}
\varepsilon_{\lambda \sigma \alpha \beta}\,
k^{\beta} \nonumber\\
&& \times 
( q_{1 \kappa} \,\delta^{\lambda}_{\;\;\mu} - q_{1}^{\lambda}\, g_{\kappa \mu})
       ( q_{2}^{\kappa} \,g_{\rho \nu} - q_{2 \rho} \,\delta^{\kappa}_{\;\;\nu} )
+ (q_{1} \leftrightarrow q_{2}, \mu \leftrightarrow \nu) \bigr];\qquad
\label{VVA_vertex}
\end{eqnarray}
see (D11) of \cite{Lebiedowicz:2020yre}.
Here the label ``bare'' is used for a vertex as derived 
from (\ref{VVA}) without a form-factor function.
The vertex function (\ref{VVA_vertex}) satisfies the relations
\begin{eqnarray}
&&\Gamma_{\mu \nu \alpha}^{(VV f_{1})}(q_{1},q_{2}) =
\Gamma_{\nu \mu \alpha}^{(VV f_{1})}(q_{2},q_{1})\,, 
\nonumber \\
&&\Gamma_{\mu \nu \alpha}^{(VV f_{1})}(q_{1},q_{2})\,
q_{1}^{\mu} = 0\,, \nonumber \\
&&\Gamma_{\mu \nu \alpha}^{(VV f_{1})}(q_{1},q_{2})\,
q_{2}^{\nu} = 0\,, \nonumber \\
&&\Gamma_{\mu \nu \alpha}^{(VV f_{1})}(q_{1},q_{2})\,
(q_{1}+q_{2})^{\alpha} = 0\,.
\label{C102}
\end{eqnarray}

For realistic applications we should multiply the 'bare' vertex
(\ref{VVA_vertex})
by a phenomenological cutoff function
(form factor)
$F_{VV f_{1}}$
which we take in the factorised ansatz
\begin{eqnarray}
F_{VV f_{1}}(q_{1}^{2},q_{2}^{2},k^{2}) = 
\tilde{F}_{V}(q_{1}^{2}) \tilde{F}_{V}(q_{2}^{2}) F_{f_{1}}(k^{2})\,.
\label{F_VVA}
\end{eqnarray}
We make the assumption that $\tilde{F}_{V}(t)$
is parametrized as
\begin{eqnarray}
\tilde{F}_{V}(q^{2}) = 
\frac{\Lambda_{V}^4}{\Lambda_{V}^4 + (q^{2} - m_{V}^2)^{2}}\,,
\label{ff_pow}
\end{eqnarray}
where the cutoff parameter $\Lambda_{V}$,
taken to be the same for both $\rho^{0}$ and $\omega$,
is a free parameter.
For the on-shell $V$ and $f_{1}$ mesons we have
$F_{VV f_{1}}(m_{V}^{2},m_{V}^{2},m_{f_{1}}^{2}) = 1$.

The vector-meson-(anti)proton vertex is \cite{Nakayama:1998zv}
\begin{eqnarray}
i\Gamma_{\mu}^{(V pp)}(p',p) = -i\Gamma_{\mu}^{(V \bar{p}\bar{p})}(p',p)
=-i g_{V pp}\, F_{V NN}(t)
\left[\gamma_{\mu} - i\frac{\kappa_{V}}{2m_{p}} \sigma_{\mu \nu} (p-p')^{\nu} 
\right]
\label{vertex_VNN}
\end{eqnarray}
with the tensor-to-vector coupling ratio, $\kappa_{V} = f_{V NN}/g_{V NN}$.
We use the following values for these coupling constants:
\begin{eqnarray}
g_{\rho pp} = 3.0\,, \quad
\kappa_{\rho} = 6.1\,, \quad
g_{\omega pp} = 9.0\,, \quad
\kappa_{\omega} = 0\,.
\label{Vpp_couplings}
\end{eqnarray}

We give a short discussion of values for the $\rho pp$
and $\omega pp$ coupling constants found in the literature.
For the the $\rho NN$ coupling constants
one finds $g_{\rho pp} = 2.63 - 3.36$ \cite{Hohler:1974ht,Nakayama:1999jx} and $\kappa_{\rho}$ 
is expected to be $\kappa_{\rho} = 6.1 \pm 0.2$ \cite{Mergell:1995bf}.
There is a considerable uncertainty 
in the $\omega NN$ coupling constants.
From Table~1 of \cite{Nakayama:1998zv} we see a broad
range of values:
$g_{\omega pp} \simeq 10$ to 21 and 
$\kappa_{\omega} \simeq -0.16$ to $+0.14$.
For example, in \cite{Mergell:1995bf} 
it was estimated $g_{\omega pp} = 20.86 \pm 0.25$
and $\kappa_{\omega}= -0.16 \pm 0.01$; see Table~3 of
\cite{Mergell:1995bf}.
Within the (full) Bonn potential 
\cite{Machleidt:1987hj}
values of $g_{\omega pp} = 15.85$ 
and $\kappa_{\omega} = 0$
are required
for a best fit to $NN$ scattering data.
In \cite{Janssen:1996kx} it was shown that
such a fairly large value of $g_{\omega pp}$
must be considered as an effective
coupling strength rather than as the intrinsic
$\omega NN$ coupling constant.
They found that the additional repulsion provided by
the correlated $\pi \rho$ exchange 
to the $NN$ interaction allows $g_{\omega NN}$ to be reduced
by about a factor 2, leading to an ``intrinsic''
$\omega NN$ coupling constant
which is more in line with the value one would
obtain from the SU(3) flavour symmetry considerations, 
$g_{\omega NN} = 3 \,g_{\rho NN} 
\cos(\Delta\theta_{V})$ \cite{Nakayama:1999jx},
where $\Delta\theta_{V} \simeq 3.7^{\circ}$ 
is the deviation from the ideal
$\omega$-$\phi$ mixing angle.
%and Appendix~A of \cite{Lebiedowicz:2019boz}.
The values of $g_{\omega pp} = 7.0-10.5$
and $\kappa_{\omega} \simeq 0$ were found to describe consistently
the $\pi N$ scattering and $\pi$ photoproduction
processes \cite{Sato:1996gk}.
The values of $g_{\omega pp} = 9.0$ 
and $\kappa_{\omega} = -0.5$ 
have been used in the analysis
of the $pp \to pp \omega$ reaction
%with the inclusion of the nucleon resonances
to reproduce the shape of the measured $\omega$ angular 
distribution; see Fig.~7 of \cite{Tsushima:2003fs}.
%and \cite{Nakayama:2000uf}.
It was shown \cite{Nakayama:2006ps}
that 
the energy dependence of the total cross section 
and the angular distribution for $pp \to pp \omega$ 
can be described rather reasonably even with a vanishing
$\kappa_{\omega}$ ($g_{\omega pp} = 9.0$, $\kappa_{\omega} = 0$); 
see Fig.~4 of \cite{Nakayama:2006ps}.
Finally we note that in \cite{Ewerz:2013kda}
the couplings of the $\omega_{\Reg}$ and $\rho_{\Reg}$ reggeons
to the proton were estimated from high-energy scattering data
and found as
\begin{eqnarray}
g_{\rho_{\Reg} pp} = 2.02 \quad {\rm and} \quad 
g_{\omega_{\Reg} pp} = 8.65\,;
\label{reggeon-proton_couplings}
\end{eqnarray}
see (3.60) and (3.62) of \cite{Ewerz:2013kda}.
Taking all these informations into account we think that
our choice (\ref{Vpp_couplings}) for the coupling constants
is quite reasonable.

The form factor $F_{VNN}(t)$ in (\ref{vertex_VNN}),
describing the $t$-dependence of the $V$-(anti)proton coupling,
can be parametrized as
\begin{eqnarray}
F_{VNN}(t) = 
\frac{\Lambda_{VNN}^{2}-m_{V}^{2}}{\Lambda_{VNN}^{2}-t}\,,
\label{F_V_M}
\end{eqnarray}
where $\Lambda_{VNN} > m_{V}$ and $t < 0$.
Please note that the form factor $F_{VNN}(t)$
is normalized to unity at $t = m_{V}^{2}$.
On the other hand, the reggeon-proton couplings
(\ref{reggeon-proton_couplings}) are defined for $t = 0$.
Since $F_{VNN}(0) < 1$ we expect that
$g_{\rho_{\Reg} pp} < g_{\rho pp}$ and
$g_{\omega_{\Reg} pp} < g_{\omega pp}$,
which is indeed the case; 
see (\ref{Vpp_couplings}) and (\ref{reggeon-proton_couplings}).

The coupling constant $g_{VV f_{1}}$
and cutoff parameters $\Lambda_{V}$ and $\Lambda_{VNN}$ 
should be adjusted to experimental data.
Examples are discussed in
Appendices~\ref{sec:appendixB} and \ref{sec:appendixC}.
There, the form factor $F_{VV f_{1}}$ (\ref{F_VVA}) 
is used for
$F_{f_{1}}(m_{f_{1}}^{2}) = 1$
and for different kinematic conditions of $\tilde{F}_{V}(q^{2})$ (\ref{ff_pow}),
that is, for spacelike ($q^{2} < 0$) and timelike ($q^{2} > 0$)
momentum transfers of the $V$ meson,
and also at $q^{2} = 0$.
In Appendix~\ref{sec:appendixB} we discuss
the radiative decays of the $f_{1}(1285)$ meson
in two ways $f_{1} \to \rho \gamma$ (\ref{1to2}) 
and $f_{1} \to (\rho^{0} \to \pi^{+} \pi^{-}) \gamma$ (\ref{1to3})
where we have
$F_{\rho \rho f_{1}}(m_{\rho}^{2},0,m_{f_{1}}^{2})$ and
$F_{\rho \rho f_{1}}(q^{2} > 0,0,m_{f_{1}}^{2})$, respectively.
In Table~\ref{tab:table1} in Appendix~\ref{sec:appendixB}
we collect our results for $g_{\rho \rho f_{1}}$
extracted from the decay rate of $f_1 \to \rho^0 \gamma$ 
using the VMD ansatz.
The process $f_1 \to \rho^0 \rho^0 \to 2 \pi^+ 2 \pi^-$,
where both $\rho^{0}$ mesons carry
timelike momentum transfers,
will be studied in detail in \cite{LNS:2020}.
For the $\gamma p \to f_{1} p$ reaction, 
discussed in Appendix~\ref{sec:appendixC}, we have
$F_{\rho \rho f_{1}}(0,q^{2} < 0,m_{f_{1}}^{2})$.
This is closer to the $VV \to f_{1}$ fusion mechanisms 
shown in Fig.~\ref{fig:diagrams} where 
both $V$ mesons have spacelike momentum transfers.
From comparison of the model 
to the $f_{1}$-meson angular distributions
of the CLAS experimental data \cite{Dickson:2016gwc}
we shall extract the cutoff parameter $\Lambda_{VNN}$
in the $V$-proton vertex (\ref{F_V_M});
see (\ref{aLam0.65})--(\ref{aLam2.0})
and Fig.~\ref{fig:B1} in Appendix~\ref{sec:appendixC}.

In the following we shall use the $VV f_{1}$ coupling (\ref{VVA})
and the corresponding vertex (\ref{VVA_vertex})--(\ref{ff_pow}) 
for our $VV \to f_{1}$ fusion processes of Fig.~\ref{fig:diagrams} 
for both:
normal off-shell vector mesons $V$
and reggeized vector mesons $V_R$.

The standard form of the vector-meson propagator 
is given e.g. in (3.2) of \cite{Ewerz:2013kda}
\begin{eqnarray}
i\Delta_{\mu \nu}^{(V)}(k) = 
i \left(-g_{\mu \nu} + \frac{k_{\mu} k_{\nu}}{k^{2} + i \epsilon} \right) 
\Delta_{T}^{(V)}(k^{2}) 
- i \frac{k_{\mu} k_{\nu}}{k^{2} + i \epsilon}
\Delta_{L}^{(V)}(k^{2})\,. 
\label{vec_mes_prop}
\end{eqnarray}
With the relations from (\ref{C102}) 
for the $VV f_{1}$ vertex (\ref{VVA_vertex})
the $k_{\mu}k_{\nu}$ term does not contribute in (\ref{amplitude_f1_pompom}).
For small values of $s_{1,2}$
and $|t_{1,2}|$ [see (\ref{2to3_kinematic})]
the simplest form of the transverse function,
$\Delta_{T}^{(V)}(t) = (t - m_{V}^{2})^{-1}$,
should be adequate.
For higher values of $s_{1}$ and $s_{2}$ 
we must take into account reggeization.
We do this, following (3.21), (3.24) 
of \cite{Lebiedowicz:2019jru},
by making in (\ref{vec_mes_prop})
the replacements
\begin{eqnarray}
\Delta_{T}^{(V)}(t_{i}) \to 
\tilde{\Delta}_{T}^{(V)}(s_{i},t_{i})=
\Delta_{T}^{(V)}(t_{i})
\left( \exp (i \phi(s_{i}))\,\frac{s_{i}}{s_{\rm thr}} 
%\alpha'_{V}  
\right)^{\alpha_{V}(t_{i})-1}\,,
\label{reggeization_2}
\end{eqnarray}
where 
\begin{eqnarray}
\phi(s_{i}) =
\frac{\pi}{2}\exp\left(\frac{s_{\rm thr}-s_{i}}{s_{{\rm thr}}}\right)-\frac{\pi}{2}\,,
\label{reggeization_aux}
\end{eqnarray}
for $i = 1$ or $2$,
and $s_{\rm thr}$ is the lowest value of $s_{i}$ 
possible here:
\begin{eqnarray}
s_{\rm thr} = (m_p+m_{f_{1}})^2\,. 
\label{sthr}
\end{eqnarray}
We use the standard linear form for the vector meson Regge trajectories
(cf., e.g., \cite{Donnachie:2002en})
\begin{eqnarray}
&&\alpha_{V}(t) = \alpha_{V}(0)+\alpha'_{V}\,t\,,\\ 
&&\alpha_{V}(0) = 0.5\,, \;\;
  \alpha'_{V} = 0.9 \; \mathrm{GeV}^{-2}\,.
\label{trajectory}
\end{eqnarray}

Our reggeized vector meson propagator, denoted by
$\tilde{\Delta}_{\mu \nu}^{(V)}(s,t)$
is obtained from (\ref{vec_mes_prop})
with the replacement
$\Delta_{T}^{(V)} \to \tilde{\Delta}_{T}^{(V)}$
from (\ref{reggeization_2}).

In the following we shall also consider 
the CEP of the $f_{1}(1285)$ with subsequent
decay into $\rho^{0} \gamma$:
\begin{eqnarray}
p(p_{a},\lambda_{a}) + p(p_{b},\lambda_{b}) \to
p(p_{1},\lambda_{1}) + 
[f_{1}(p_{34}) \to \rho^{0}(p_{3},\lambda_{3}) + \gamma(p_{4},\lambda_{4})]
+ p(p_{2},\lambda_{2}) \qquad \quad
\label{2to4_reaction_f1}
\end{eqnarray}
with $p_{34} = p_{3} + p_{4}$.
Here
$p_{3}, p_{4}$ and $\lambda_{3} = 0, \pm 1$, $\lambda_{4} = \pm 1$
denote the four-momenta and helicities of the $\rho^{0}$ meson
and the photon, respectively.

The amplitude for the reaction (\ref{2to4_reaction_f1})
can be written as in (\ref{amplitude_f1_pompom}) but
with the replacements
\begin{eqnarray}
s_{1} &\to& \tilde{s}_{1} = (p_{1} + p_{34})^{2}\,,\nonumber \\
s_{2} &\to& \tilde{s}_{2} = (p_{2} + p_{34})^{2}\,,\nonumber \\
%s_{\rm thr} &\to& \tilde{s}_{\rm thr} = (m_p + m_{\rho})^2\,,\nonumber \\
(\epsilon^{(f_{1})}_{\alpha}(\lambda_{f_{1}}))^* &\to&
\frac{e}{\gamma_{\rho}}
%i\Delta^{(f_{1})\,\alpha \alpha'}(p_{34})\, 
%i\Gamma^{(VV f_{1})}_{\rho \sigma \alpha'}(-p_{3},-p_{4})\,
\Delta_{T}^{(f_{1})}(p_{34}^{2})\,
\Gamma^{(\rho\rho f_{1})}_{\rho \sigma \alpha}(-p_{3},-p_{4})\,
(\epsilon^{(\rho)\,\rho }(\lambda_{3}))^*
(\epsilon^{(\gamma)\,\sigma}(\lambda_{4}))^*\,.
\label{2to4_replacement}
\end{eqnarray}
Here $\epsilon^{(\rho)}$ and $\epsilon^{(\gamma)}$
are the polarisation vectors of $\rho^{0}$ and $\gamma$, respectively,
and $\Delta_{T}^{(f_{1})}$ is the transverse part
of the $f_{1}$ propagator which has a structure analogous
to (\ref{vec_mes_prop}).
The factor $e/\gamma_{\rho}$ comes from the $\rho$-$\gamma$
transition vertex; see (3.23)--(3.25) of \cite{Ewerz:2013kda}.

In practical calculations we introduce 
in the $\rho \rho f_{1}$ vertex the form factor
$F_{f_{1}}(p_{34}^{2})$ [see~(\ref{F_VVA})]
for the virtual $f_{1}$ meson
\begin{eqnarray}
F_{f_{1}}(p_{34}^{2}) =
\exp{ \left( \frac{-(p_{34}^{2}-m_{f_{1}}^{2})^{2}}{\Lambda_{f_{1}}^{4}} \right)}\,,
\quad \Lambda_{f_{1}} = 1\;{\rm GeV}\,.
\label{F_f1_ff}
\end{eqnarray}
In (\ref{2to4_replacement}) we shall use a simple Breit-Wigner ansatz
for the $f_{1}$ meson propagator
\begin{eqnarray}
\Delta_{T}^{(f_{1})}(p_{34}^{2}) = 
\frac{1}{p_{34}^{2} - m_{f_{1}}^{2} + i m_{f_{1}} \Gamma_{f_{1}}}\,.
\label{P_f1}
\end{eqnarray}
The mass and total width of $f_{1}$ meson
from \cite{Zyla:2020zbs} are
\begin{eqnarray}
m_{f_{1}} &=& (1281.9 \pm 0.5) \;\,{\rm MeV}\,, 
\label{f1_PDG_mass}\\
\Gamma_{f_{1}} &=& (22.7 \pm 1.1) \;\,{\rm MeV}\,.
\label{f1_PDG}
\end{eqnarray}
We note that the mass of $1281.0 \pm 0.8$~MeV 
measured in the CLAS experiment \cite{Dickson:2016gwc}
is in very good agreement with the PDG average 
value (\ref{f1_PDG_mass}). 
The total width measured by the CLAS Collaboration is however smaller than the value (\ref{f1_PDG}):
\begin{eqnarray}
%m_{f_{1}} &=& (1281.0 \pm 0.8) \;\,{\rm MeV}\,, \nonumber \\
\Gamma_{f_{1}} &=& (18.4 \pm 1.4) \;\,{\rm MeV}\,.
\label{f1_CLAS}
\end{eqnarray}

For the proton-antiproton collisions we can write
\begin{eqnarray}
{\cal M}_{p\bar{p} \to p\bar{p} f_{1}(1285)} = 
{\cal M}_{p\bar{p} \to p\bar{p} f_{1}(1285)}^{(\omega \omega \; {\rm fusion})} +
{\cal M}_{p\bar{p} \to p\bar{p} f_{1}(1285)}^{(\rho \rho \; {\rm fusion})} \,.
\label{sum_amplitudes_ppbar_f1}
\end{eqnarray}
Then the corresponding amplitudes
are as in (\ref{amplitude_f1_pompom})
but with the replacement
\begin{eqnarray}
\bar{u}(p_{2}, \lambda_{2}) 
i\Gamma^{(V pp)}_{\mu_{2}}(p_{2},p_{b}) 
u(p_{b}, \lambda_{b})
& \to &
\bar{v}(p_{b}, \lambda_{b})
i\Gamma^{(V \bar{p} \bar{p})}_{\mu_{2}}(p_{2},p_{b}) 
v(p_{2}, \lambda_{2})
\nonumber \\
&& =
-\bar{u}(p_{2}, \lambda_{2}) 
i\Gamma^{(V pp)}_{\mu_{2}}(p_{2},p_{b}) 
u(p_{b}, \lambda_{b}) \,.
\label{repl1_ppbar_f1}
\end{eqnarray}
Using the $V$-(anti)proton coupling (\ref{vertex_VNN})
in the $VV$-fusion amplitudes
we obtain
\begin{eqnarray}
{\cal M}_{p\bar{p} \to p\bar{p} f_{1}(1285)}^{(VV  \; {\rm fusion})} = 
- {\cal M}_{pp \to pp f_{1}(1285)}^{(VV  \; {\rm fusion})}\,.
\label{amp_sign_ppbar_f1}
\end{eqnarray}
%

%----------------------------------------------------------------------
\subsection{Background processes to the $\rho \rho$ and $\rho \gamma$ channels of the $f_{1}$ decay in CEP}
\label{sec:bakground}
%----------------------------------------------------------------------

The main decay modes of the $f_{1}(1285)$ are \cite{Zyla:2020zbs}
$4 \pi$, $\eta \pi \pi$, $K \bar{K} \pi$, and $\rho^{0} \gamma$.
If the $f_{1}$ is to be identified and measured in CEP
in any one of these channels one will have to consider background processes
giving the same final state, for instance, $pp 4\pi$.
Therefore, in this section we discuss two background reactions:
CEP of $4 \pi$ via $\rho^{0} \rho^{0}$ in the continuum
and CEP of $\rho^{0} \gamma$ in the continuum.

\begin{figure}[!ht]
\includegraphics[width=6.cm]{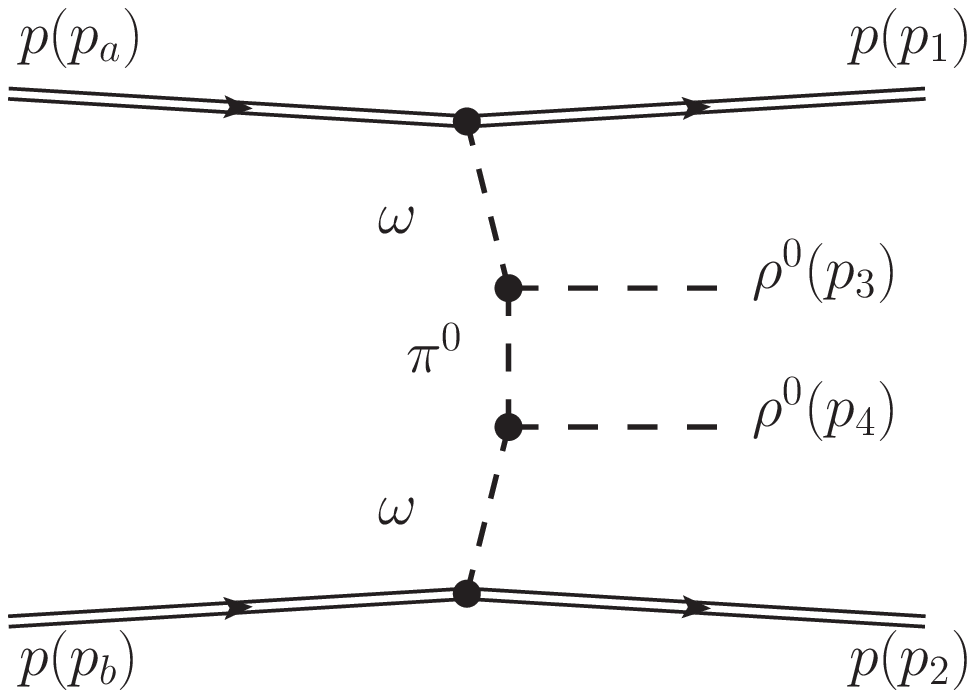}
\includegraphics[width=6.cm]{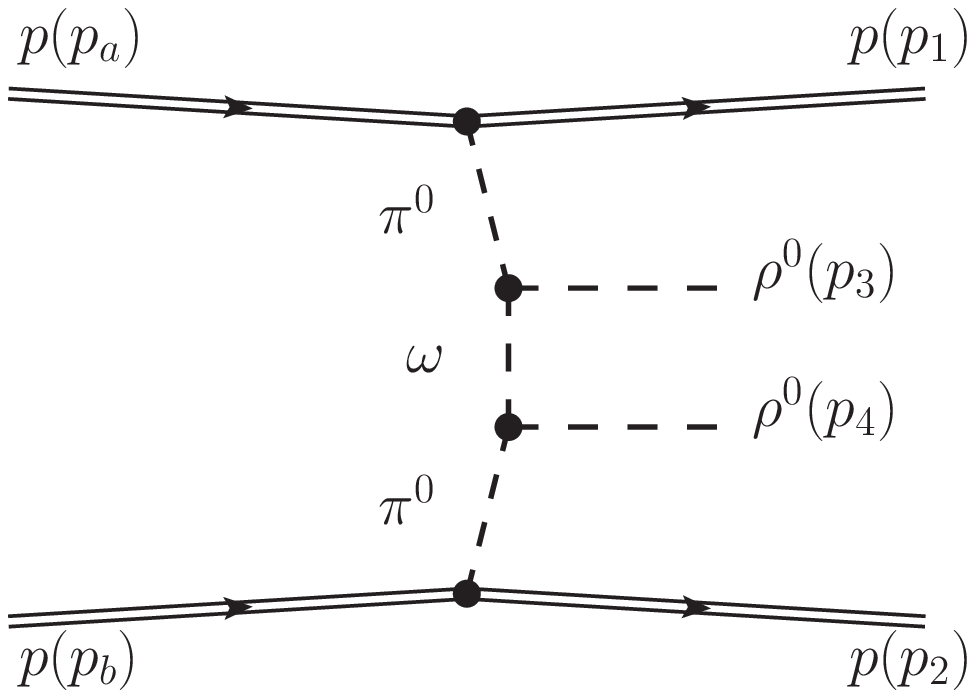}
\caption{
Diagrams for exclusive continuum $\rho^{0} \rho^{0}$ production 
in proton-proton collisions. There are also the diagrams 
with $p_{3} \leftrightarrow p_{4}$.}
\label{fig:diagrams_rhorho}
\end{figure}

First we discuss the exclusive production of 
$\rho^{0} \rho^{0}$ in proton-proton collisions,
\begin{eqnarray}
p(p_{a},\lambda_{a}) + p(p_{b},\lambda_{b}) \to
p(p_{1},\lambda_{1}) + \rho^{0}(p_{3},\lambda_{3}) 
+ \rho^{0}(p_{4},\lambda_{4}) + p(p_{2},\lambda_{2}) \,,
\label{2to4_reaction_rhorho}
\end{eqnarray}
where 
$p_{3,4}$ and $\lambda_{3,4} = 0, \pm 1$ 
denote the four-momenta and helicities of the $\rho^{0}$ mesons, respectively.
In Fig.~\ref{fig:diagrams_rhorho} we show the diagrams
for two mechanisms which will contribute 
to the reaction (\ref{2to4_reaction_rhorho}) at low energies,
$\omega \omega$ and $\pi^{0} \pi^{0}$ fusion.
%-------------------------------------------------------------
%-------------------------------------------------------------

There can also be $\rho\rho$ fusion 
with exchange of an intermediate $\sigma \equiv f_{0}(500)$ meson
and the $\sigma \sigma$ fusion with $\rho^{0}$ exchange.
From the Bonn potential \cite{Machleidt:1987hj,Oh:2003aw}
we get for the squared coupling constant 
$g_{\sigma pp}^{2}/4\pi \simeq 6.0$ 
which is smaller than $g_{\pi pp}^{2}/4\pi \simeq 14.0$.
Moreover, we can expect that 
$|g_{\sigma \rho \rho}| \ll |g_{\rho \omega \pi}|$.
Due to large form-factor uncertainties and the poorly known
$\sigma \rho \rho$ coupling
we neglect these contributions in our present study.
Other contributions may be due to the exchanges of
the $f_{2}(1270)$ meson ($f_2$-$\rho^0$-$f_2$ 
or $\rho^0$-$f_2$-$\rho^0$) and 
the neutral $a_{2}(1320)$ meson ($a_2$-$\omega$-$a_2$ 
or $\omega$-$a_2$-$\omega$).
For the $f_{2} \rho \rho$ and $a_{2} \omega \rho$ couplings 
one could use the rather well known
couplings from (3.55), (3.56), (7.29)--(7.34)
and (3.57), (3.58), (7.38)--(7.43) of \cite{Ewerz:2013kda}, respectively.
Since the $f_{2} pp$ coupling, taking it equal to $f_{2 \Reg} pp$
from (3.49), (3.50) of \cite{Ewerz:2013kda}, is rather large,
the $f_2$-$\rho^0$-$f_2$ fusion may give a large background contribution.
Since $g_{\omega_{\Reg} pp} > g_{a_{2 \Reg} pp}$,
see (3.52) and (3.60) of \cite{Ewerz:2013kda},
and
the $a_{2} \omega \rho$ couplings have values similar
to the $f_{2} \rho \rho$ couplings the $\omega$-$a_2$-$\omega$
contribution may also be potentially important.
However, we expect that the tensor meson propagator(s) 
will reduce the cross section for these processes.

At higher energies the pomeron plus $f_{2}$ reggeon
($\Pom$ + $f_{2 \Reg}$) fusion
[($\Pom$ + $f_{2 \Reg}$)-$\rho^0$-($\Pom$ + $f_{2 \Reg}$)]
and $\rho^{0}$ fusion with $\Pom$ + $f_{2 \Reg}$ exchange
[$\rho^0$-($\Pom$ + $f_{2 \Reg}$)-$\rho^0$] will be important,
probably the dominant processes; see
\cite{Lebiedowicz:2016zka}.
We expect that these processes will give only a small contribution
in the threshold region, of interest for us here.
Therefore, we shall neglect also these mechanisms in the following.

With the assumption, motivated above, 
that the diagrams of Fig.~\ref{fig:diagrams_rhorho}
represent the dominant reaction mechanisms in the threshold region, 
the continuum amplitude for the reaction (\ref{2to4_reaction_rhorho}) 
can be written as
\begin{eqnarray}
{\cal M}_{pp \to pp \rho^{0} \rho^{0}}^{(\rho\rho \;{\rm continuum})} =
{\cal M}_{pp \to pp \rho^{0} \rho^{0}}^{(\omega \omega  \; {\rm fusion})} +
{\cal M}_{pp \to pp \rho^{0} \rho^{0}}^{(\pi\pi  \; {\rm fusion})}\,.
\label{pp_rhorho}
\end{eqnarray}

The $\omega \omega$- and $\pi \pi$-fusion amplitudes 
(\ref{pp_rhorho})
are given by
\begin{eqnarray}
&&{\cal M}_{\lambda_{a}\lambda_{b}\to\lambda_{1}\lambda_{2}\lambda_{3}\lambda_{4}}^{(\omega \omega \; {\rm fusion})}
= (-i) 
(\epsilon^{(\rho)\rho_{3}}(\lambda_{3}))^*
(\epsilon^{(\rho)\rho_{4}}(\lambda_{4}))^*
\nonumber \\
&&\qquad \times 
\bar{u}(p_{1}, \lambda_{1}) 
i\Gamma^{(\omega pp)}_{\mu_{1}}(p_{1},p_{a}) 
u(p_{a}, \lambda_{a})\nonumber \\
&&
\qquad \left. \times 
\bigl[
i\tilde{\Delta}^{(\omega)\, \mu_{1} \nu_{1}}(s_{13},t_{1})\,
i\Gamma^{(\rho \omega \pi)}_{\rho_{3} \nu_{1}}(p_{3},q_{1})\,
i\Delta^{(\pi)}(p_{t})\,
i\Gamma^{(\rho \omega \pi)}_{\rho_{4} \nu_{2}}(p_{4},q_{2})\,
i\tilde{\Delta}^{(\omega)\,\nu_{2} \mu_{2} }(s_{24},t_{2})
\right.
\nonumber \\
&&\qquad \left. + \;\;
i\tilde{\Delta}^{(\omega)\, \mu_{1} \nu_{1}}(s_{14},t_{1})\,
i\Gamma^{(\rho \omega \pi)}_{\rho_{4} \nu_{1}}(p_{4},q_{1})\,
i\Delta^{(\pi)}(p_{u})\,
i\Gamma^{(\rho \omega \pi)}_{\rho_{3} \nu_{2}}(p_{3},q_{2})\,
i\tilde{\Delta}^{(\omega)\,\nu_{2} \mu_{2} }(s_{23},t_{2}) \bigr]
\right.
\nonumber \\
&&\qquad
\times 
\bar{u}(p_{2}, \lambda_{2}) 
i\Gamma^{(\omega pp)}_{\mu_{2}}(p_{2},p_{b}) 
u(p_{b}, \lambda_{b}) \,,
\label{amplitude_omeome}
\end{eqnarray}
\begin{eqnarray}
&&{\cal M}_{\lambda_{a}\lambda_{b}\to\lambda_{1}\lambda_{2}\lambda_{3}\lambda_{4}}^{(\pi\pi \; {\rm fusion})}
= (-i) 
(\epsilon^{(\rho)\rho_{3}}(\lambda_{3}))^*
(\epsilon^{(\rho)\rho_{4}}(\lambda_{4}))^*
\nonumber \\
&&\qquad \times 
\bar{u}(p_{1}, \lambda_{1}) 
i\Gamma^{(\pi pp)}(p_{1},p_{a}) 
u(p_{a}, \lambda_{a})\nonumber \\
&&
\qquad \left. \times 
\bigl[
i\Delta^{(\pi)}(q_{1})\,
i\Gamma^{(\rho \omega \pi)}_{\rho_{3} \nu_{1}}(p_{3},-p_{t})\,
i\tilde{\Delta}^{(\omega)\nu_{1}\nu_{2}}(s_{34},p_{t}^{2})\,
i\Gamma^{(\rho \omega \pi)}_{\rho_{4} \nu_{2}}(p_{4},p_{t})\,
i\Delta^{(\pi)}(q_{2})
\right.
\nonumber \\
&&\qquad \left. + \;\;
i\Delta^{(\pi)}(q_{1})\,
i\Gamma^{(\rho \omega \pi)}_{\rho_{4} \nu_{1}}(p_{4},p_{u})\,
i\tilde{\Delta}^{(\omega)\nu_{1}\nu_{2}}(s_{34},p_{u}^{2})\,
i\Gamma^{(\rho \omega \pi)}_{\rho_{3} \nu_{2}}(p_{3},-p_{u})\,
i\Delta^{(\pi)}(q_{2}) \bigr]
\right.
\nonumber \\
&&\qquad
\times 
\bar{u}(p_{2}, \lambda_{2}) 
i\Gamma^{(\pi pp)}(p_{2},p_{b}) 
u(p_{b}, \lambda_{b}) \,,
\label{amplitude_pipi}
\end{eqnarray}
where $s_{ij} = (p_{i} + p_{j})^{2}$,
$p_{t} = p_{a} - p_{1} - p_{3}$,
$p_{u} = p_{4} - p_{a} + p_{1}$.
In the formulas above the $\epsilon^{(\rho)}_{\mu}$'s denote
the polarisation vectors of the outgoing $\rho^{0}$ mesons.
The standard pion propagator
$i\Delta^{(\pi)}(k) = i/(k^{2} - m_{\pi}^{2})$ is used in the calculations.
The reggeized vector meson propagator, denoted by
$\tilde{\Delta}_{\mu \nu}^{(V)}(s_{ij},t_{i})$
is obtained from (\ref{vec_mes_prop})
with the replacement
$\Delta_{T}^{(V)} \to \tilde{\Delta}_{T}^{(V)}$
from (\ref{reggeization_2}) and (\ref{reggeization_aux})
and with the relevant $s_{ij}$, $s_{\rm thr}$, 
and $t_{i}$, the four-momentum transfer squared,
in the $p \rho^{0}$ and $\rho^{0} \rho^{0}$ subsystems.

With $k', \mu$ and $k,\nu$ the four-momentum and vector index
of the outgoing $\rho^{0}$ and incoming $\omega$ meson, respectively,
and $k'-k$ the four-momentum of the pion 
the $\rho \omega \pi$ vertex,
including form factor, reads\footnote{The effective
Lagrangian is as given in (1) of \cite{Nakayama:2006ps}
taking into account that we use the opposite sign convention
for $\varepsilon_{\mu \nu \rho \sigma}$ 
($\varepsilon_{0123} = +1$).}
\newline
\hspace*{0.5cm}\includegraphics[width=145pt]{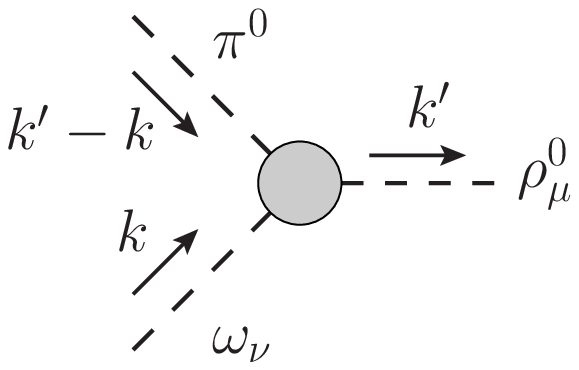}
\begin{eqnarray}
i\Gamma_{\mu \nu}^{(\rho \omega \pi)}(k',k) 
= -i \,\frac{g_{\rho \omega \pi}}{\sqrt{m_{\rho} m_{\omega}}}\, 
\varepsilon_{\mu \nu \rho \sigma} k'^{\rho} k^{\sigma}\,
F(k'^{2},k^{2},(k'-k)^{2})\,,
\label{rho_ome_pi}
\end{eqnarray}
where $g_{\rho \omega \pi} \simeq \pm 10$
\cite{Nakayama:1998zv,Nakayama:1999jx,Nakayama:2006ps}.
We note that the value of $g_{\rho \omega \pi} = +10$,
has been extracted in \cite{Nakayama:1999jx} 
from the measured $\omega \to \pi^{0} \gamma$ 
radiative decay rate 
and the positive sign from the analysis of pion photoproduction
reaction in conjunction with the VMD assumption.
%The sign of this coupling constant is determined by the sign
%of the $\pi \omega \gamma$ coupling constant which, in turn,
%has been fixed from the analysis of pion photoproduction
%in the 1~GeV energy region.
In \cite{Nakayama:2006ps} it was found that
the data for the reaction $pp \to pp \omega$
strongly favour a negative sign of the coupling constant 
$g_{\rho \omega \pi}$.
In our case, the sign of $g_{\rho \omega \pi}$
does not matter as this coupling occurs twice in the amplitudes
(\ref{amplitude_omeome}) and (\ref{amplitude_pipi}).

We use a factorized ansatz for the form factor
\begin{eqnarray}
F(k'^{2},k^{2},(k'-k)^{2}) = 
F_{\rho}(k'^{2})F_{\omega}(k^{2})F_{\pi}((k'-k)^{2})\,.
\label{rho_ome_pi_ff}
\end{eqnarray}
The form factor (\ref{rho_ome_pi_ff}) should be normalised as 
$F(0,m_{\omega}^{2},m_{\pi}^{2}) = 1$, 
consistent with the kinematics at which the coupling constant 
$g_{\rho \omega \pi}$ is determined.
This is the $\omega \to \pi^{0} \gamma$ reaction where
$\omega$ and $\pi^{0}$ are on shell and the virtual $\rho^{0}$
which gives the $\gamma$ has mass zero.
Following \cite{Nakayama:1999jx} we take
\begin{eqnarray}
&&F_{V}(t) = 
\frac{\Lambda_{MV}^{2}-x m_{V}^{2}}{\Lambda_{MV}^{2}-t}\,,
\label{F_V}\\
&&F_{\pi}(t) = 
\frac{\Lambda_{M\pi}^{2}-m_{\pi}^{2}}{\Lambda_{M\pi}^{2}-t}\,,
\label{F_pi}
\end{eqnarray}
where $x = 0$ for $V = \rho$ and 
$x = 1$ for $V = \omega$.
In this way $F_{\rho}(t)$ in (\ref{F_V}) is normalised at $t = 0$
and $F_{\omega}(t)$ at $t = m_{\omega}^{2}$.
We assume for the cutoff parameters that
they are equal to a common value
$\Lambda_{M} \equiv \Lambda_{M \omega} 
= \Lambda_{M \rho} = \Lambda_{M \pi}$.
Following \cite{Nakayama:1999jx} 
we take $\Lambda_{M} = 1.45$~GeV.
Smaller values of the cutoff parameters,
$\Lambda_{M \rho} = \Lambda_{M \pi} = 1.0$~GeV,
were used in \cite{Tsushima:2003fs} (see Table~I there).
Also a dipole form factor $F_{V}(t)$ in (\ref{F_V})
was considered; 
see \cite{Nakayama:2000uf,Kaptari:2004sd} and 
Table~II of \cite{Tsushima:2003fs}.

For the $\pi^{0}$-(anti)proton vertex we have 
[see (3.4) of \cite{Lebiedowicz:2016ryp}]
\begin{eqnarray}
i\Gamma^{(\pi pp)}(p',p) = -i\Gamma^{(\pi \bar{p} \bar{p})}(p',p) 
= -\gamma_{5} g_{\pi pp} \, F_{\pi NN}((p'-p)^{2})\,.
\label{PSNN}
\end{eqnarray}
We take $g_{\pi pp} = \sqrt{4 \pi \times 14.0}$ 
and the form factor $F_{\pi NN}(t)$ as in (\ref{F_pi}) 
with the replacement 
$\Lambda_{M \pi} \to \Lambda_{\pi NN}$.
We take $\Lambda_{\pi NN} = 1.0$~GeV;
see the discussion in \cite{Janssen:1996kx,Nakayama:1998zv}.
%From the Bonn model $\Lambda_{\pi NN} = 1.3$~GeV; 
%see Table~4 of \cite{Machleidt:1987hj}.

Likewise, the monopole form factor (\ref{F_V_M})
in the $V pp$ vertex (\ref{vertex_VNN})
is assumed with the cutoff parameter
$\Lambda_{V NN}$.
We take $\Lambda_{V NN} = 0.9$~GeV and $1.35$~GeV 
in accordance with (\ref{aLam1.0}) and (\ref{aLam0.65}), respectively.

Taking into account the statistical factor 
$\frac{1}{2}$ due to the identity
of the two $\rho^{0}$ mesons in (\ref{2to4_reaction_rhorho})
we get for the amplitude squared
\begin{eqnarray}
\frac{1}{2}
\abs{{\cal M}_{pp \to pp \rho^{0} \rho^{0}}}^{2}
&=&\frac{1}{2} \frac{1}{4} \sum_{\rm{spins}}
\abs{{\cal M}_{\lambda_{a}\lambda_{b}\to\lambda_{1}\lambda_{2}\lambda_{3}\lambda_{4}}}^{2} \nonumber\\
&=&\frac{1}{8} \sum_{\rm{spins}}
\left({\cal M}_{\lambda_{a}\lambda_{b}\to\lambda_{1}\lambda_{2}\lambda_{3}\lambda_{4}}\right)^{*}
{\cal M}_{\lambda_{a}\lambda_{b}\to\lambda_{1}\lambda_{2}\lambda_{3}\lambda_{4}}\,.
\label{amplitude_squared_rhorho}
\end{eqnarray}

Now we discuss the proton-antiproton collisions.
Here the amplitudes of the $\rho\rho$ continuum 
via $\omega \omega$ and $\pi \pi$ fusion
can be treated as in (\ref{amplitude_omeome}) and (\ref{amplitude_pipi}) 
but with the replacements (\ref{repl1_ppbar_f1}) and 
\begin{eqnarray}
\bar{u}(p_{2}, \lambda_{2}) 
i\Gamma^{(\pi pp)}(p_{2},p_{b}) 
u(p_{b}, \lambda_{b})
& \to &
\bar{v}(p_{b}, \lambda_{b})
i\Gamma^{(\pi \bar{p} \bar{p} )}(p_{2},p_{b}) 
v(p_{2}, \lambda_{2})
\nonumber \\
&& = \bar{u}(p_{2}, \lambda_{2}) 
i\Gamma^{(\pi pp)}(p_{2},p_{b}) 
u(p_{b}, \lambda_{b})
\,,
\label{repl1}
\end{eqnarray}
respectively.
Using (\ref{repl1_ppbar_f1}) and (\ref{repl1}) we obtain
\begin{eqnarray}
&&{\cal M}_{p\bar{p} \to p\bar{p} \rho^{0} \rho^{0}}^{(\omega \omega  \; {\rm fusion})} = -
{\cal M}_{pp \to pp \rho^{0} \rho^{0}}^{(\omega \omega  \; {\rm fusion})}\,,
\label{amp_sign} \\
&&{\cal M}_{p\bar{p} \to p\bar{p} \rho^{0} \rho^{0}}^{(\pi \pi  \; {\rm fusion})} =
{\cal M}_{pp \to pp \rho^{0} \rho^{0}}^{(\pi \pi  \; {\rm fusion})}\,.
\label{amp_sign_pipi}
\end{eqnarray}

As will be discussed in the following, from 
the $\pi^+ \pi^-\pi^+ \pi^-$ channel it may be
rather difficult to extract the $f_{1}(1285)$ signal.
Another decay channel worth considering is $\rho^0 \gamma$.

Therefore, now we discuss the exclusive production of the
$\rho^{0} \gamma$ continuum in proton-proton collisions,
\begin{eqnarray}
p(p_{a},\lambda_{a}) + p(p_{b},\lambda_{b}) \to
p(p_{1},\lambda_{1}) + \rho^{0}(p_{3},\lambda_{3}) + \gamma(p_{4},\lambda_{4}) + p(p_{2},\lambda_{2})
\label{2to4_reaction_rhogamma}
\end{eqnarray}
with
$p_{4}$ and $\lambda_{4} = \pm 1$ 
the four-momentum and helicities of the photon.

%-------------------------------------------------------------
\begin{figure}[!ht]
(a)\includegraphics[width=6.cm]{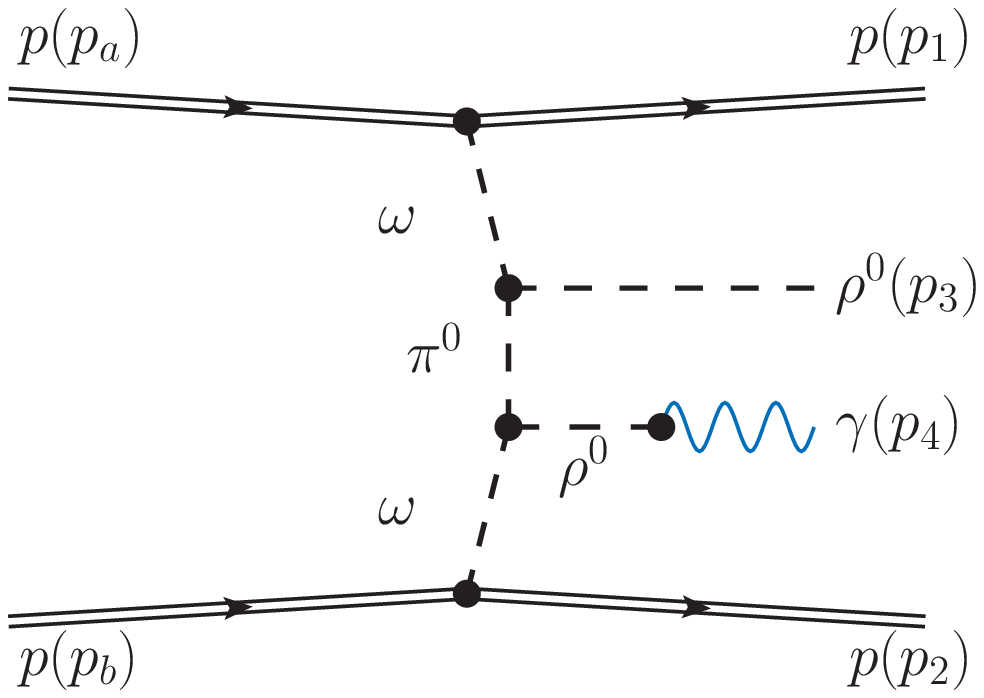}
   \includegraphics[width=6.cm]{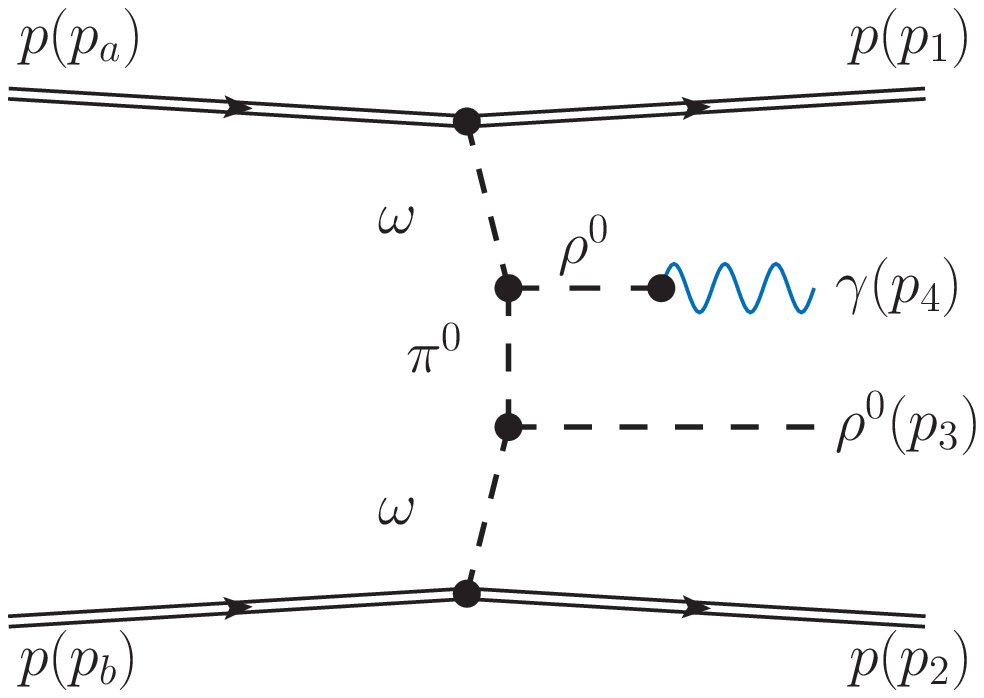}
(b)\includegraphics[width=6.cm]{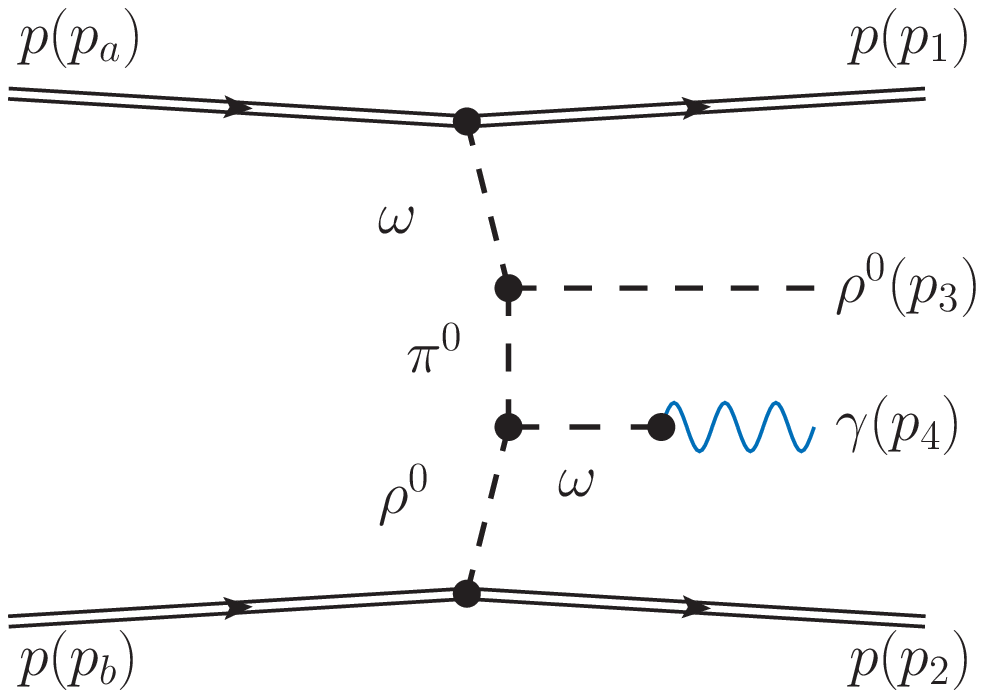}
   \includegraphics[width=6.cm]{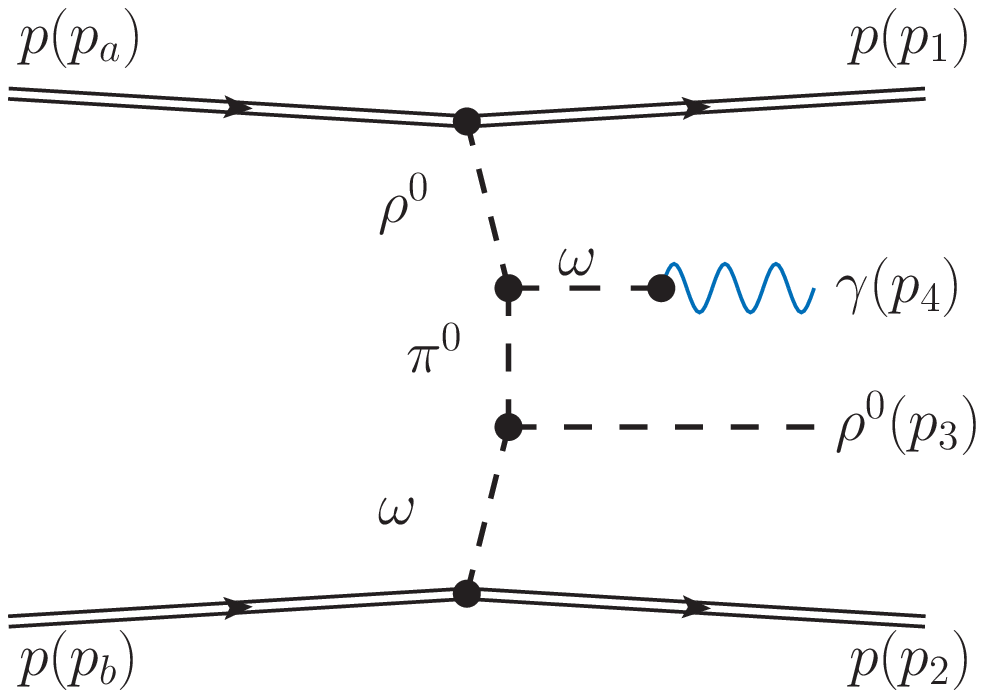}
(c)\includegraphics[width=6.cm]{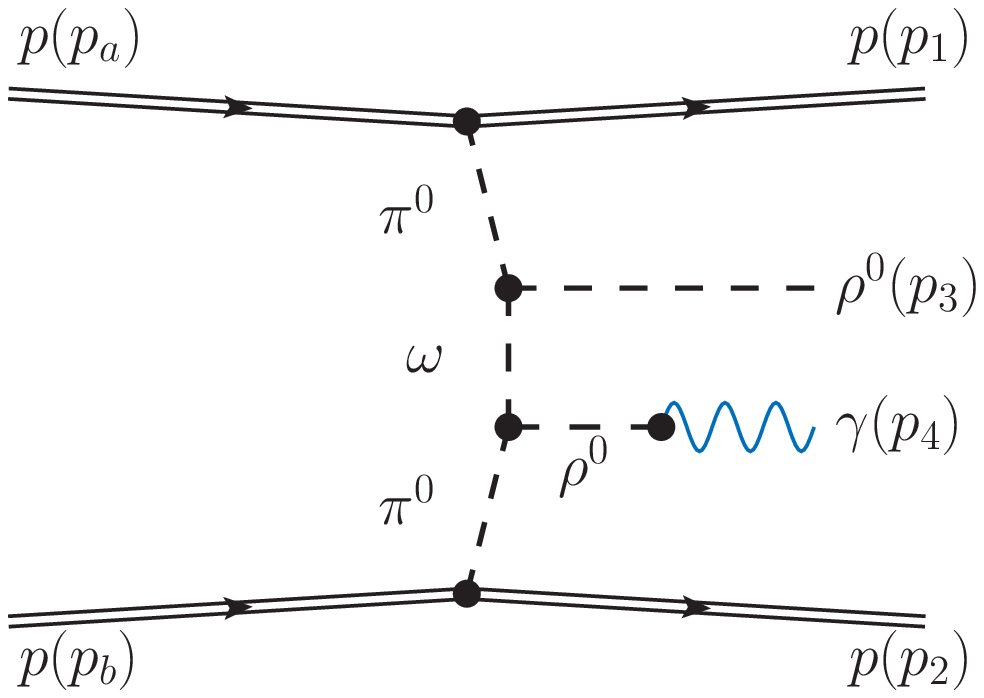}
   \includegraphics[width=6.cm]{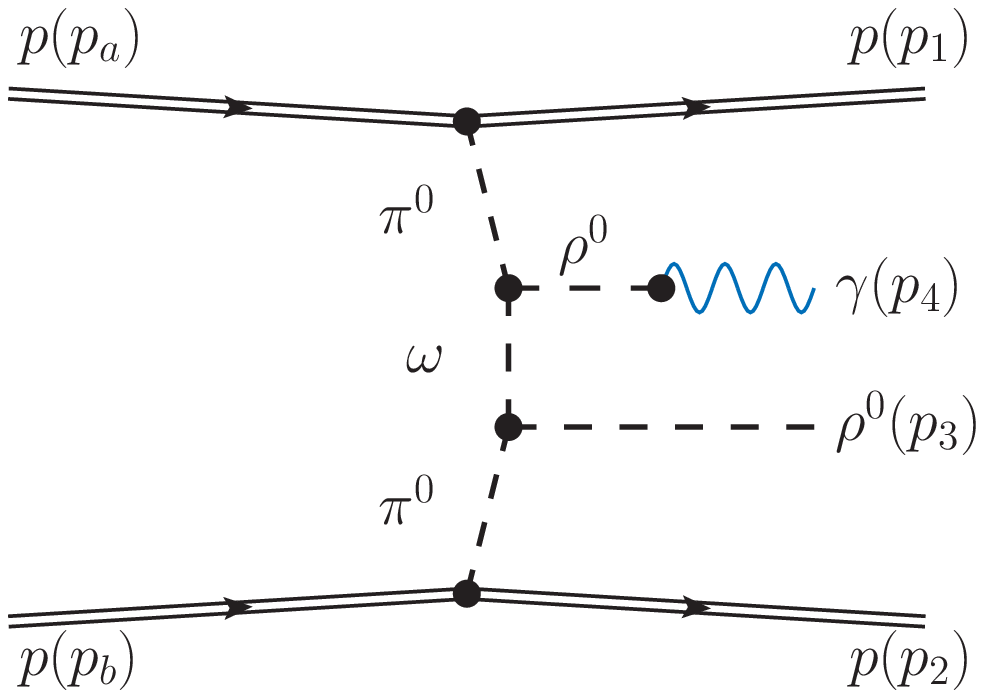}
\caption{
Diagrams for continuum $\rho^{0} \gamma$ production:
(a) $\omega \omega$ fusion,
(b) $\omega \rho$ and $\rho \omega$ fusion,
(c) $\pi \pi$ fusion.}
\label{fig:diagrams_rhogam}
\end{figure}
%-------------------------------------------------------------

In order to calculate the amplitude for the reaction (\ref{2to4_reaction_rhogamma})
we use the standard VMD model with the $\gamma V$ couplings
as given in (3.23)--(3.25) of \cite{Ewerz:2013kda}.
We shall consider the diagrams shown in Fig.~\ref{fig:diagrams_rhogam}.
The result is as follows:
\begin{eqnarray}
{\cal M}_{pp \to pp \rho^{0} \gamma}^{(\rho\gamma \;{\rm continuum})} =
{\cal M}_{pp \to pp \rho^{0} \gamma}^{(\omega \omega \; {\rm fusion})} +
{\cal M}_{pp \to pp \rho^{0} \gamma}^{(\omega \rho \; {\rm fusion})} +
{\cal M}_{pp \to pp \rho^{0} \gamma}^{(\pi \pi \; {\rm fusion})} \,.
\label{pp_rhogam}
\end{eqnarray}
We could also have $\pi \eta$ and $\pi \sigma$ fusion contributions.
For these we have to replace in the left (right)
diagram in Fig.~\ref{fig:diagrams_rhogam}(c) the lower (upper)
particles $(\pi^{0},\rho^{0})$ by $(\eta,\omega)$ or $(\sigma,\omega)$, respectively.
Discussing first $\pi \eta$ fusion we note that
the couplings $\eta pp$ and $\omega \omega \eta$
are smaller than those 
of $\pi^{0} pp$ and $\rho \omega \pi$ \cite{Nakayama:1998zv}. 
In addition, the $\eta$ exchange is suppressed 
relative to 
the $\pi^{0}$ exchange because of the heavier mass occurring in the propagator.
Another mechanism is the $\pi \sigma$ fusion
involving the $\sigma pp$ and $\sigma \omega \omega$ 
vertices.
However, here $g_{\sigma \omega \omega} \sim 0.5$ 
\cite{Nakayama:1998zv} is extremely small.
Moreover, the $\omega \to \gamma$ 
transition coupling is much smaller
than the $\rho \to \gamma$ one; see (\ref{A5}).
Therefore, we neglect the $\pi \eta$ and $\pi \sigma$
contributions in our considerations.

Thus, we are left with the 
($\omega$ + $\rho^{0}$)-$\pi^{0}$-$\omega$, 
$\omega$-$\pi^{0}$-($\omega$ + $\rho^{0}$),
and $\pi^{0}$-$\omega$-$\pi^{0}$
contributions, which we shall treat 
in a way similar to (\ref{amplitude_omeome}) and (\ref{amplitude_pipi}).
As an example, 
the ${\cal M}_{pp \to pp \rho^{0} \gamma}^{(\omega \omega \; {\rm fusion})}$
amplitude can be written as in (\ref{amplitude_omeome})
with the following replacement:
\begin{eqnarray}
&& \epsilon^{(\rho)\rho_{4}}(\lambda_{4} = 0, \pm 1) \to \frac{e}{\gamma_{\rho}}\epsilon^{(\gamma)\rho_{4}}(\lambda_{4} = \pm 1)\,.
\label{rhogam_aux_1}
\end{eqnarray}

In the case of the diagrams 
with the $\omega \to \gamma$ transition,
the outgoing $\omega$ has four-momentum squared 
$p^{2} = 0$.
Since nothing is known about the form factor
at the $\rho \omega \pi$ vertex where
both the $\pi^{0}$ and $\rho^{0}$ 
are off their mass-shell,
we assume in (\ref{rho_ome_pi}) 
the form factor (\ref{rho_ome_pi_ff}) as
$F(m_{\rho}^{2},0,m_{\pi}^{2}) = 1$
which is consistent with (\ref{F_V}) and (\ref{F_pi}).

The $\rho\gamma$-continuum processes in 
proton-antiproton collisions
can be treated in a completely analogous way to the
$\rho\gamma$-continuum processes in 
proton-proton collisions but
with the appropriate replacements
given by (\ref{repl1_ppbar_f1}) and (\ref{repl1}).

%-----------------------------------
\section{Numerical results}
\label{sec:num_results}
%-----------------------------------

We start by showing the integrated cross section 
for the exclusive reaction $pp \to pp f_{1}(1285)$
as a function of collision energy $\sqrt{s}$
from threshold to 8~GeV.
Note that due to (\ref{amp_sign_ppbar_f1}) 
the cross sections and distributions
for the $VV$-fusion mechanism are equal for $pp$
and $p \bar{p}$ scattering for the same kinematical values.

\begin{figure}[!ht]
\includegraphics[width=8.1cm]{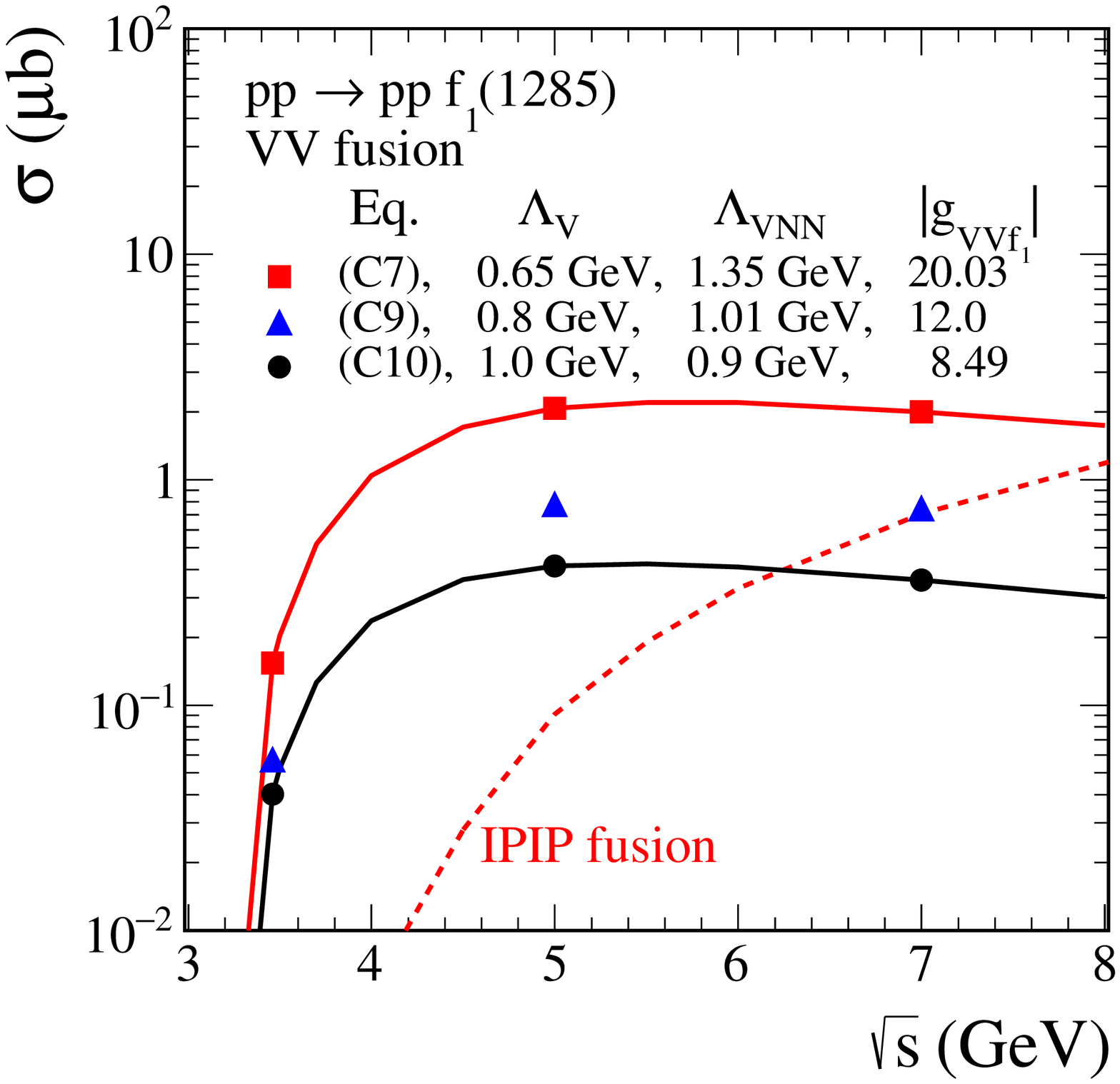}
\includegraphics[width=8.1cm]{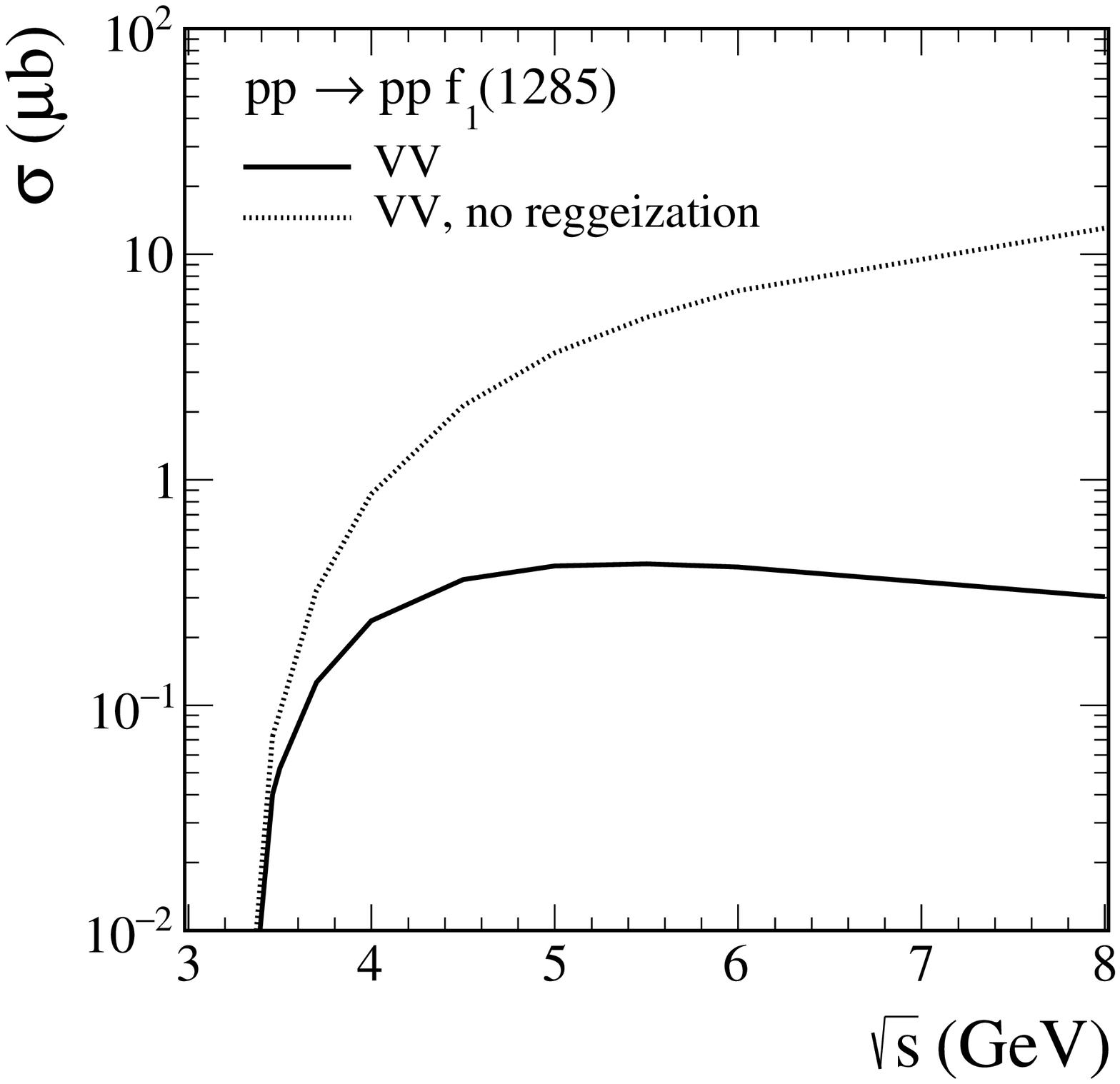}
\caption{
Integrated cross section for the $p p \to p p f_{1}(1285)$
reaction as a function of collision energy $\sqrt{s}$ for 
$VV \to f_{1}(1285)$ fusion with different parameters
from Eqs.~(\ref{aLam0.65}), (\ref{aLam0.8}), 
and (\ref{aLam1.0}).
We show also the pomeron-pomeron fusion mechanism 
(red dashed line).
In the right panel, the solid line 
is for the parameters of (\ref{aLam1.0})
and the reggeized propagators 
$\tilde{\Delta}_{T}^{(V)}$,
the dotted line corresponds to the result for 
the standard vector-meson propagators 
$\Delta_{T}^{(V)}$, i.e. without reggeization; 
see (\ref{reggeization_2})--(\ref{sthr}).
No rescattering effects are included here.}
\label{fig:sig_tot_W}
\end{figure}

In Fig.~\ref{fig:sig_tot_W} we show results for
for the $VV$-fusion contributions ($V = \rho, \omega$) for
different parameters given by (\ref{aLam0.65}),
(\ref{aLam0.8}) and (\ref{aLam1.0}) in Appendix~\ref{sec:appendixC}.
We assume $g_{\omega \omega f_{1}} = g_{\rho \rho f_{1}}
\equiv g_{VV f_{1}}$; see (\ref{A9}).
The cross section first rises from the threshold 
$\sqrt{s_{\rm thr}} = 2 m_p + m_{f_1}$ 
to $\sqrt{s} \approx 5$~GeV (PANDA energy range), 
where it starts to decrease towards higher energies. 
The region of fast growth of the cross section 
is related to the fast opening of the phase space,
while the reggeization is responsible for the decreasing part.
Without the reggeization the cross section would continue to grow.
The reggeization, calculated 
according to (\ref{reggeization_2})--(\ref{sthr}),
reduces the cross section by a factor of 1.8 already 
for the HADES c.m. energy $\sqrt{s} = 3.46$~GeV.
%As the energy $\sqrt{s}$ increases, the role of the effect increases.
For comparison we also show
the high-energy contribution of the $\Pom \Pom \to f_{1}(1285)$ fusion 
(see the red dashed line) 
with parameters fixed in \cite{Lebiedowicz:2020yre};
see Eq.~(3.7) there.

At near-threshold energies one should consider
final state interactions (FSI)
between the two produced protons; 
see e.g.~\cite{Nakayama:1998zv,Kaptari:2004sd}.
But the effect is sizeable only for extremely
small excess energies of tens of MeV:
$Q_{\rm exc} = \sqrt{s} - \sqrt{s_{\rm thr}}$.
In our case, we have $Q_{\rm exc} > 300$~MeV
and this FSI effect can be neglected.

We remind the reader that our calculation of the $VV$-fusion processes
should only be applied at energies $\sqrt{s} \lesssim 8$~GeV.
In the intermediate energy range also other processes like 
$f_{2 \Reg} f_{2 \Reg}$ fusion must be considered;
see the discussion in Appendix~D of \cite{Lebiedowicz:2020yre}.

The salient feature of the results shown in Fig.~\ref{fig:sig_tot_W}
is the high sensitivity of the $VV$-fusion cross section
to the different sets of parameters.
In our procedure of extracting the coupling constant 
$g_{VV f_{1}}$
and the form-factor cutoff parameters
from the CLAS data [see Appendices~\ref{sec:appendixB} and \ref{sec:appendixC}]
the dominant sensitivity is on $g_{VV f_{1}}$,
not on the form factors.
%Our procedure reduces the uncertainty
%from rather poorly known form factors.
Also the form of reggeization used in our model,
according to (\ref{reggeization_2})--(\ref{trajectory}),
affects the size of the cross section.
With the parameter values of (\ref{aLam1.0}) we get 
\begin{eqnarray}
&&{\rm for}\; \sqrt{s} = 3.46 \;{\rm GeV}: \;
\sigma_{pp \to pp f_{1}}
=  40.22\; {\rm nb} \,,
\label{2to3_f1_3.46_L1.0}\\
%&& \;
%\sigma_{pp \to pp f_{1}}^{\rm no \; reggeization}
%=  72.75\; {\rm nb} \,,
%\label{2to3_f1_3.46_L1.0_noreggez}
%\end{eqnarray}
%
%\begin{eqnarray}
&&{\rm for}\; \sqrt{s} = 5.0 \;{\rm GeV}: \;
\sigma_{p\bar{p} \to p\bar{p} f_{1}}
=  413.85\; {\rm nb} \,.
\label{2to3_f1_5.0_L1.0}
%&& \;
%\sigma_{p\bar{p} \to p\bar{p} f_{1}}^{\rm no \; reggeization}
%=  3645.04\; {\rm nb} \,.
%\label{2to3_f1_5.0_L1.0_noreggez}
\end{eqnarray}
With the parameter values of (\ref{aLam0.65}) we get
\begin{eqnarray}
&&{\rm for}\; \sqrt{s} = 3.46 \;{\rm GeV}: \;
\sigma_{pp \to pp f_{1}}
=  153.52\; {\rm nb} \,,
\label{2to3_f1_3.46_L0.65}\\
%&& \;
%\sigma_{pp \to pp f_{1}}^{\rm no \; reggeization}
%=  271.13\; {\rm nb} \,,
%\label{2to3_f1_3.46_L0.65_noreggez}
%\end{eqnarray}
%
%\begin{eqnarray}
&&{\rm for}\; \sqrt{s} = 5.0 \;{\rm GeV}: \;
\sigma_{p\bar{p} \to p\bar{p} f_{1}}
=  2071.43\; {\rm nb} \,.
%=  2.07\; \mu{\rm b} \,.
\label{2to3_f1_5.0_L0.65}
%&& \;
%\sigma_{p\bar{p} \to p\bar{p} f_{1}}^{\rm no \; reggeization}
%=  15932.61\; {\rm nb} \,.
%=  15.93\; \mu{\rm b} \,.
%\label{2to3_f1_5.0_L0.65_noreggez}
\end{eqnarray}
%
%Results without the reggeization effect
%are only meant to give quantitative information.

As mentioned above, the different numbers in (\ref{2to3_f1_3.46_L1.0}) 
and (\ref{2to3_f1_5.0_L1.0}) compared to (\ref{2to3_f1_3.46_L0.65}) 
and (\ref{2to3_f1_5.0_L0.65}) reflect mainly 
the different couplings $g_{VV f_{1}}$.
Indeed, from (\ref{2to3_f1_3.46_L0.65}) and (\ref{2to3_f1_3.46_L1.0})
we get for the cross section ratio 3.8,
from (\ref{2to3_f1_5.0_L0.65}) and (\ref{2to3_f1_5.0_L1.0}) we get 5.0,
and from (\ref{aLam0.65}) and (\ref{aLam1.0}) we get
for the ratio of the coupling constants squared 5.6,
not far from the two numbers above.

%----------------------------------------------------------------
%----------------------------------------------------------------

\begin{figure}[!ht]
\includegraphics[width=8.1cm]{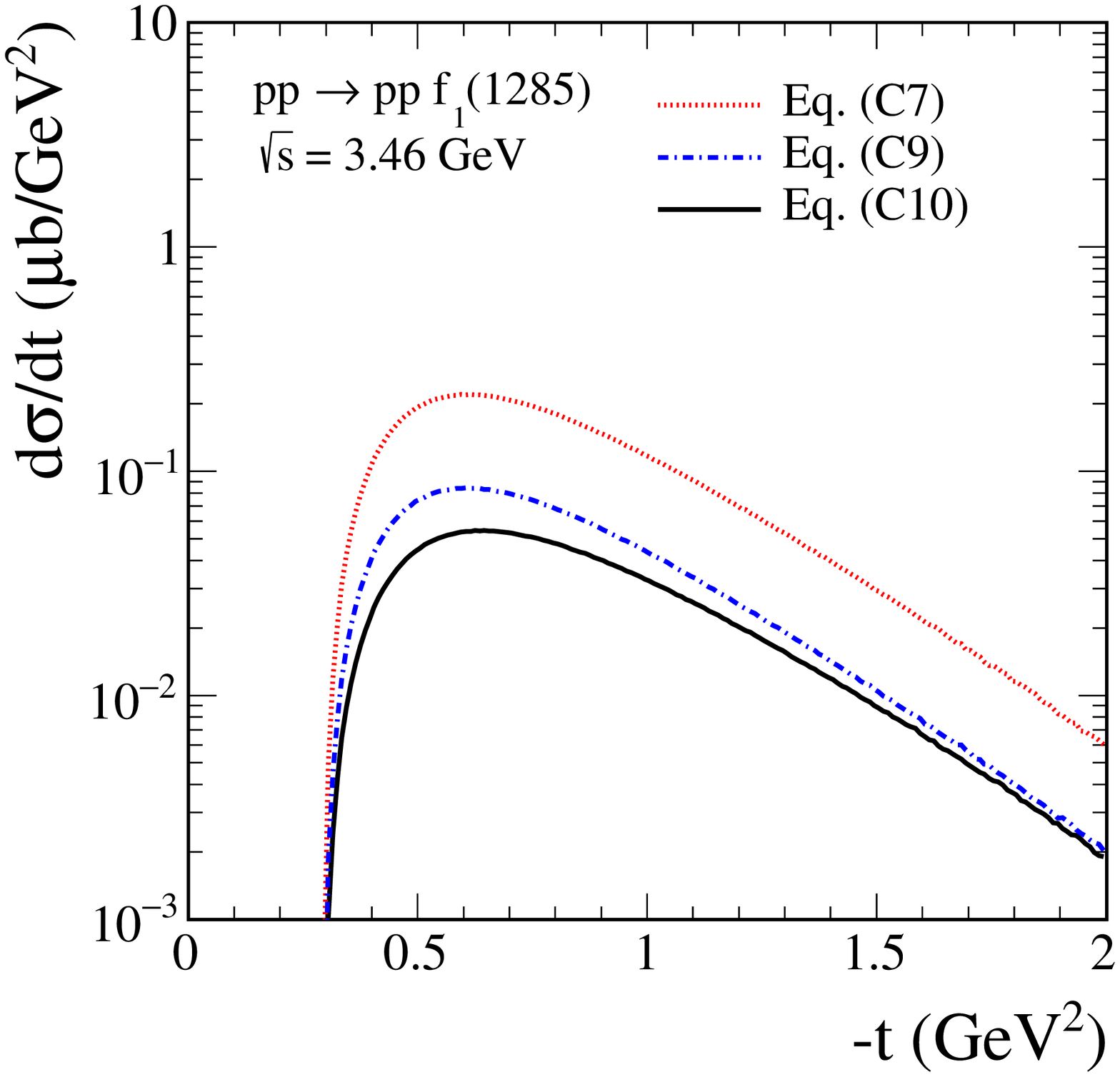}
\includegraphics[width=8.1cm]{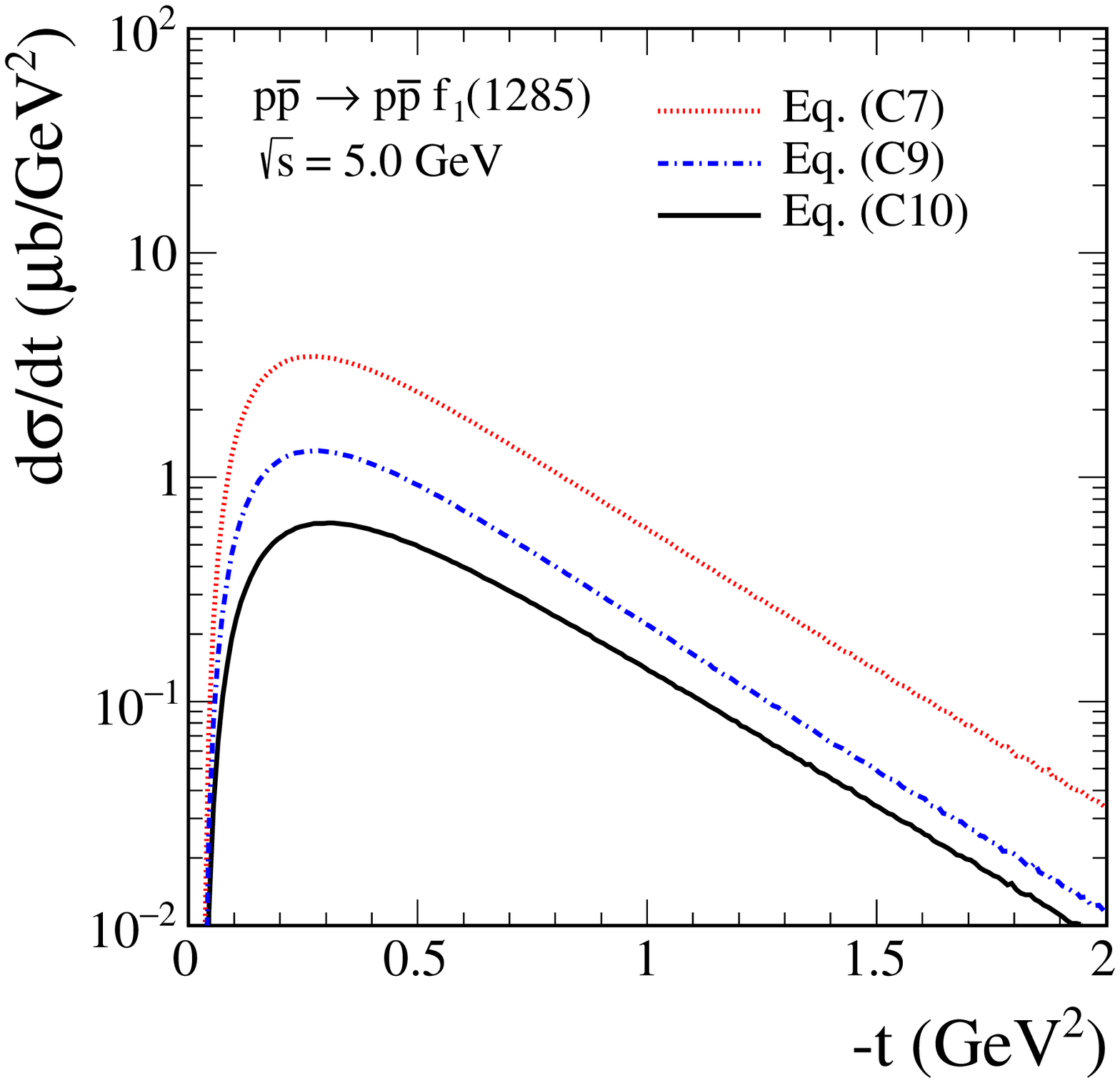}
\caption{
Distributions in $-t$ for $\sqrt{s} = 3.46$ and 5.0~GeV.
Results for different parameters (\ref{aLam0.65}),
(\ref{aLam0.8}), and (\ref{aLam1.0}) are shown.
In the calculations we take the $Vpp$ coupling constants 
from (\ref{Vpp_couplings}).}
\label{fig:HADES_taux}
\end{figure}

In Fig.~\ref{fig:HADES_taux} we show the distributions
in the four-momentum transfer squared from one of the proton vertices
[we have $t = t_{1}$ or $t_{2}$, cf.~(\ref{2to3_kinematic})]
for $\sqrt{s} = 3.46$~GeV (HADES) and 5.0~GeV (PANDA).
One can observe that $d\sigma/dt$ decreases
rapidly at forward scattering $|t| \to |t|_{\rm min}$, where
$|t|_{\rm min} \simeq 0.3$~GeV$^{2}$ at $\sqrt{s} = 3.46$~GeV.
At near threshold energy the values of small 
$|t_1|$ and $|t_2|$ are not accessible kinematically.
The maximum of $d\sigma/dt$ appears
at $-t_{1,2} \simeq 0.65$~GeV$^2$ for 
the parameter values of (\ref{aLam1.0})
and at $-t_{1,2} \simeq 0.77$~GeV$^2$ for 
those of (\ref{aLam1.5}).
The close-to-threshold production of the $f_{1}$ meson, 
therefore, probes corresponding form factors, 
(\ref{F_VVA}), (\ref{ff_pow}) and (\ref{F_V_M}), 
at relatively large values of $|t_1|$ and $|t_2|$,
far from their on mass-shell values at $t_{1,2} = m_{V}^{2}$ 
where they were normalised.
Thus, the $VV$-fusion cross section 
is very sensitive to the choice of the form factors.
Therefore the HADES and PANDA experiments have a good opportunity 
to study physics of large four-momentum transfer squared.
%---------------------------------------------------------------------
%---------------------------------------------------------------------

\begin{figure}[!ht]
\includegraphics[width=8.cm]{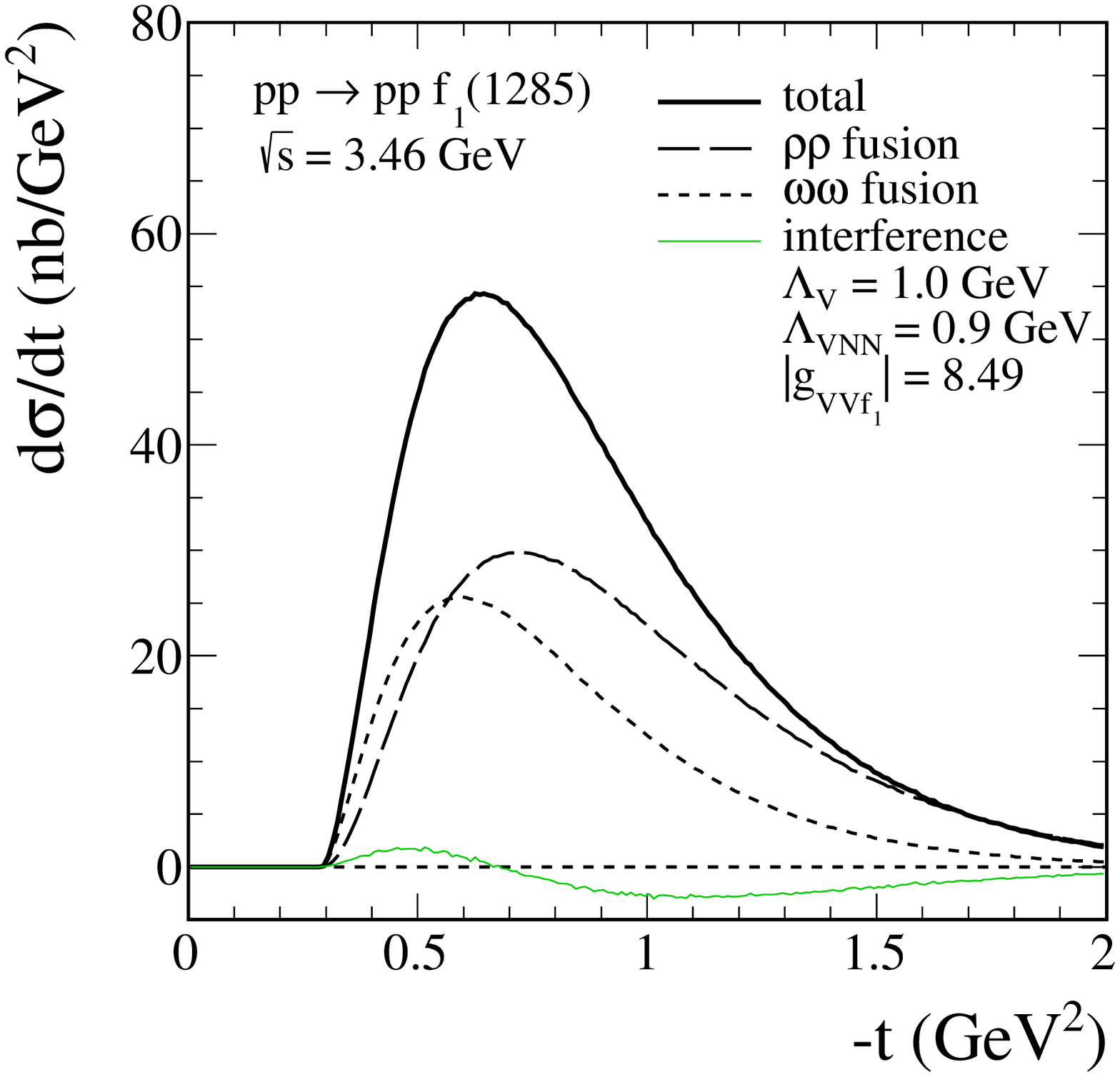}
\includegraphics[width=8.cm]{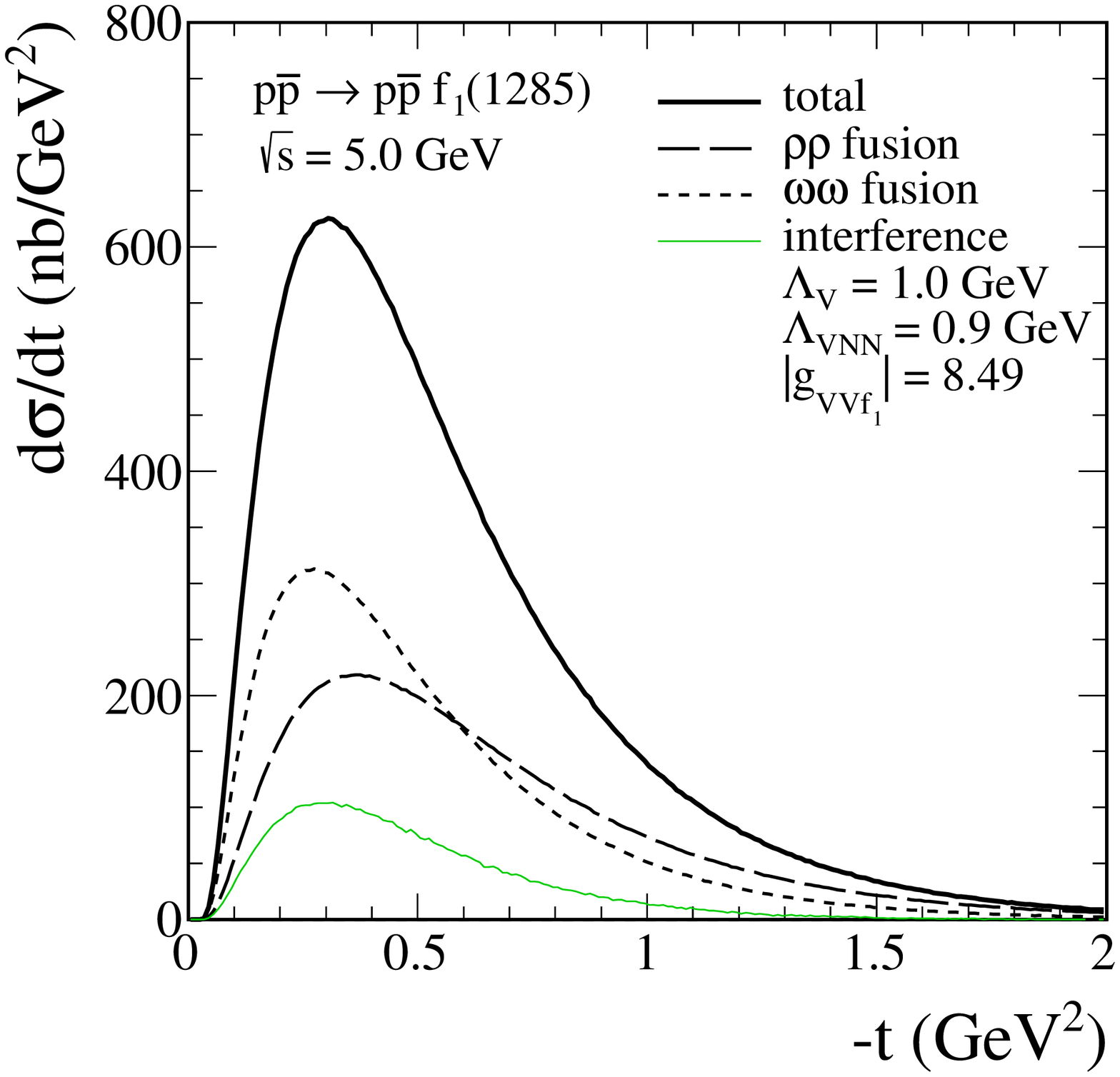}
\includegraphics[width=8.cm]{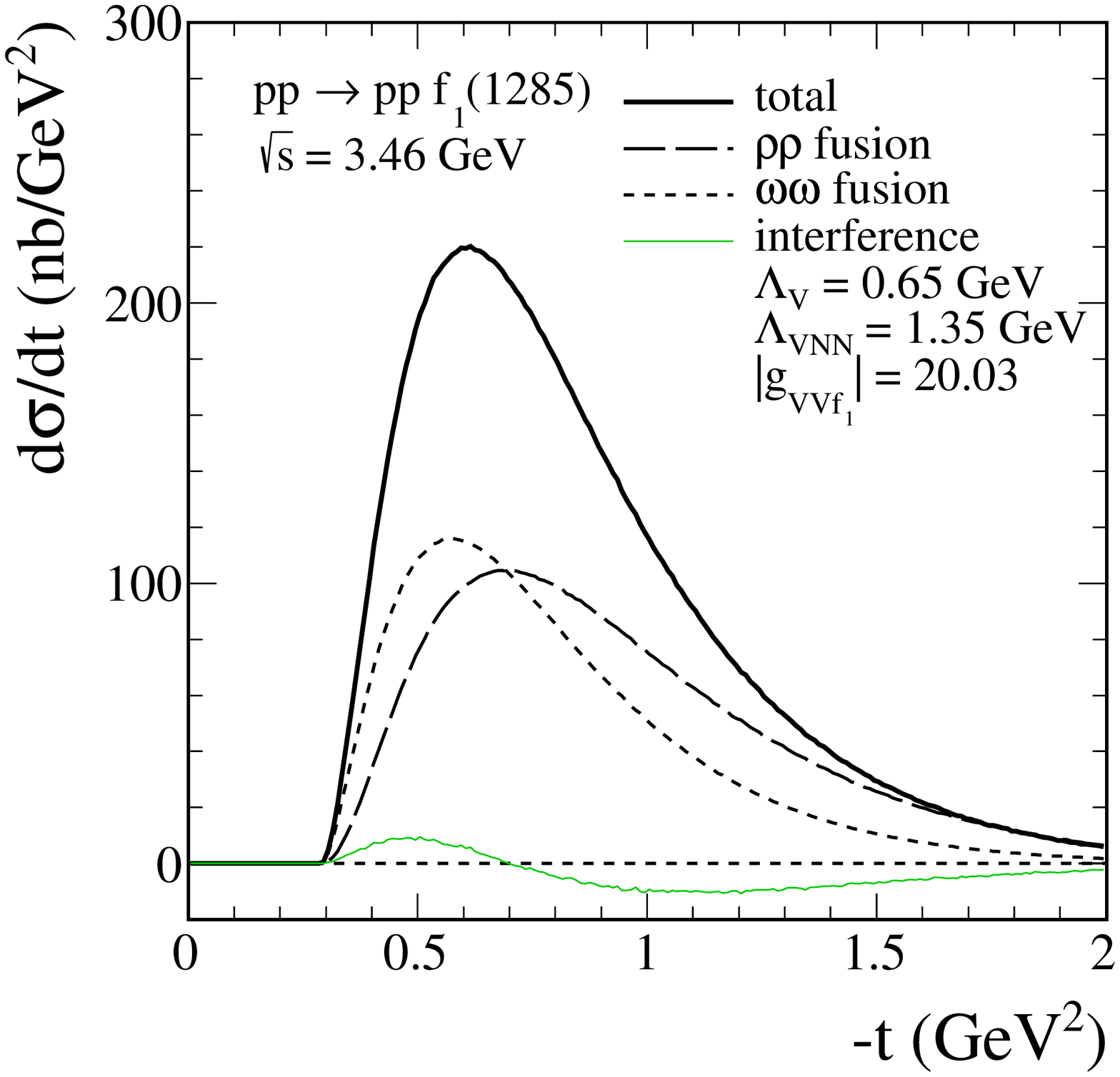}
\includegraphics[width=8.cm]{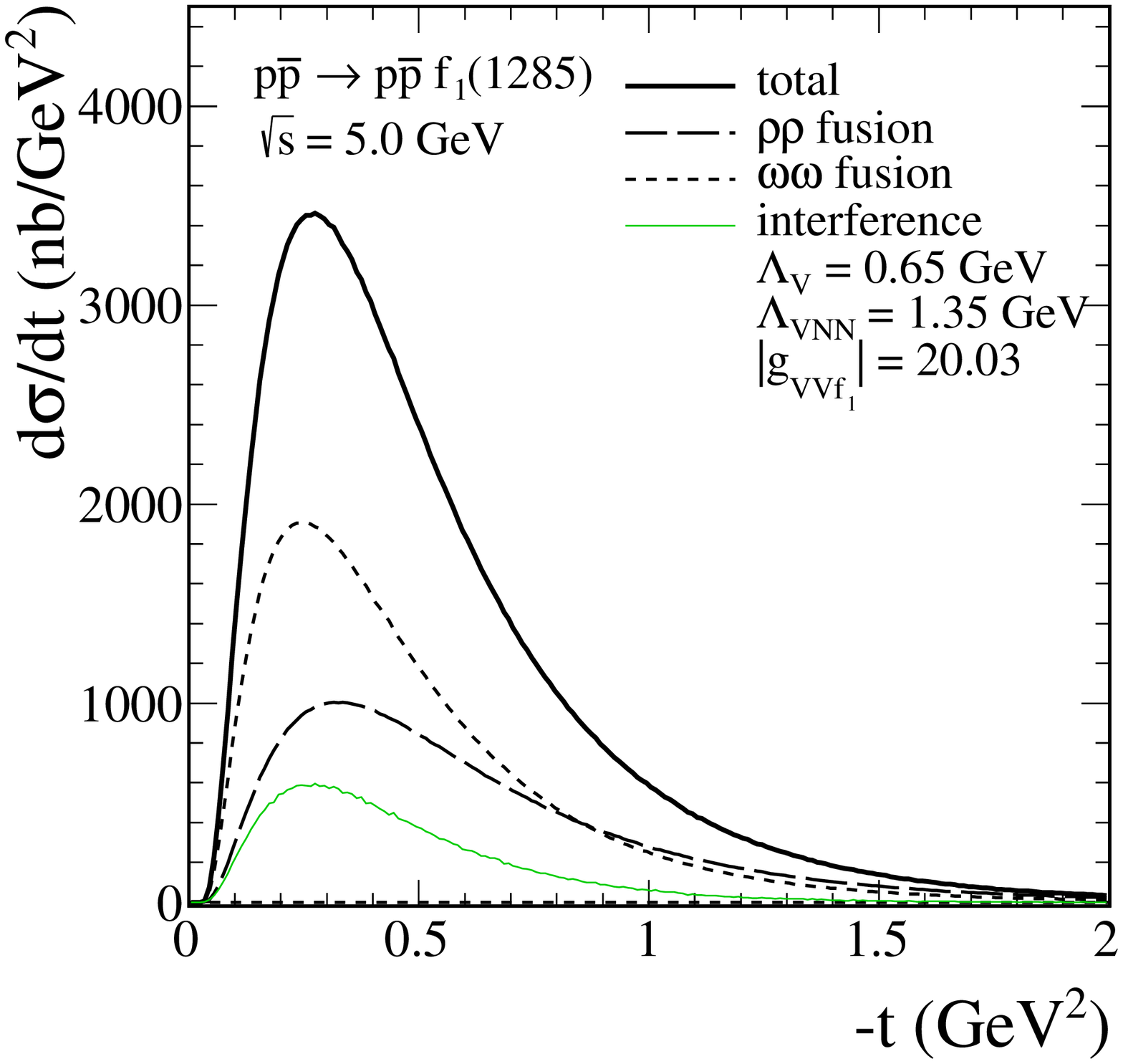}
\caption{Distributions in $-t$ for $\sqrt{s} = 3.46$
(left panels) and 5.0~GeV (right panels).
In the calculations we take the parameters given in (\ref{Vpp_couplings}), (\ref{aLam1.0}) and (\ref{aLam0.65}).
The results shown on the top panels correspond to (\ref{aLam1.0})
and those on the bottom panels are for (\ref{aLam0.65}).}
\label{fig:HADES_t}
\end{figure}

In Fig.~\ref{fig:HADES_t} we present
the contributions for the $\omega \omega$- 
and $\rho \rho$-fusion processes separately
and their coherent sum (total).
The interference term is shown also (see the green solid line).
Both processes play roughly similar role.
For large values of $|t_1|$ and $|t_2|$, 
in spite of $g_{\rho pp} < g_{\omega pp}$ (\ref{Vpp_couplings}),
the spin-flip term of the $\rho^{0}$-proton coupling is important.
For $\sqrt{s} = 5$~GeV the $\omega \omega$-fusion contribution
is the dominant process for $|t_{1,2}| \lesssim 0.5$~GeV$^{2}$.
There one can see also a large constructive interference effect.
%---------------------------------------------------------------------
%---------------------------------------------------------------------

\begin{figure}[!ht]
\includegraphics[width=8.cm]{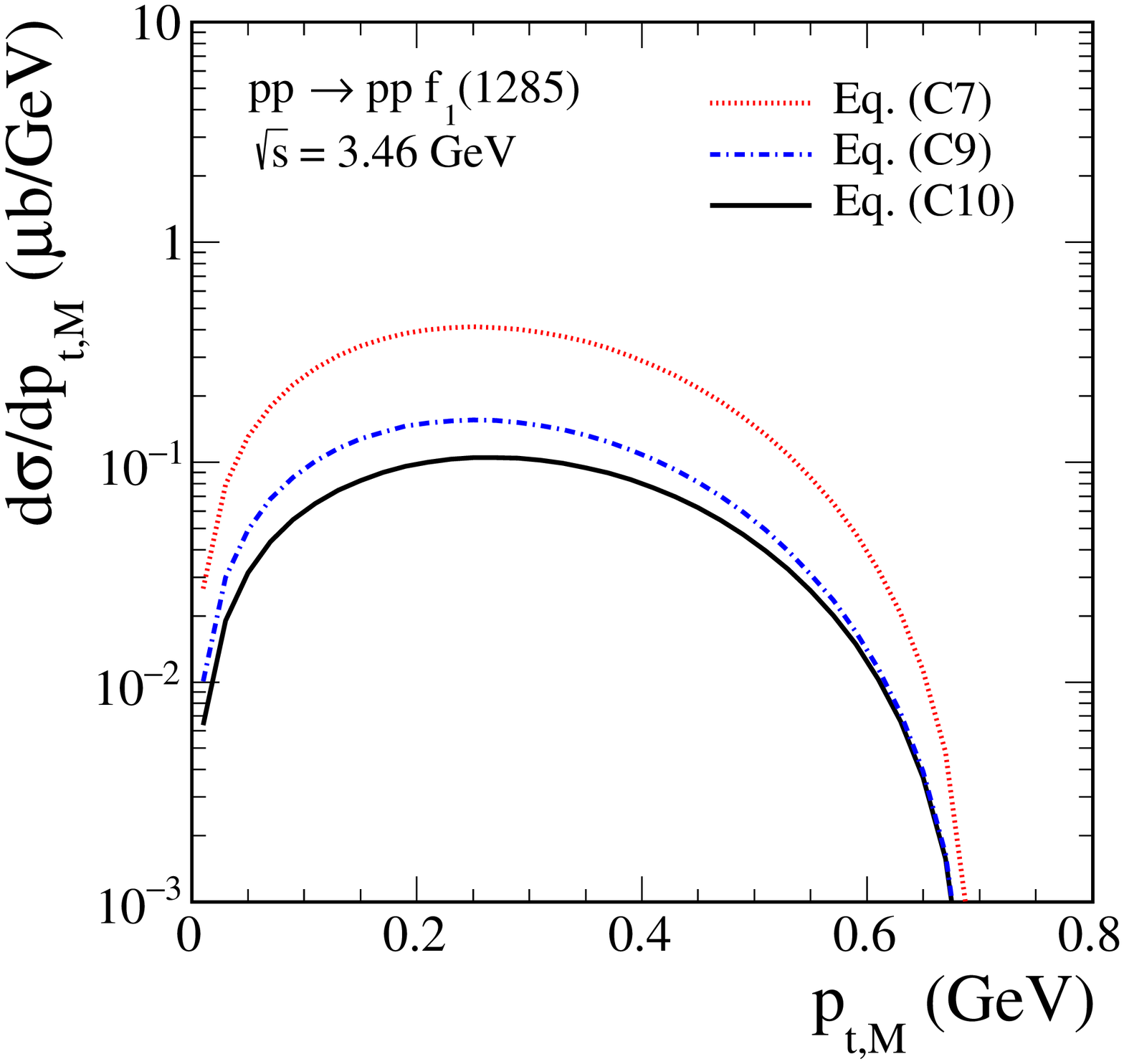}
\includegraphics[width=8.cm]{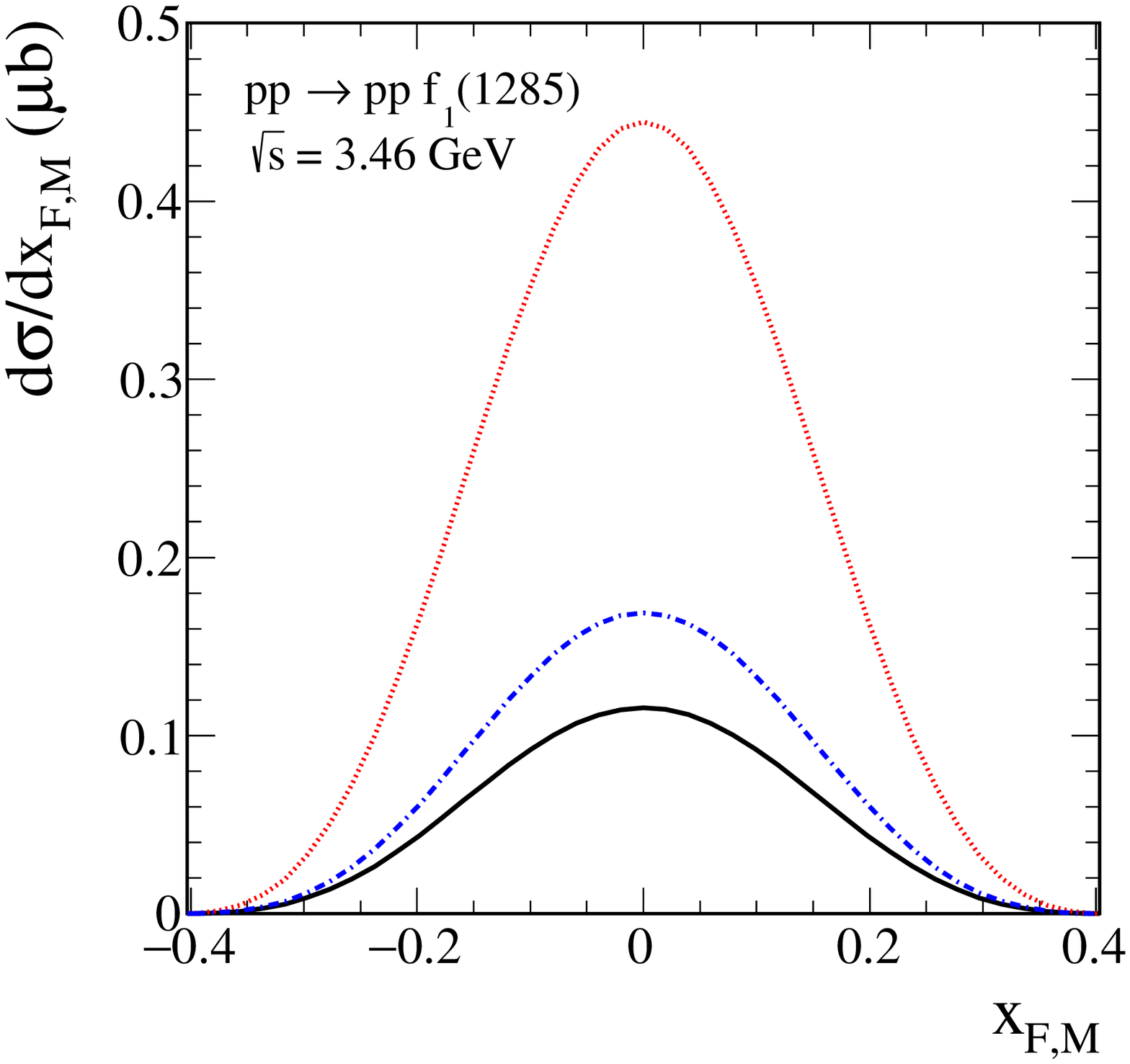}
\includegraphics[width=8.cm]{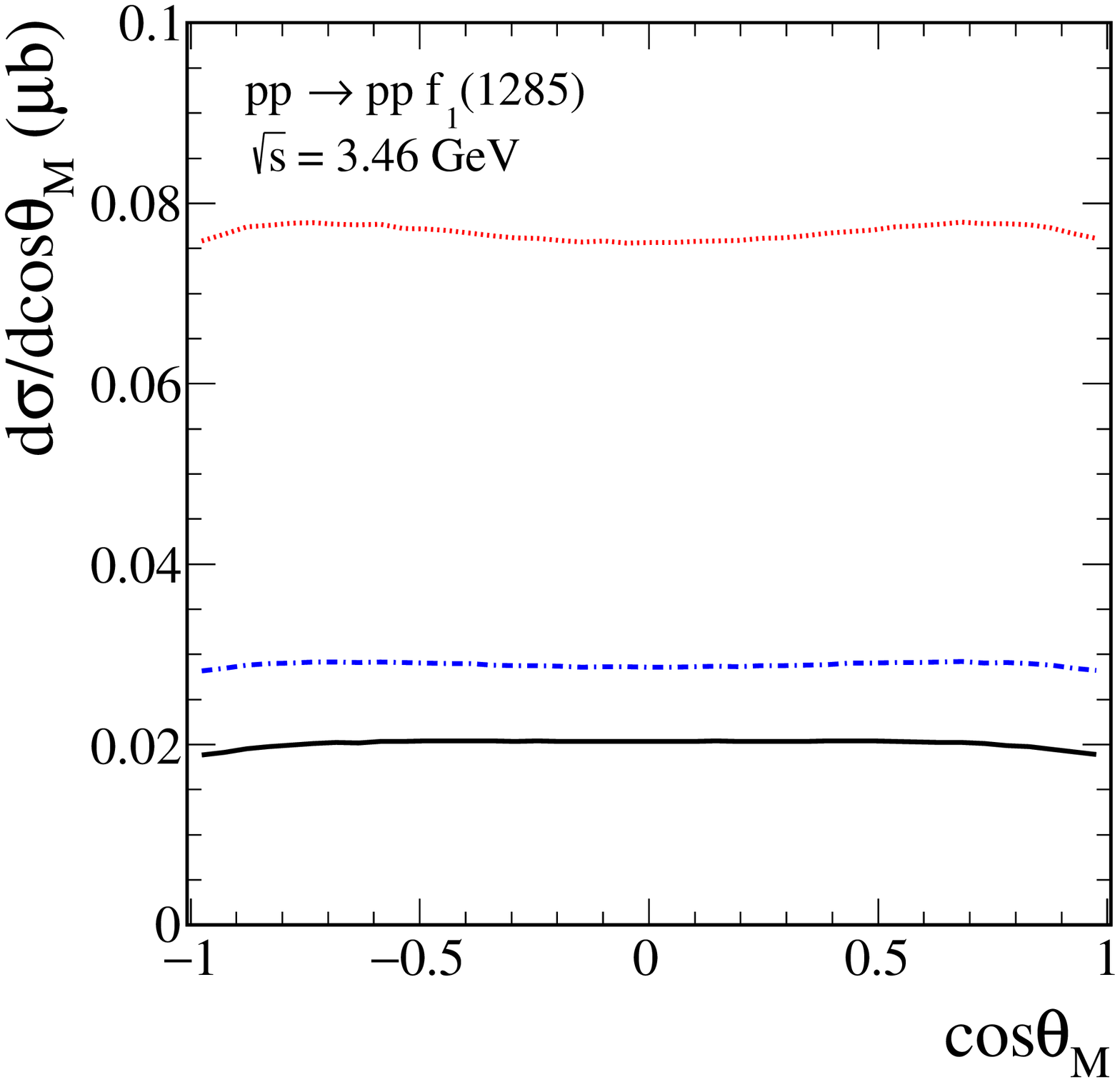}
\includegraphics[width=8.cm]{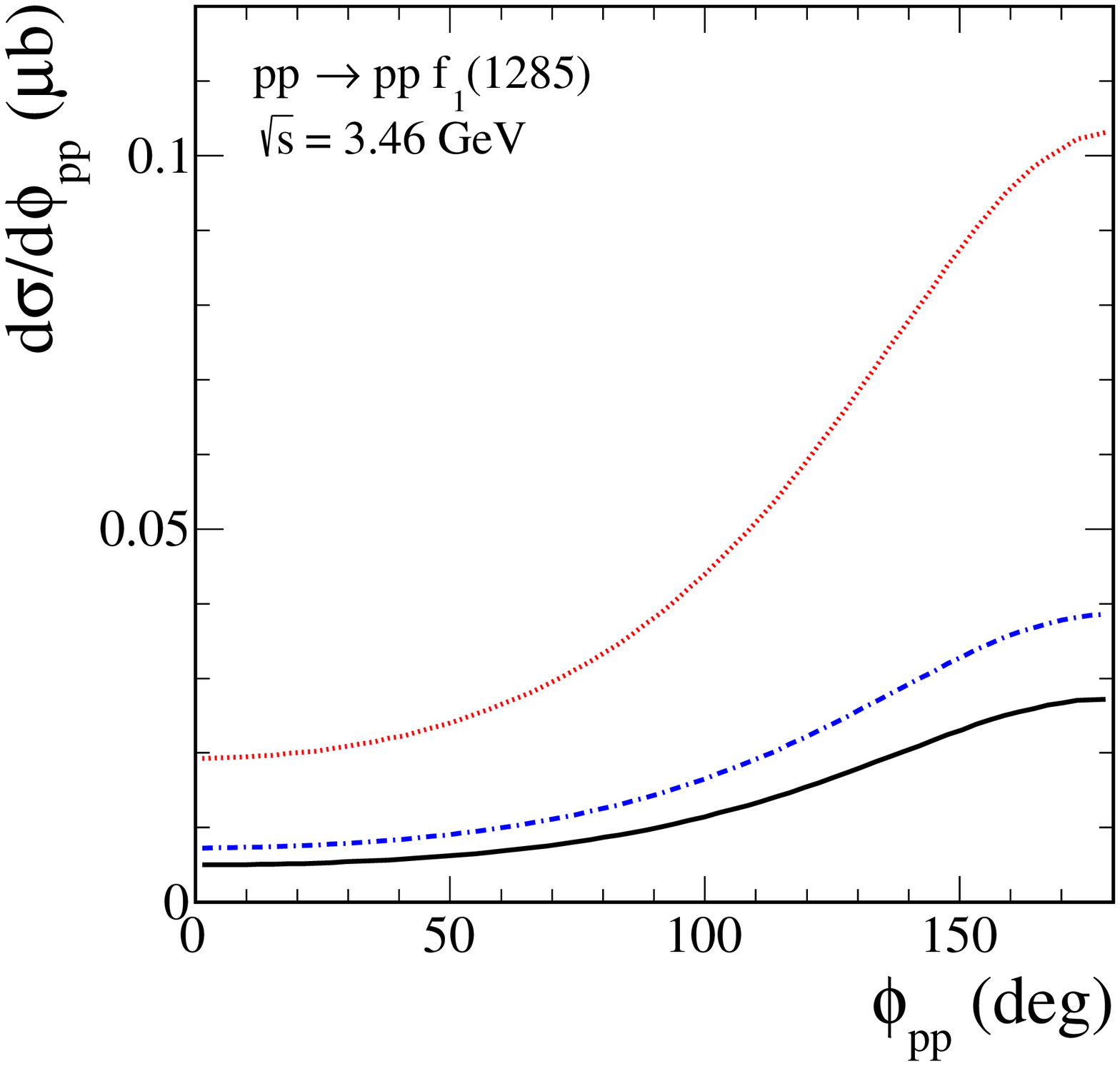}
\caption{
Several differential distributions for the reaction
$p p \to p p f_{1}(1285)$ at $\sqrt{s} = 3.46$~GeV
relevant for the HADES experiment.
The meaning of the lines is as in Fig.~\ref{fig:HADES_taux} (left panel).}
\label{fig:HADES}
\end{figure}

\begin{figure}[!ht]
\includegraphics[width=8.cm]{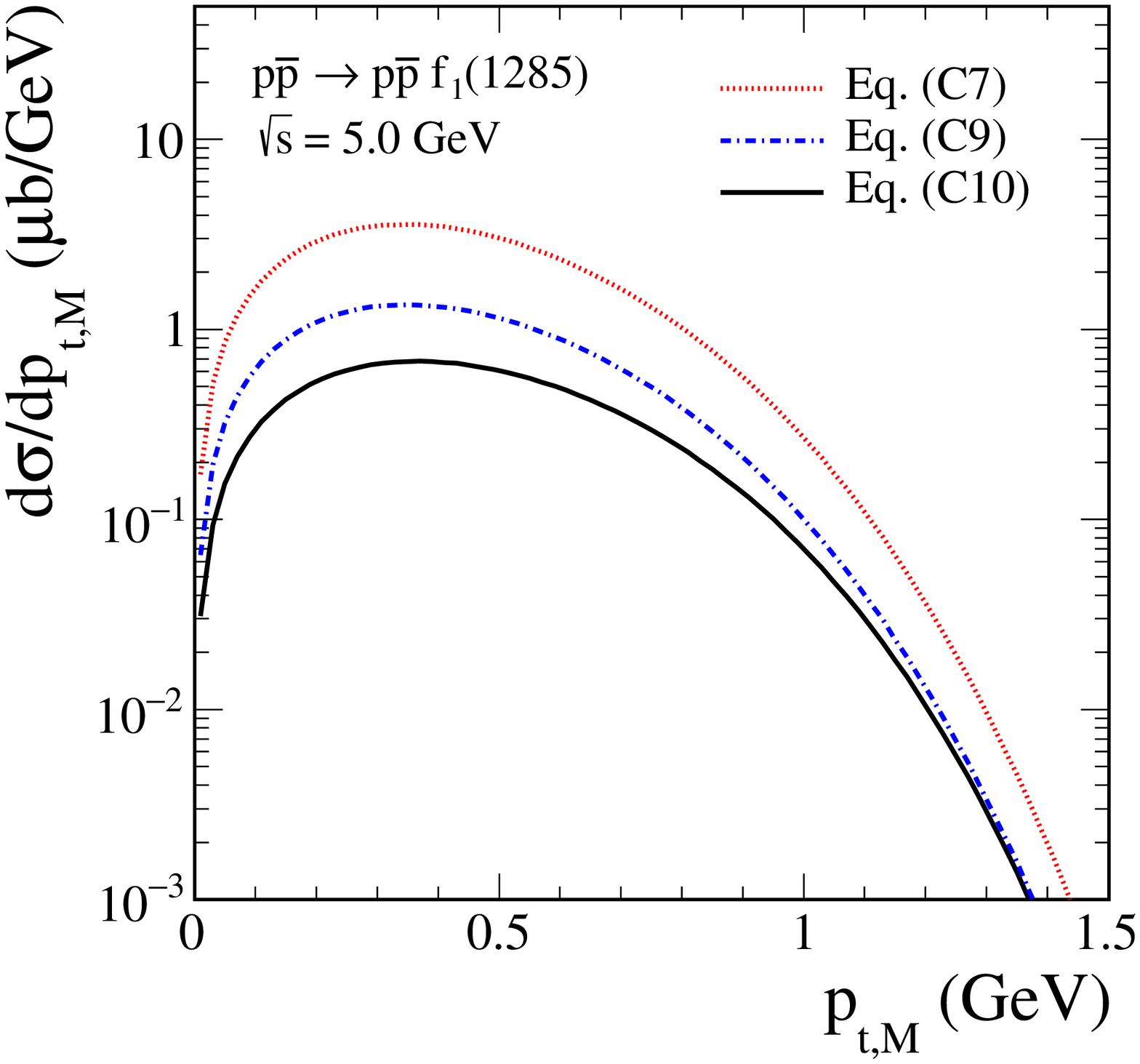}
\includegraphics[width=8.cm]{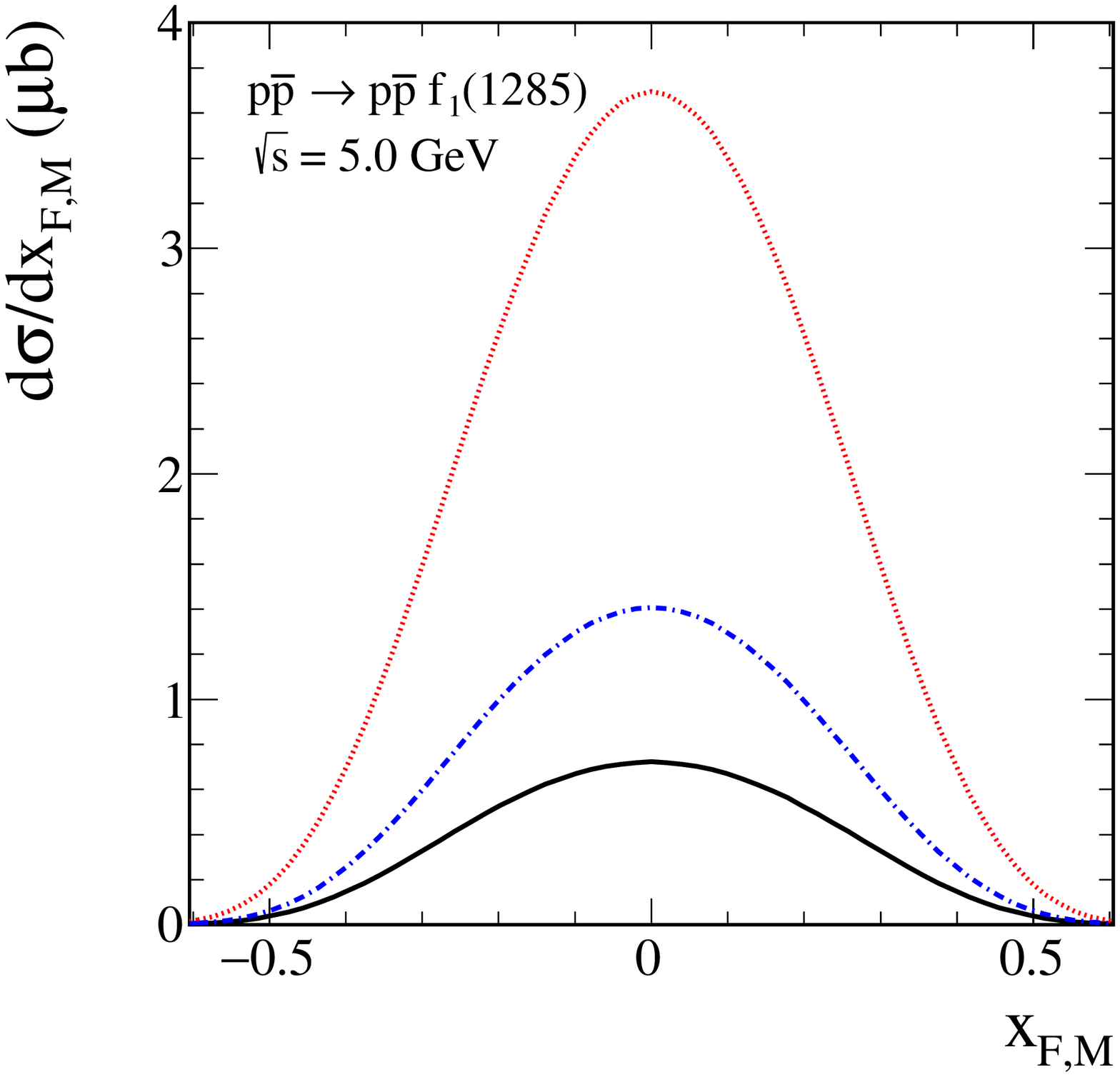}
\includegraphics[width=8.cm]{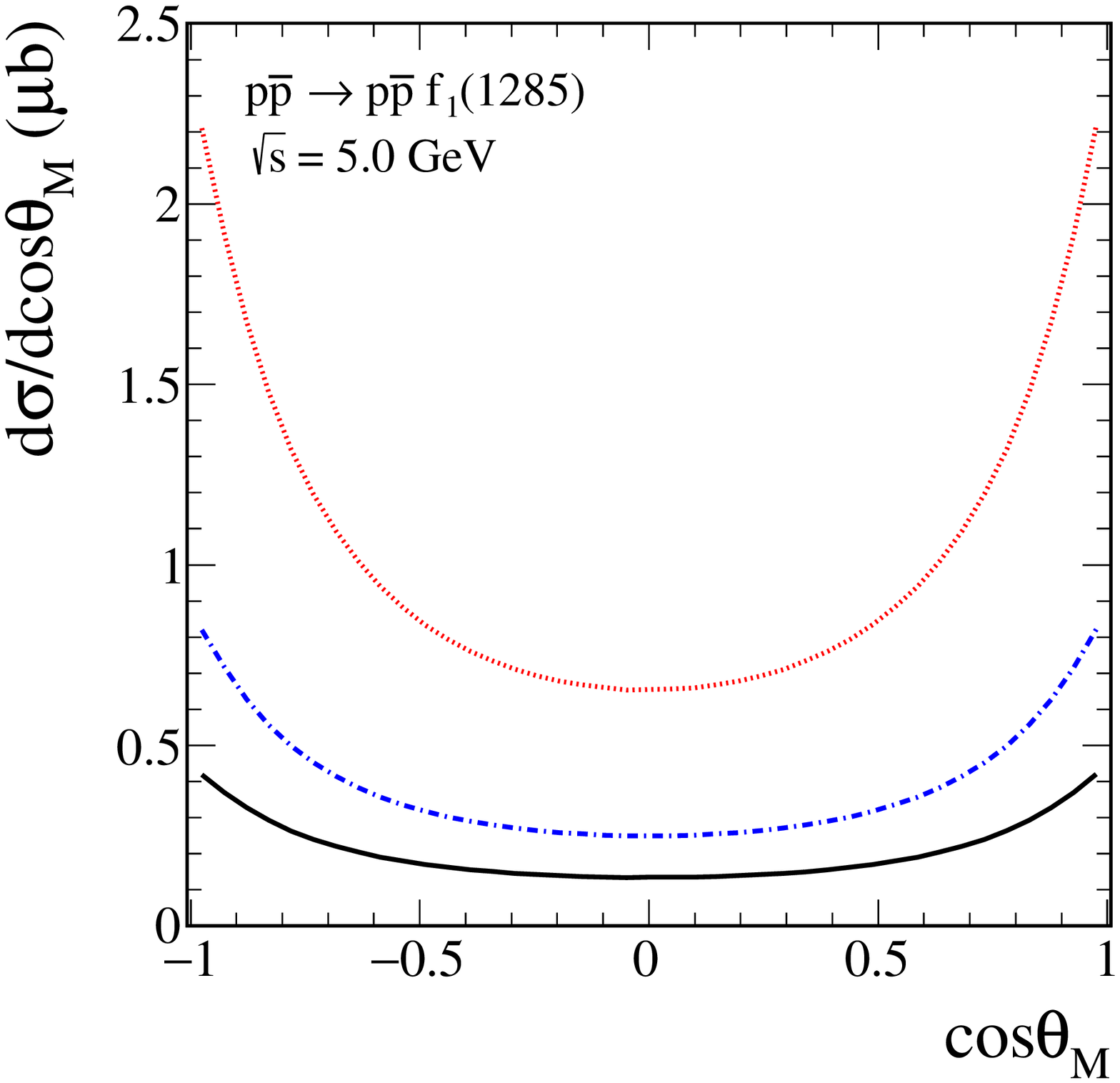}
\includegraphics[width=8.cm]{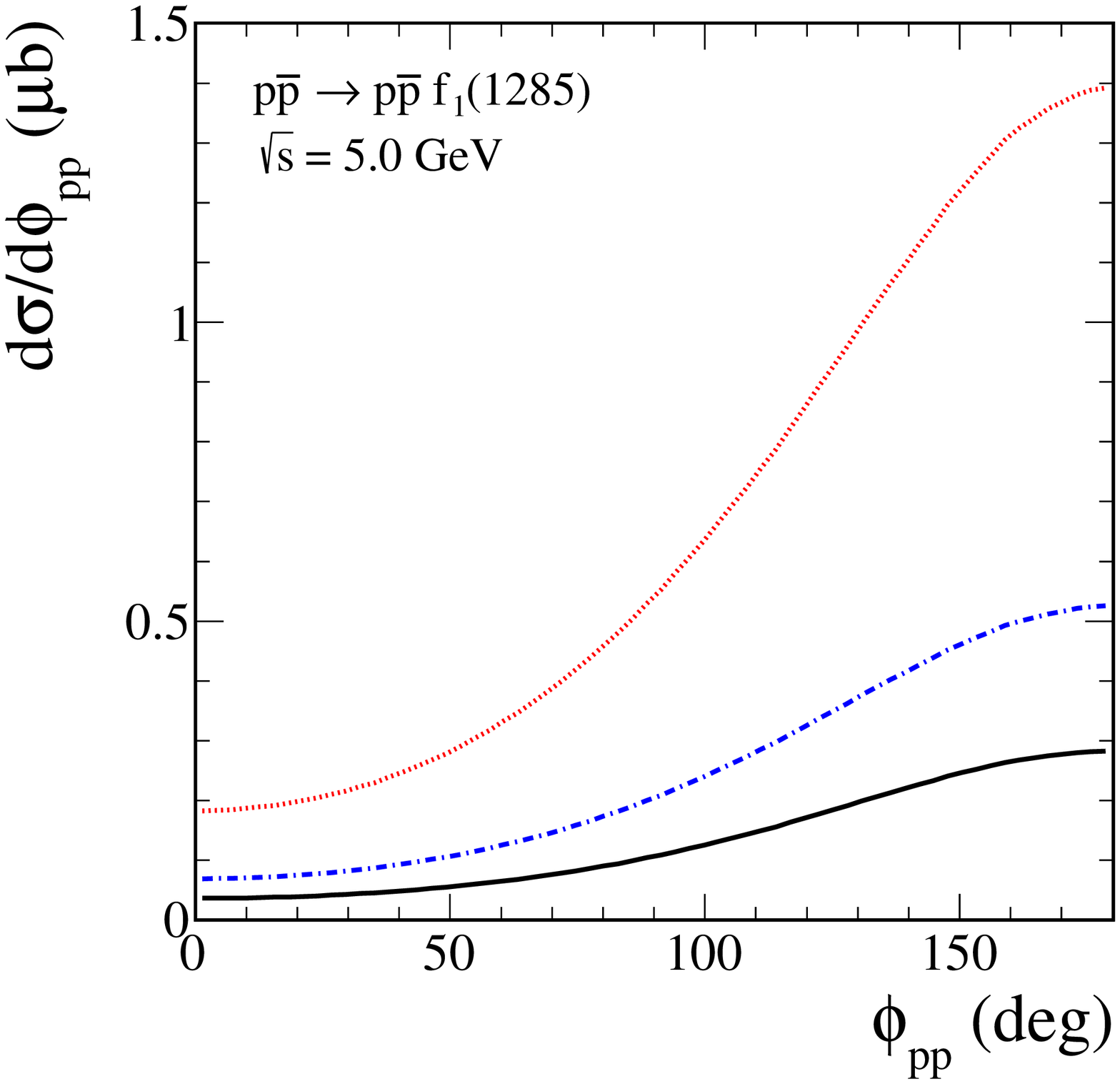}
\caption{
Several differential distributions for the reaction
$p \bar{p} \to p \bar{p} f_{1}(1285)$ at $\sqrt{s} = 5.0$~GeV
relevant for the PANDA experiment.
The meaning of the lines is as in Fig.~\ref{fig:HADES_taux} (right panel).}
\label{fig:PANDA}
\end{figure}

\begin{figure}[!ht]
\includegraphics[width=8.cm]{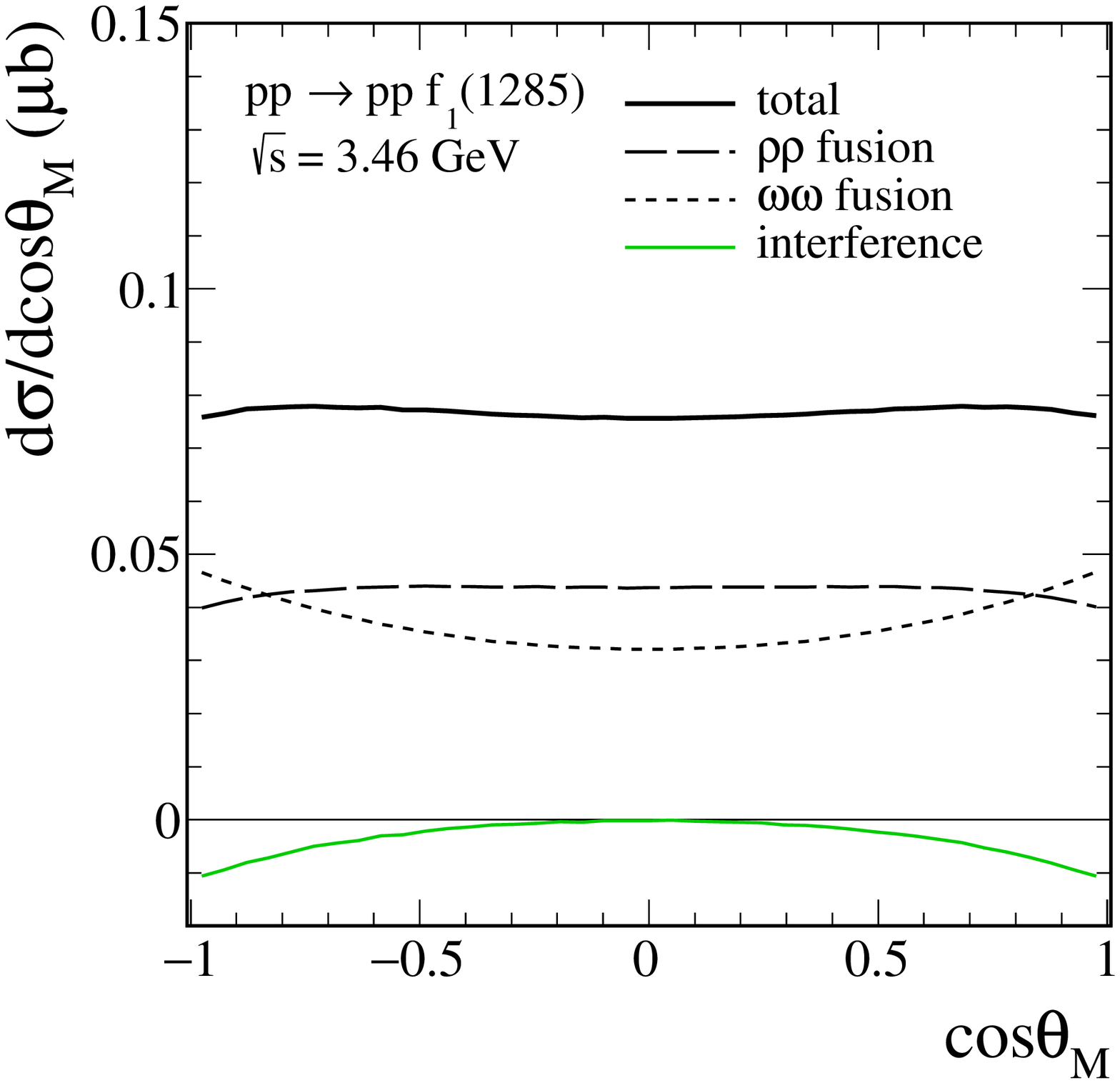}
\includegraphics[width=8.cm]{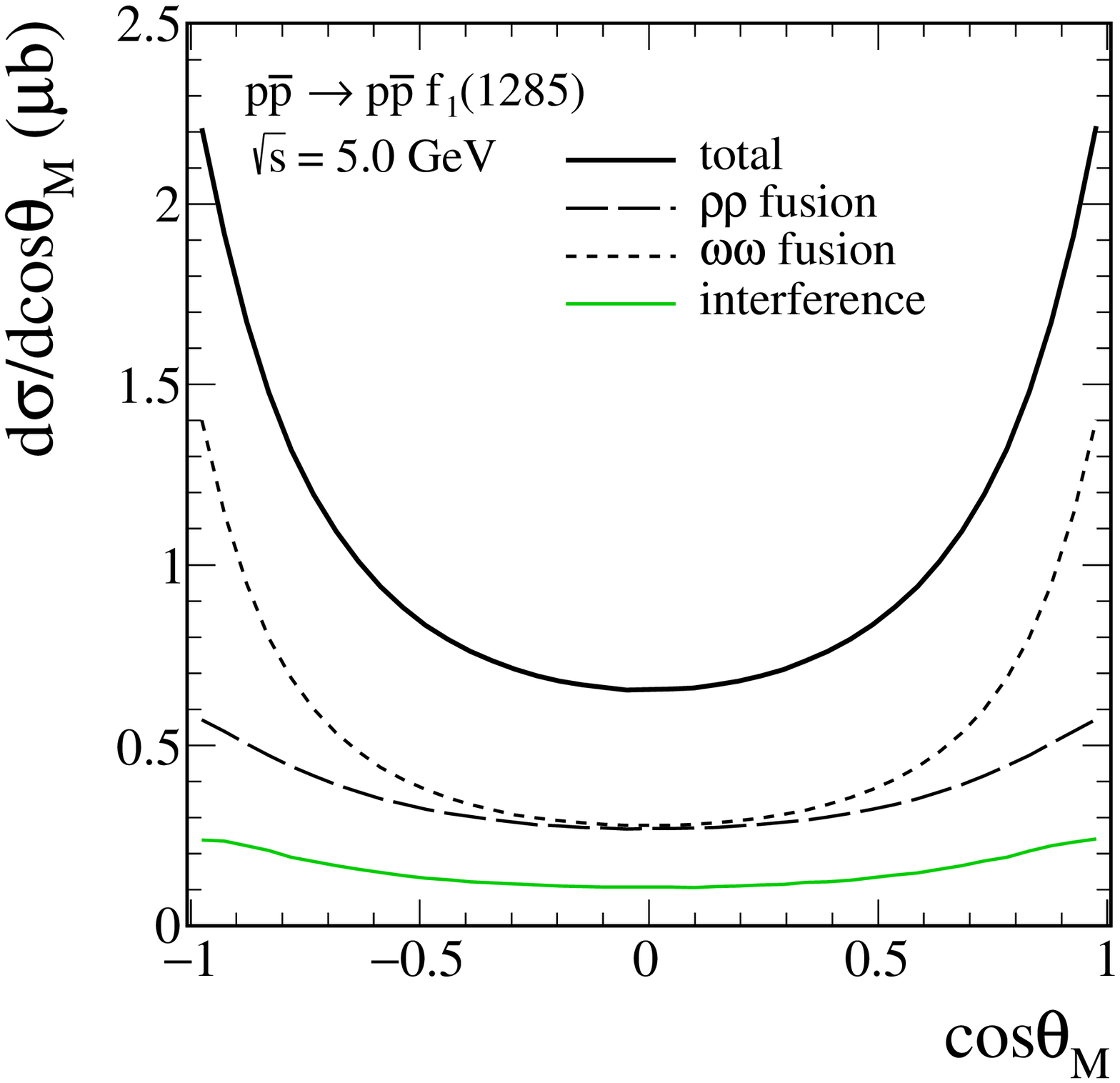}
\caption{Distributions in $\cos\theta_{M}$ for 
$\sqrt{s} = 3.46$~GeV (left panel) and
$\sqrt{s} = 5.0$~GeV (right panel).
The meaning of the lines is as 
in the bottom panels of Fig.~\ref{fig:HADES_t}.
Results for the parameter values of
(\ref{Vpp_couplings}) and (\ref{aLam0.65}) are shown.}
\label{fig:z3}
\end{figure}

In Figs.~\ref{fig:HADES} and \ref{fig:PANDA} 
we show several differential
distributions for the reaction 
$pp \to pp f_{1}(1285)$
for $\sqrt{s} = 3.46$~GeV relevant 
for the HADES experiment and for the reaction
$p\bar{p} \to p\bar{p} f_{1}(1285)$ 
for $\sqrt{s} = 5.0$~GeV relevant 
for the PANDA experiment, respectively.
We show the distributions in the transverse momentum 
of the $f_{1}(1285)$ meson,
in $x_{F,M}$, 
the Feynman variable of the meson, 
in the $\cos\theta_{M}$ where $\theta_{M}$
is the angle between $\bk$ and $\bpa$ in the c.m. frame,
and in $\phi_{pp}$, the azimuthal angle between 
the transverse momentum vectors $\bpta$, $\bptb$ 
of the outgoing nucleons in the c.m. frame.
We predict a strong preference for the outgoing
nucleons to be produced with their transverse momenta
being back-to-back ($\phi_{pp} \approx \pi$).
The distributions in $\cos\theta_{M}$
for the energies $\sqrt{s} = 3.46$~GeV and
$\sqrt{s} = 5.0$~GeV have a different shape.
This is explained in Fig.~\ref{fig:z3}.
One can observe from Figs.~\ref{fig:HADES_t} and \ref{fig:z3}
that the $\omega \omega$- and $\rho \rho$-fusion 
processes have different kinematic dependences. 
With increasing energy $\sqrt{s}$ 
the averages of $|t_{1}|$ and $|t_{2}|$ decrease
(damping by form factors), 
hence the $\omega \omega$ contribution 
becomes more important.
%---------------------------------------------------------------------
%---------------------------------------------------------------------
%---------------------------------------------------------------------
%---------------------------------------------------------------------

%---------------------------------------------------------------------
%---------------------------------------------------------------------

Now we turn to the $pp \to pp (f_{1}(1285) \to \rho^{0} \gamma)$
reaction and the discussion of background processes.

\begin{figure}[!ht]
\includegraphics[width=8.1cm]{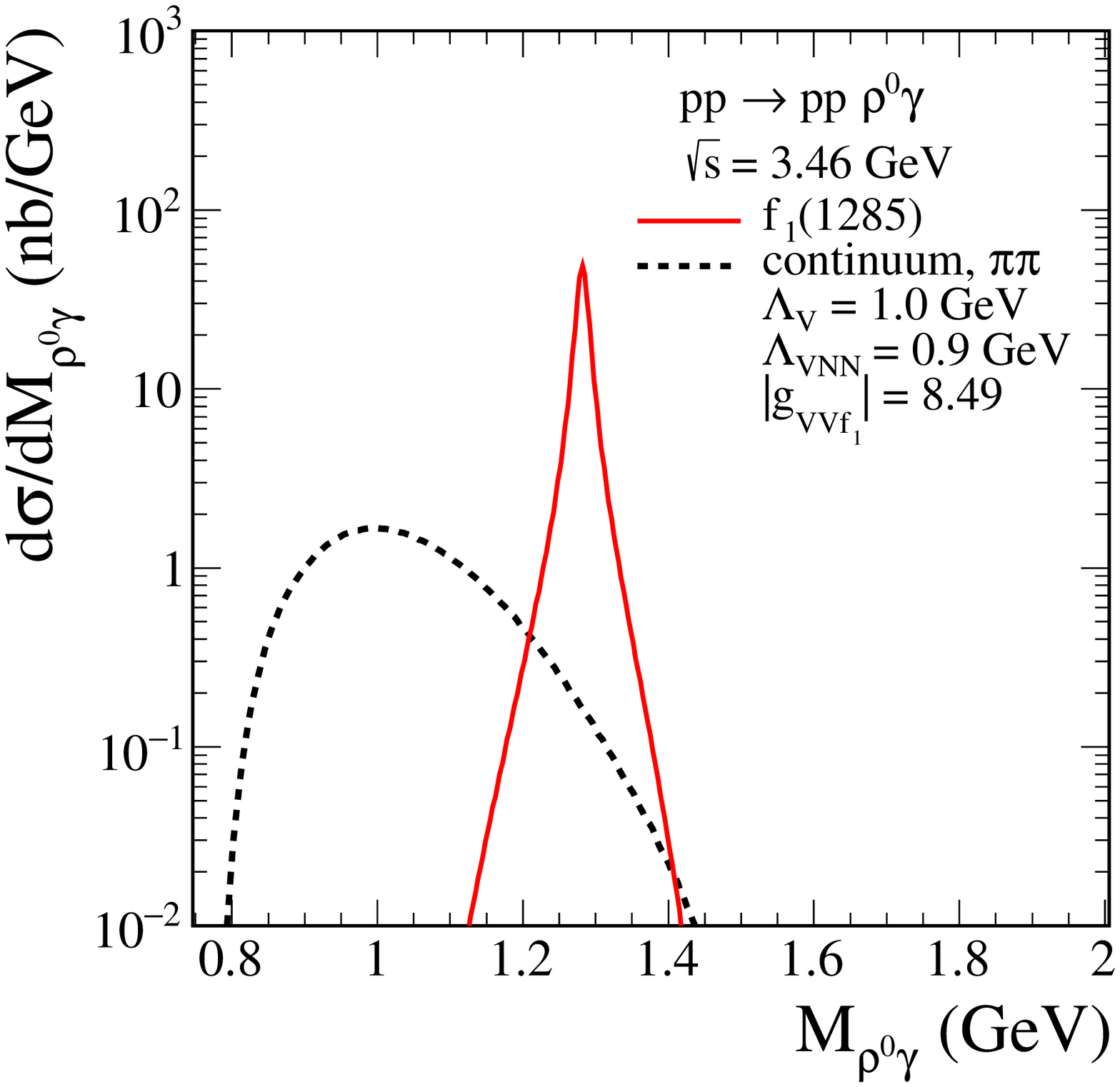}
\includegraphics[width=8.1cm]{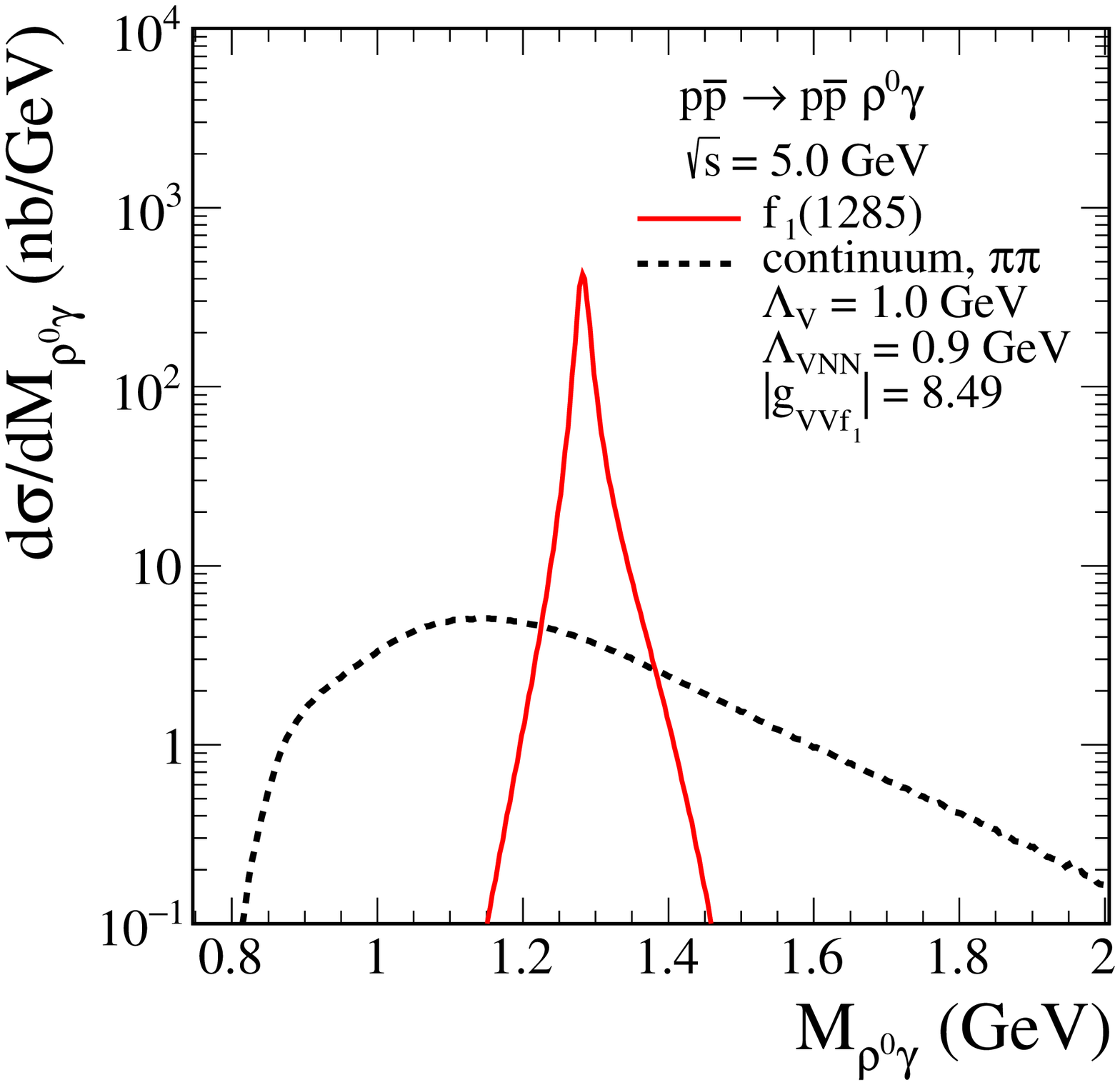}\\
\includegraphics[width=8.1cm]{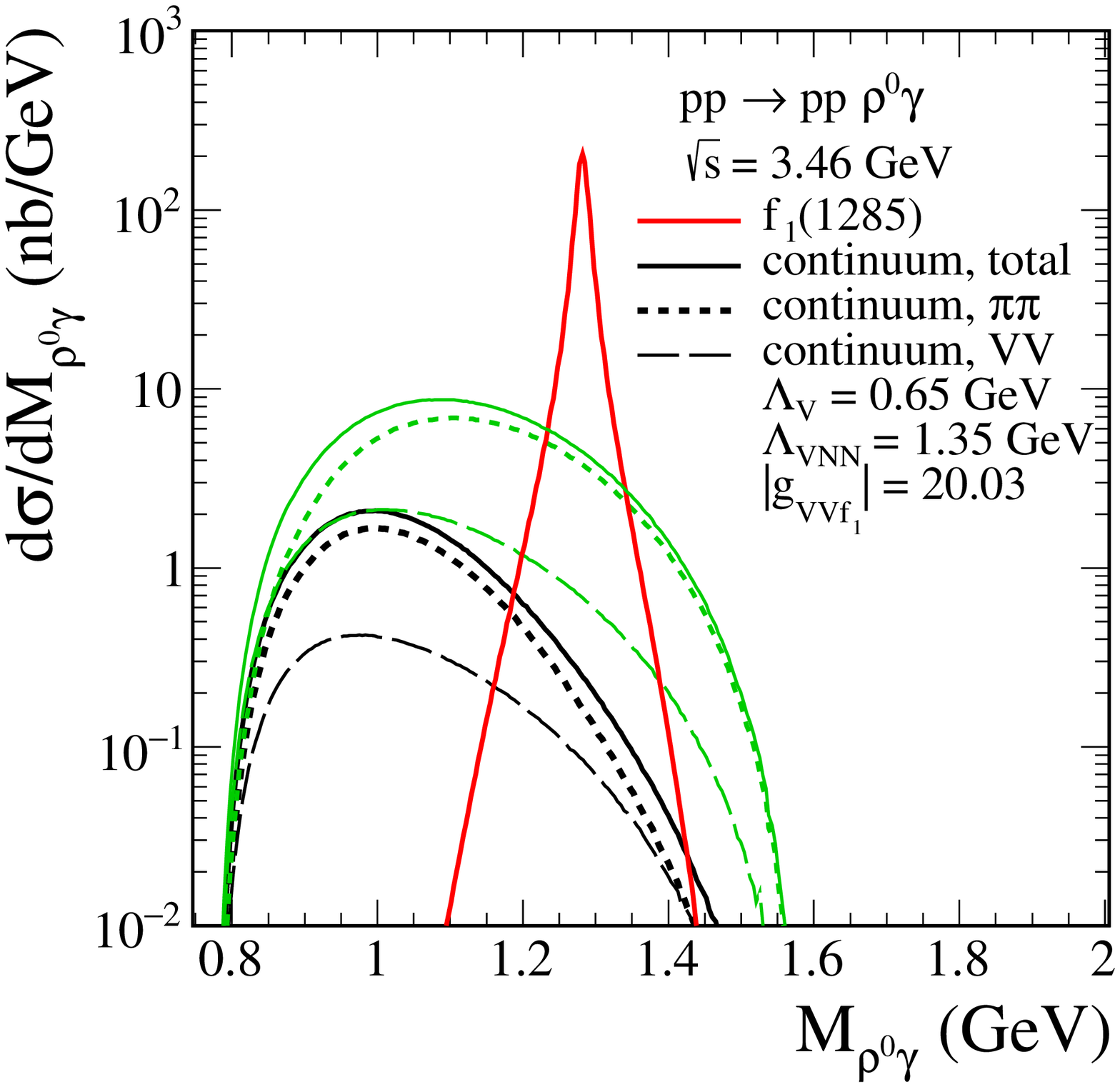}
\includegraphics[width=8.1cm]{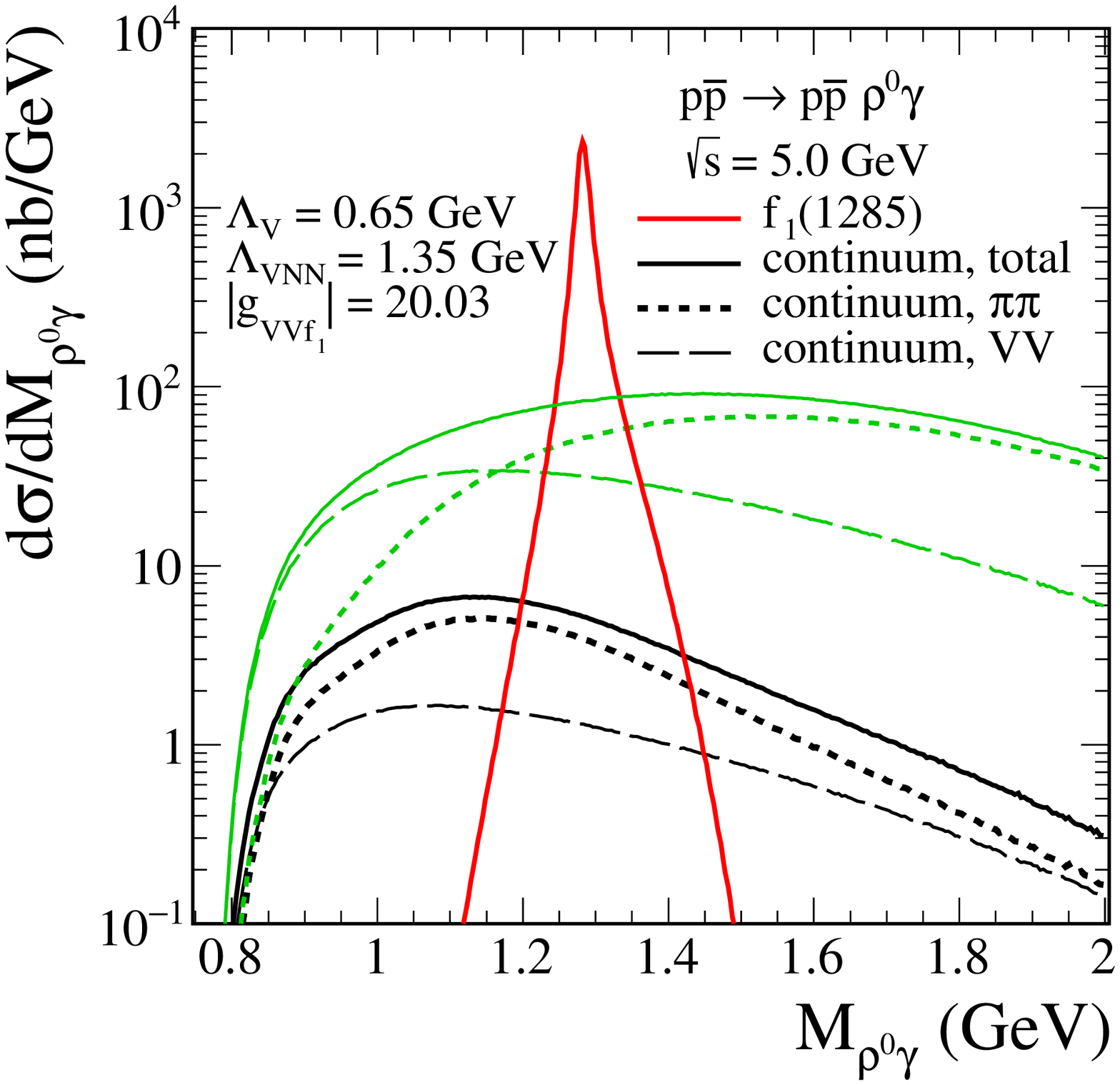}
\caption{
Invariant mass distributions of the $\rho^{0} \gamma$ system
for the HADES experiment (left panels) 
and the PANDA experiment (right panels).
The $VV \to f_{1}(1285)$ resonance term and 
continuum processes via $\pi \pi$ and $VV$ fusion are shown.
Results in the top panels are for
the parameter values of (\ref{aLam1.0}),
while in the bottom panels for those of (\ref{aLam0.65}).
In the calculations we take $\Lambda_{M} = 1.45$~GeV and
$\Lambda_{\pi NN} = 1.0$~GeV.
In the bottom panels, the green lines
(upper lines for the same type) 
correspond to the continuum processes 
without reggeization.}
\label{fig:M_rhogam}
\end{figure}

In Fig.~\ref{fig:M_rhogam} we show the invariant mass distributions
of the $\rho^0 \gamma$ system
at $\sqrt{s} = 3.46$~GeV (HADES experiment)
and at $\sqrt{s} = 5.0$~GeV (PANDA experiment).
The red solid line represents the
$f_1(1285)$ signal via the $\rho\rho$ and $\omega \omega$ 
fusion processes while the other lines represent the background
corresponding to the processes via the $VV$ ($\omega \omega$, 
$\rho \omega$) and $\pi \pi$ fusion 
shown by the diagrams of Fig.~\ref{fig:diagrams_rhogam}.
Results for the two sets of parameters
(\ref{aLam1.0}) and (\ref{aLam0.65}) which correspond to
the top and bottom panels, respectively, are shown.
In the $f_{1}$ meson propagator (\ref{P_f1}) 
we take $\Gamma_{f_{1}} = 18.4$~MeV 
measured in the CLAS experiment; see (\ref{f1_CLAS}).
For the set of parameters (\ref{aLam1.0}) 
the $VV$-continuum contribution,
due to the small value of $\Lambda_{VNN}$,
turns out to be negligible.
The situation changes when we use the parameter set of (\ref{aLam0.65}). 
But still the $\pi \pi$-continuum contribution 
is larger than the $VV$-continuum contribution.
In both cases the $f_1(1285)$ resonance is clearly visible,
even without the reggeization effects in the continuum processes.
This result makes us rather optimistic that
an experimental study of the $f_{1}$ in the $\rho^{0} \gamma$
decay channel should be possible.

In our calculations we find practically
no interference effects between
the $\pi \pi$ and $VV$ fusion contributions
in the continuum.
For our exploratory study we have neglected
interference effects between the background 
$\rho^{0} \gamma$ and the signal 
$f_{1} \to \rho^{0} \gamma$ processes. 
We have also neglected the background processes
due to bremsstrahlung of $\gamma$ and $\rho^{0}$
from the nucleon lines.
For an analysis of real data these effects should be
included or at least estimated.
But this goes beyond the scope of our present paper.

%---------------------------------------------------------------------
%---------------------------------------------------------------------

Now we wish to discuss the integrated cross sections 
for the reactions $pp \to pp (f_{1} \to \rho^{0} \gamma)$ 
and $p\bar{p} \to p\bar{p} (f_{1} \to \rho^{0} \gamma)$
treated with exact $2 \to 4$ kinematics.
In our calculation we took into account the reggeization effects
according to (\ref{reggeization_2})--(\ref{trajectory})
and the replacements given in (\ref{2to4_replacement}).
We consider two sets of parameters,
(\ref{aLam1.0}) and (\ref{aLam0.65}),
extracted from the CLAS data.
With the parameter values of (\ref{aLam1.0}) we get 
\begin{eqnarray}
&&{\rm for}\; \sqrt{s} = 3.46 \;{\rm GeV}: \;
\sigma_{pp \to pp (f_{1} \to \rho^{0} \gamma)}
%\times {\cal BR}(f_{1}(1285) \to \pi^{+}\pi^{-}\pi^{+}\pi^{-}) 
%=  0.99\; {\rm nb} \,,
= 1.26\; {\rm nb} \,,
\label{f1_3.46_L1.0_rho} \\
&&{\rm for}\; \sqrt{s} = 5.0 \;{\rm GeV}: \;
\sigma_{p\bar{p} \to p\bar{p} (f_{1} \to \rho^{0} \gamma)}
%\times {\cal BR}(f_{1}(1285) \to \pi^{+}\pi^{-}\pi^{+}\pi^{-}) 
%=  9.09\; {\rm nb} \,.
=  11.45\; {\rm nb} \,.
\label{f1_5.0_L1.0_rho}
\end{eqnarray}
With the parameter values of (\ref{aLam0.65}) we get 
\begin{eqnarray}
&&{\rm for}\; \sqrt{s} = 3.46 \;{\rm GeV}: \;
\sigma_{pp \to pp (f_{1} \to \rho^{0} \gamma)}
%\times {\cal BR}(f_{1}(1285) \to \pi^{+}\pi^{-}\pi^{+}\pi^{-}) 
%= 4.25\; {\rm nb} \,,
= 5.38\; {\rm nb} \,,
%= 5.41\; {\rm nb} \,, for \Lambda_{f_{1}} = 1.2~GeV
\label{f1_3.46_L0.65_rho} \\
&&{\rm for}\; \sqrt{s} = 5.0 \;{\rm GeV}: \;
\sigma_{p\bar{p} \to p\bar{p} (f_{1} \to \rho^{0} \gamma)}
%\times {\cal BR}(f_{1}(1285) \to \pi^{+}\pi^{-}\pi^{+}\pi^{-}) 
%= 49.90 \; {\rm nb} \,.
= 62.86\; {\rm nb} \,.
%= 63.49\; {\rm nb} \,.
\label{f1_5.0_L0.65_rho}
\end{eqnarray}
The results (\ref{f1_3.46_L1.0_rho})--(\ref{f1_5.0_L0.65_rho})
are for $\Gamma_{f_{1}} = 18.4$~MeV; see (\ref{f1_CLAS}).
We have checked, that if we take
for the cutoff parameter $\Lambda_{f_{1}} = 1.2$~GeV 
in (\ref{F_f1_ff})
that the cross sections will increase slightly,
by about $1.6\, \%$.

Now we compare the above results 
with those estimated as
\begin{eqnarray}
\sigma_{pp \to pp (f_{1} \to \rho^{0} \gamma)}
= \sigma_{pp \to pp f_{1}}
\times {\cal BR}(f_{1}(1285) \to \rho^{0} \gamma)
\label{xs_f1_rhogamma_BR}
\end{eqnarray}
with the corresponding values 
of the $2 \to 3$ cross sections from
(\ref{2to3_f1_3.46_L1.0})--(\ref{2to3_f1_5.0_L0.65}) 
and
${\cal BR}(f_{1} \to \rho^{0} \gamma)$
from CLAS (\ref{Gamma_rhogamma_CLAS}).
For the parameter set (\ref{aLam1.0})
we get
\begin{eqnarray}
&&{\rm for}\; \sqrt{s} = 3.46 \;{\rm GeV}: 
\;\sigma_{pp \to pp (f_{1} \to \rho^{0} \gamma)} =
1.00^{+0.28}_{-0.36}\; {\rm nb}\,,
\label{xs_f1_rhogamma_H_1.0}\\
&&{\rm for}\; \sqrt{s} = 5.0 \;{\rm GeV}: 
\;\sigma_{p\bar{p} \to p\bar{p} (f_{1} \to \rho^{0} \gamma)} =
10.35^{+2.89}_{-3.72} \; {\rm nb}\,.
\label{xs_f1_rhogamma_P_1.0}
\end{eqnarray}
For the parameter set (\ref{aLam0.65})
we get
\begin{eqnarray}
&&{\rm for}\; \sqrt{s} = 3.46 \;{\rm GeV}: 
\;\sigma_{pp \to pp (f_{1} \to \rho^{0} \gamma)} =
3.84^{+1.08}_{-1.38}\; {\rm nb}\,,
\label{xs_f1_rhogamma_H}\\
&&{\rm for}\; \sqrt{s} = 5.0 \;{\rm GeV}: 
\;\sigma_{p\bar{p} \to p\bar{p} (f_{1} \to \rho^{0} \gamma)} =
51.79^{+14.50}_{-18.64} \; {\rm nb}\,.
\label{xs_f1_rhogamma_P}
\end{eqnarray}
The errors in (\ref{xs_f1_rhogamma_H_1.0})--(\ref{xs_f1_rhogamma_P})
come from the uncertainty of ${\cal BR}(f_{1} \to \rho^{0} \gamma)$; see (\ref{Gamma_rhogamma_CLAS}).
The results (\ref{f1_3.46_L1.0_rho})--(\ref{f1_5.0_L0.65_rho})
are larger than the corresponding central values of 
(\ref{xs_f1_rhogamma_H_1.0})--(\ref{xs_f1_rhogamma_P}).

For $p \bar{p}$ at $\sqrt{s} = 5.0$~GeV 
we obtain about 10 times larger cross section 
than for $pp$ at $\sqrt{s} = 3.46$~GeV;
see (\ref{f1_5.0_L0.65_rho}) and (\ref{f1_3.46_L0.65_rho}). respectively.
Thus we predict a large cross section 
for the exclusive axial-vector $f_{1}(1285)$ production 
compared to the continuum processes considered 
in the $\rho^{0} \gamma$ channel.

In Table~\ref{tab:table0} we have collected integrated 
cross sections in nb
for the continuum processes considered.
These numbers were obtained for
$g_{\rho \omega \pi} = 10.0$,
$\Lambda_{M} = 1.45$~GeV in (\ref{rho_ome_pi})--(\ref{F_pi}),
$\Lambda_{VNN} = 1.35$~GeV in (\ref{F_V_M}), and
$\Lambda_{\pi NN} = 1.0$~GeV in (\ref{PSNN}).
The reggeization effects were included.
We can observe very small numbers
for the production of $\rho^{0}\rho^{0}$ 
at $\sqrt{s} = 3.46$~GeV
which is caused by the threshold behaviour of the process
(the assumption of a fixed $\rho^{0}$-meson mass
of $m_{\rho} = 0.775$~GeV in the calculation)
and limited phase space.
%%%%%%%%%%%%%%%%%%%%%%%%%%%%%%%%%%%%%%%%
\begin{table}[!ht]
\centering
\caption{The integrated cross sections in nb 
for the continuum processes
in proton-(anti)proton collisions.
We show results for the $VV$- and $\pi\pi$-fusion
contributions separately and for their coherent sum (``total'').}
\label{tab:table0}
\begin{tabular}{|l|c|c|c|c|}
\hline
\multirow{2}{1cm}{\;\;\;\;Reaction} &
\multirow{2}{1.7cm}{$\sqrt{s}$ (GeV)} &
\multicolumn{3}{c|}{$\sigma$ (nb)}\\
\cline{3-5} & & $VV$ fusion & $\pi\pi$ fusion & total\\
\hline
$pp \to pp \rho^{0}\rho^{0}$ & 3.46 
%& $\Lambda_{VNN} = 1.35$~GeV
& $0.6 \times 10^{-3}$ & $6.7 \times 10^{-3}$ & $7.3 \times 10^{-3}$ \\
\hline 
$p\bar{p} \to p\bar{p} \rho^{0}\rho^{0}$ & 5.0 
%& $\Lambda_{VNN} = 1.35$~GeV
& 156.24 & 1823.25 & 1979.50 \\
\hline
$pp \to pp \rho^{0}\gamma$ & 3.46 
%& $\Lambda_{VNN} = 1.35$~GeV
& 0.13 & 0.45 & 0.58 \\
\hline 
$p\bar{p} \to p\bar{p} \rho^{0}\gamma$ & 5.0 
%& $\Lambda_{VNN} = 1.35$~GeV
& 1.06 & 2.49 & 3.56 \\
\hline 
\end{tabular}
\end{table}
%%%%%%%%%%%%%%%%%%%%%%%%%%%%%%%%%%%%%%%%

Now we compare the cross section 
for the $\rho \rho$ continuum 
from Table~\ref{tab:table0}
to the cross section for the $f_{1}(1285)$ signal according to
\begin{eqnarray}
\sigma_{pp \to pp (f_{1} \to 2 \pi^{+} 2 \pi^{-})}
= \sigma_{pp \to pp f_{1}} \times
{\cal BR}(f_{1} \to 2 \pi^{+} 2 \pi^{-})\,
\label{xs_f1_rhorho}
\end{eqnarray}
with $\sigma_{pp \to pp f_{1}}$ from Fig.~\ref{fig:sig_tot_W}
and a branching ratio
\begin{eqnarray}
{\cal BR}(f_{1}(1285) \to 2\pi^{+}2\pi^{-}) 
= (10.9 \pm 0.6) \%
\label{BR_f1_1285_4pi}
\end{eqnarray}
from \cite{Zyla:2020zbs}.
Taking into account the values of 
$\sigma_{pp \to pp f_{1}}$
in (\ref{xs_f1_rhorho})
that correspond to (\ref{aLam1.0}) we get
\begin{eqnarray}
{\rm for}\; \sqrt{s} &=& 3.46 \;{\rm GeV}:\;
\sigma_{pp \to pp (f_{1} \to 2 \pi^{+} 2 \pi^{-})} 
= 4.38 \; {\rm nb} \,,
\label{f1_3.46_4pi} \\
{\rm for}\; \sqrt{s} &=& 5.0 \;{\rm GeV}:\;
\sigma_{p\bar{p} \to p\bar{p} (f_{1} \to 2 \pi^{+} 2 \pi^{-})} 
= 45.11 \; {\rm nb} \,.
\label{f1_5.0_4pi}
\end{eqnarray}
With (\ref{aLam0.65}) we get
\begin{eqnarray}
{\rm for}\; \sqrt{s} &=& 3.46 \;{\rm GeV}:\;
\sigma_{pp \to pp (f_{1} \to 2 \pi^{+} 2 \pi^{-})} 
= 16.73 \; {\rm nb} \,,
\label{f1_3.46_4pi_aLam0.65} \\
{\rm for}\; \sqrt{s} &=& 5.0 \;{\rm GeV}:\;
\sigma_{p\bar{p} \to p\bar{p} (f_{1} \to 2 \pi^{+} 2 \pi^{-})} 
= 225.79 \; {\rm nb} \,.
\label{f1_5.0_4pi_aLam0.65}
\end{eqnarray}
%
%Without the reggeization, 
%the above cross sections for $\sqrt{s} = 3.46$~GeV
%will be larger approximately by a factor of 1.8.
These roughly estimated results show that,
for the cases treated here,
the background processes considered 
in the $\rho^{0} \rho^{0}$ channel
(see Table~\ref{tab:table0})
can be important only for $\sqrt{s} = 5.0$~GeV
in the $p \bar{p}$ case.

The reaction $pp \to pp \rho^{0} \rho^{0}$ 
is treated technically as a $2 \to 4$ process.
A better approach would be to consider
the $p p \to p p \pi^+ \pi^-\pi^+ \pi^-$ reaction, as a $2 \to 6$ process. 
This is however beyond the scope of the present study. 
In addition, as will be discussed in the following, the background 
for the $p p \to p p \pi^+ \pi^-\pi^+ \pi^-$ reaction measured 
long ago by the bubble chamber experiment \cite{Alexander:1967zz} 
was found to be much larger than 
the result for the continuum terms (``total'') 
presented in Table~\ref{tab:table0}.

%--------------------------------------------------
\section{HADES and PANDA experiments}
\label{sec:HADES_exp}
%--------------------------------------------------

The HADES (High Acceptance Dielectron Spectrometer)
is a magnetic spectrometer located at the SIS18 accelerator 
in the Facility for Antiproton and Ion Research (FAIR)  
in Darmstadt (Germany) \cite{Agakishiev:2009am}. 
It is a versatile detector allowing measurement of charged hadrons 
(pions, kaons and protons), leptons (electrons and positrons) 
originating from various reactions on fixed proton or nuclear targets 
in the energy regime of a few $A \cdot {\rm GeV}$.
The spectrometer covers the polar angle region $18^\circ < \theta < 80^\circ$ 
and features almost complete azimuthal coverage w.r.t. the beam axis. 
The detector has been recently upgraded by a large area electromagnetic 
calorimeter and a forward detector (for a recent review 
see \cite{Adamczewski-Musch:2021rlv}) extending the coverage to very 
forward region ($0.5^\circ < \theta < 7.5^\circ$). 
These upgrades allow to measure hadron decays involving photons 
and significantly improve acceptance for protons and hyperons 
which at these energies are emitted to large extent in forward directions. 

The spectrometer is specialized for electron-positron pair detection 
but it also provides excellent hadron (pion, kaon, proton)-identification capabilities. 
It has a low material budget and consequently features 
an excellent invariant mass resolution for electron-positron pairs 
of $\Delta M/M \approx 2.5 \,\%$ in the $\rho/\omega/\phi$ vector meson
mass region.

The PANDA (antiProton ANnihilations at DArmstadt) detector 
is currently under construction at FAIR.
PANDA will utilise a beam of antiprotons, provided by 
the High Energy Storage Ring (HESR),
and with its almost full solid-angle coverage will be a detector 
for precise measurements in hadron physics.
HESR will deliver antiprotons with momenta from 1.5~GeV/$c$ up to 15~GeV/$c$ 
(which corresponds to $\sqrt{s} \simeq 2.25-5.47$~GeV)
impinging on a cluster jet or pallet proton target placed in PANDA. 
The scientific programme of PANDA is very broad 
and includes charmonium and hyperon spectroscopy, 
elastic proton form-factor measurements, 
searches of exotic states and studies of in-medium hadron properties 
(for a recent review of stage-one experiments see \cite{Fischer:2021kcr}).

The luminosity of both detectors are comparable and 
are at the level of $L = 10^{31} {\rm cm}^{-2}s^{-1}$ 
(after first years of operation and completion of the detector 
PANDA will increase it by one order of magnitude). 
 
For the count rate estimates and signal to background considerations
for the $f_1$ meson production we will use 
the properties of the HADES detector. 
This presents a ``worse case'' scenario.
As it was shown in previous sections cross sections for the meson production 
in proton-proton interactions are about a factor 10 lower than 
for the proton-antiproton case. 
Furthermore, the PANDA detector features also larger acceptance 
for the reaction of multi-particle finals and presents better opportunities 
for the studies discussed in this work. 
On the other hand HADES will measure proton-proton reactions 
at the c.m.~energy $\sqrt{s} = 3.46$~GeV 
(proton beam energy $E_{\rm kin} = 4.5$~GeV) already in 2021. 
Hence it will provide first valuable experimental results to verify our 
model predictions.

%--------------------------------------------------------------
\subsection{Simulation for $2\pi^+ 2\pi^-$ and $\pi^+\pi^-\eta$ decay channels}
\label{sec:simulation}
%--------------------------------------------------------------

We have considered production of the $f_1(1285)$ meson
in proton-proton reactions 
and its decay into final states with four charged pions reconstructed 
in the HADES detector. 
For the $f_{1}$ production cross section we have assumed 
$\sigma_{f_{1}} = 150$~nb [estimate using the C7 parameter set; 
see (\ref{aLam0.65}) and (\ref{2to3_f1_3.46_L0.65})].

Two reaction channels were simulated:
\begin{eqnarray}
&&\bullet \quad p+p \to p+p+f_{1} (\to 2\pi^+2\pi^-) \;
{\rm with} \; {\cal BR}(f_{1}(1285) \to 2\pi^{+}2\pi^{-}) 
= 10.9 \,\%\,, \qquad
\label{HADES_4pi}\\
&&\bullet \quad p+p \to p+p+f_{1} (\to \pi^+\pi^-\eta) \;
{\rm with} \; {\cal BR}(f_{1}(1285) \to \pi^+\pi^-\eta) 
= 35 \,\%\,.
\label{HADES_pipieta}
\end{eqnarray}
In the second case the $\eta$ meson is reconstructed 
via the $\eta \to \pi^+\pi^-\pi^0$ decay channel, 
hence the final state has also four charged pions.
The neutral pion from the $\eta$ decay can be reconstructed 
via missing mass technique or via two photon decay. 
However, the latter case has smaller total reconstruction efficiency 
(see below for details). 
  
The $f_1(1285)$ meson decay into four charged pions 
has been simulated using the {\textsc PLUTO} event generator 
\cite{Frohlich:2007bi,Frohlich:2009sv,Frohlich:2009eu}.
For the meson reconstruction four pions from the decay and at least 
one final state proton have been demanded in the analysis to establish 
exclusive channel identification. 
The HADES acceptance and reconstruction efficiencies 
for protons and pions 
have been parametrized as a function of the polar 
and azimuthal angles and the momentum. 
Furthermore, a momentum resolution $\Delta p/p = 2\,\%$ of the spectrometer 
for charged tracks has been taken into account in the simulation, 
as described in \cite{Agakishiev:2009am}.
        
For the $pp \to pp 2\pi^+ 2\pi^-$ reaction 
a total cross section $\sigma_{back} = (227 \pm 23)\,\mu$b
has been measured; see Table~I of \cite{Alexander:1967zz}.
This reaction was measured in \cite{Alexander:1967zz} 
at slightly higher energies $E_{\rm kin} = 4.64$~GeV 
(corresponding to proton beam momentum $P = 5.5~{\rm GeV}/c$ or 
$\sqrt{s} \simeq 3.5$~GeV).\footnote{In \cite{Alexander:1967zz} 
a four-pion invariant-mass
histogram was shown for the $pp \to pp 2\pi^+ 2\pi^-$ reaction
[see Fig.~28~(a) therein].
No attempt has been made to analyse this channel there
and only an upper limit on the resonance
(there $f^{*}(1250) \to 2\pi^+ 2\pi^-$)
production cross section, $\sigma < 15~\mu$b, was estimated.}

We tried to understand the large background 
in the $\pi^+ \pi^- \pi^+ \pi^-$ channel.
We analysed a few contributions due to double nucleon excitations.
We considered the following processes:
\begin{eqnarray}
&&pp \to N(1440) N(1440) \quad {\rm via} \; \pi^{0},\; \sigma\,,
\label{pp_1440_1440}\\
&&pp \to N(1440) N(1535) \quad {\rm via} \; \pi^{0}\,,
\label{pp_1440_1535}\\
&&pp \to N(1535) N(1535) \quad {\rm via} \; \pi^{0},\; \eta,\; \rho^{0}\,.
\label{pp_1535_1535}
\end{eqnarray}
Both resonances have considerable branching fraction to
the $N \pi \pi$ channel and the $N(1535)$
to the $N \eta$ channel; see PDG \cite{Zyla:2020zbs}.
In our evaluation (estimation) 
we used effective Lagrangians and relevant parameters from \cite{Ouyang:2009kv}.
These parameters were found in \cite{Ouyang:2009kv}
to describe the total cross section
for the the $pp \to pn \pi^{+}$ reaction
measured in the close-to-threshold region.
The coupling constants and the cutoff parameters
in the monopole form factors 
used in the calculation are the following ones:
\begin{eqnarray}
&&g_{N(1440)N\sigma}^{2}/4 \pi = 3.20\,, \quad \;\,
\Lambda_{N(1440)N\sigma} = 1.1\;{\rm GeV}\,, \nonumber \\
&&g_{N(1440)N\pi}^{2}/4 \pi = 0.51\,, \quad \;\,
\Lambda_{N(1440)N\pi} = 1.3\;{\rm GeV}\,, \nonumber \\
&&g_{N(1535)N\pi}^{2}/4 \pi = 0.037\,, \quad 
\Lambda_{N(1535)N\pi} = 1.3\;{\rm GeV}\,, \nonumber \\
&&g_{N(1535)N\eta}^{2}/4 \pi = 0.34\,, \quad \;\;
\Lambda_{N(1535)N\eta} = 1.3\;{\rm GeV}\,, \nonumber \\
&&g_{N(1535)N\rho}^{2}/4 \pi = 0.097\,, \quad 
\Lambda_{N(1535)N\rho} = 1.3\;{\rm GeV}\,.
\label{pp_NN_parameters}
\end{eqnarray}
Similar values were also taken in \cite{Cao:2009ea}
for the $pn \to d \phi$ reaction.
To describe the total cross sections of the $p N \to N N \pi \pi$ 
and $\bar{p} N \to \bar{N} N \pi \pi$ reactions
measured in the near-threshold region
the cutoff parameters
$\Lambda_{N^{*}NM} = 1.0$~GeV were assumed 
in \cite{Cao:2010km,Cao:2010ji}.
Therefore, our estimates 
for the reactions (\ref{pp_1440_1440})--(\ref{pp_1535_1535}) 
with the parameters given in (\ref{pp_NN_parameters}) 
should be treated rather as an upper limit.

There is a question about the role of the $\eta'$ exchange
in the reaction (\ref{pp_1535_1535}). 
For example, in \cite{Cao:2008st} 
sub-threshold resonance-dominance of the $N(1535)$ was assumed 
with $g_{N(1535)N\eta'}^{2} / 4 \pi = 1.1$
to describe both the $\pi N \to \eta' N$ 
and $NN \to NN \eta'$ cross section data.
However, it was shown in \cite{Huang:2012xj}
that the $N(1535)$ contribution 
is not necessary in these processes 
(see Figs.~9--14 of \cite{Huang:2012xj})
or, at least, its significant role 
(significant coupling strength of $N(1535) \to \eta' N$ used
in \cite{Sibirtsev:2003ng,Cao:2008st}) was precluded.

For energy $\sqrt{s} = 3.5$~GeV we get the cross section
for the $pp \to N(1440) N(1440)$ reaction of the order of 0.8~mb.
With the input from \cite{Morsch:1992vj,AlvarezRuso:1997mx,Hernandez:1998ra},
$g_{N(1440)N\sigma}^{2}/4 \pi = 1.33$ and
$\Lambda_{N(1440)N\sigma} = 1.7\;{\rm GeV}$,
we get even smaller cross section 
by about $30\,\%$. 
For the $pp \to N(1440) N(1535)$ reaction 
we get the cross section of 10~$\mu$b 
and for the $pp \to N(1535) N(1535)$ reaction about 7~$\mu$b.
So we conclude that the double excitation of the $N(1440)$ resonances 
via the $\sigma$-meson exchange
is probably the dominant mechanism of this type 
in the $pp \to pp 2\pi^+ 2\pi^-$ reaction.
This is due to the large $N(1440) N \sigma$ coupling.
Taking
${\cal BR}(N(1440) \to p \pi^{+} \pi^{-}) = 0.1$
we get 
$\sigma_{pp \to N^{*}N^{*} \to pp 2\pi^{+} 2 \pi^{-}} \simeq 80$~$\mu$b.
This background is much higher than that for the $\omega \omega$- and 
$\pi \pi$-fusion mechanisms considered in Sec.~\ref{sec:num_results}; 
see Table~\ref{tab:table0}.

%For the background $pp \to pp 2\pi^+ 2\pi^-$ reaction with  
%a total cross section $\sigma_{back} = 227 \,\mu$b
%has been assumed;
%see Table~I of \cite{Alexander:1967zz}.

The background channel was simulated assuming
multi-pion production 
via two intermediate charged baryon resonances, 
each of them decaying into two pion final states.
Since the exact production mechanism is not known
we have assumed production of two $N(1440)$.
We take the total cross section 
for the background in the four pion
channel to be $\sigma^{4\pi}_{back} = 227 \,\mu$b \cite{Alexander:1967zz}.
For the signal we take 
$\sigma^{4\pi}_{f_{1}} = 
\sigma_{f_{1}} \times {\cal BR}(f_{1}(1285) \to 2\pi^{+}2\pi^{-}) = 16$~nb [see (\ref{f1_3.46_4pi_aLam0.65})].
A total reconstruction efficiency $\epsilon = 2\, \%$ 
for the $f_1$ decay in four charged pions has been estimated 
and the signal is hardly visible on the top of the background. 
We conclude that it will be difficult to see a peak
on the four-pion continuum without additional cuts. 

Now we wish to consider the $pp \to pp \pi^+ \pi^- \pi^+ \pi^- \pi^0$ reaction.
In Table~\ref{tab:xsection} we have collected the cross sections
which we use in the simulations.
We take the total cross section for the continuum background 
in the five pion channel to be 
$\sigma^{5\pi}_{back} = 88 \,\mu$b \cite{Alexander:1967zz},
which seems to be rather an upper limit for the background.
Taking into account both processes 
(\ref{pp_1440_1535}) and (\ref{pp_1535_1535})
we estimate the cross section 
in the $pp \pi^+ \pi^- \eta$ final state 
of the order of 0.8~$\mu$b
to be compared to 53~nb
for the signal $p p \to p p f_1 (\to \pi^+ \pi^- \eta)$.
We include
${\cal BR}(\eta \to \pi^{+} \pi^{-} \pi^{0}) = 0.23$
to get the
$pp \pi^+ \pi^- \pi^+ \pi^- \pi^0$ final state.
For the $pp \to pp \pi^{+}\pi^{-}\omega(\to \pi^{+} \pi^{-} \pi^{0})$ contribution
we assume about $0.07 \,\mu$b 
taking 
${\cal BR}(\omega \to \pi^{+} \pi^{-} \pi^{0}) = 0.89$ 
\cite{Zyla:2020zbs}.
The narrow width of the $\eta$ meson 
allows to impose an extra mass cut on the $\pi^+\pi^-\pi^0$ invariant mass 
and suppresses the multi-pion background efficiently.
The reconstruction of this decay channel 
has a smaller efficiency ($\epsilon = 0.8\, \%$)
compared to the $2 \pi^+ 2 \pi^-$ decay channel.
%%%%%%%%%%%%%%%%%%%%%%%%%%%%%%%%%%%%%%%%
\begin{table}[!ht]
\centering
\caption{Contributions and cross sections used in the simulations
of the $pp \to pp 2\pi^{+} 2\pi^{-} \pi^{0}$ reaction.}
\label{tab:xsection}
\begin{tabular}{|l|c|l|}
\hline
Contribution & Cross section ($\mu b$)
&      \\
\hline
(1) $pp \to pp \pi^{+} \pi^{-}\pi^{+} \pi^{-} \pi^{0}$ & 88
& $\sigma = (88 \pm 14)$~$\mu$b \cite{Alexander:1967zz}, $P = 5.5$~GeV/$c$\\
\hline
(2) $pp \to pp \pi^{+} \pi^{-} \eta (\to \pi^{+} \pi^{-} \pi^{0})$ & $0.18$ 
& estimates via two $N^{*}$ resonances,\\
 & 
& see (\ref{pp_1440_1535}) and (\ref{pp_1535_1535})\\
\hline
(3) $pp \to pp \pi^{+} \pi^{-} \omega (\to \pi^{+} \pi^{-}\pi^{0})$ & 0.07
& $\sigma = (0.09 \pm 0.03)$~$\mu$b \cite{Danieli:1971tf}\\
 & 
& for $pp \to pp \pi^{+} \pi^{-} \omega$ at $P = 6.92$~GeV/$c$\\
\hline
(4) $pp \to pp f_{1} [\to \pi^{+} \pi^{-} \eta (\to \pi^{+} \pi^{-}\pi^{0})]$ & $0.012$ 
& $\sigma = (3.2 - 12.4)$~nb, see (\ref{2to3_f1_3.46_L1.0}) and (\ref{2to3_f1_3.46_L0.65})\\
\hline 
\end{tabular}
\end{table}
%%%%%%%%%%%%%%%%%%%%%%%%%%%%%%%%%%%%%%%%

\begin{figure}[!ht]
(a)\includegraphics[width=7.7cm]{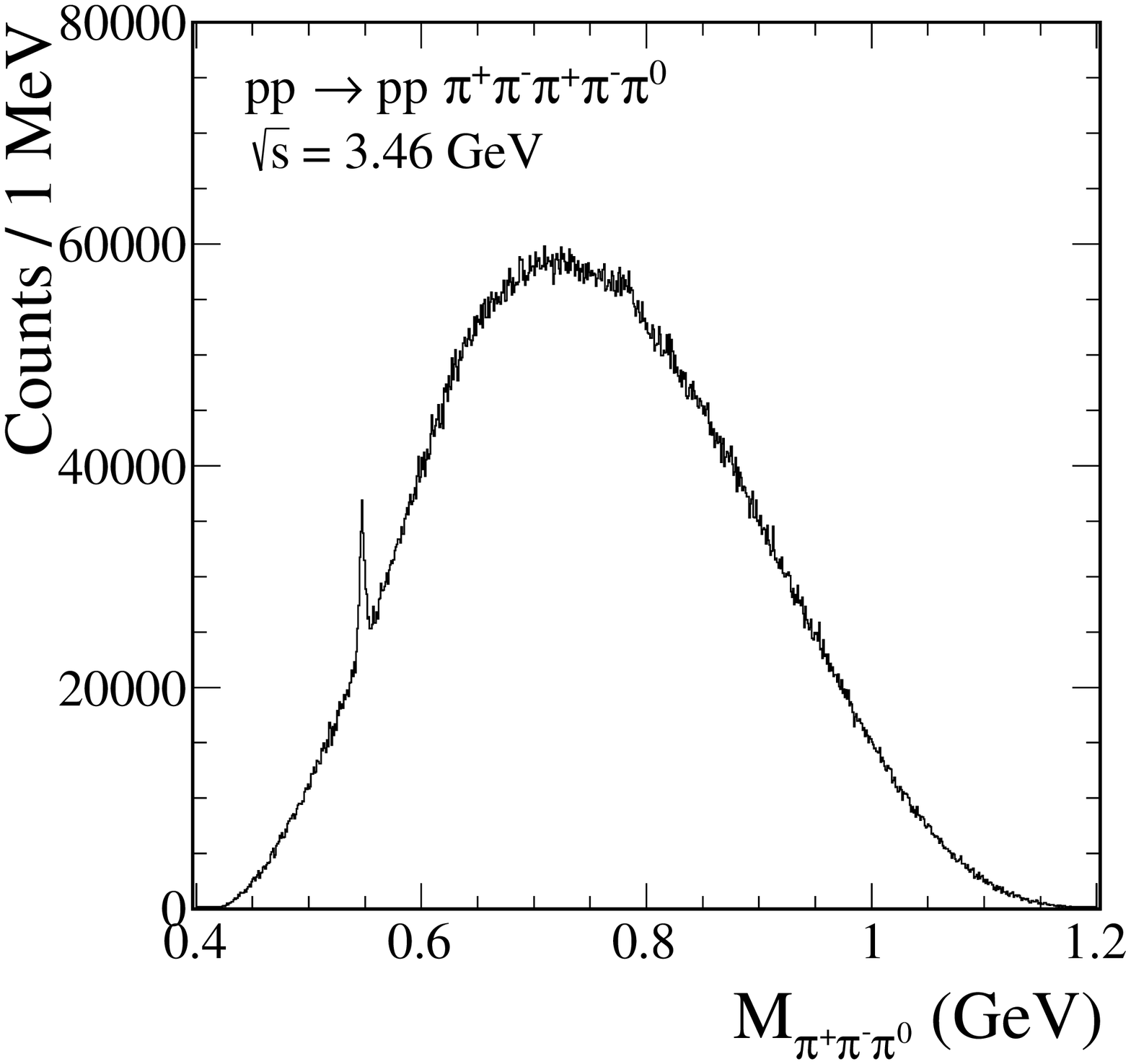}
(b)\includegraphics[width=7.7cm]{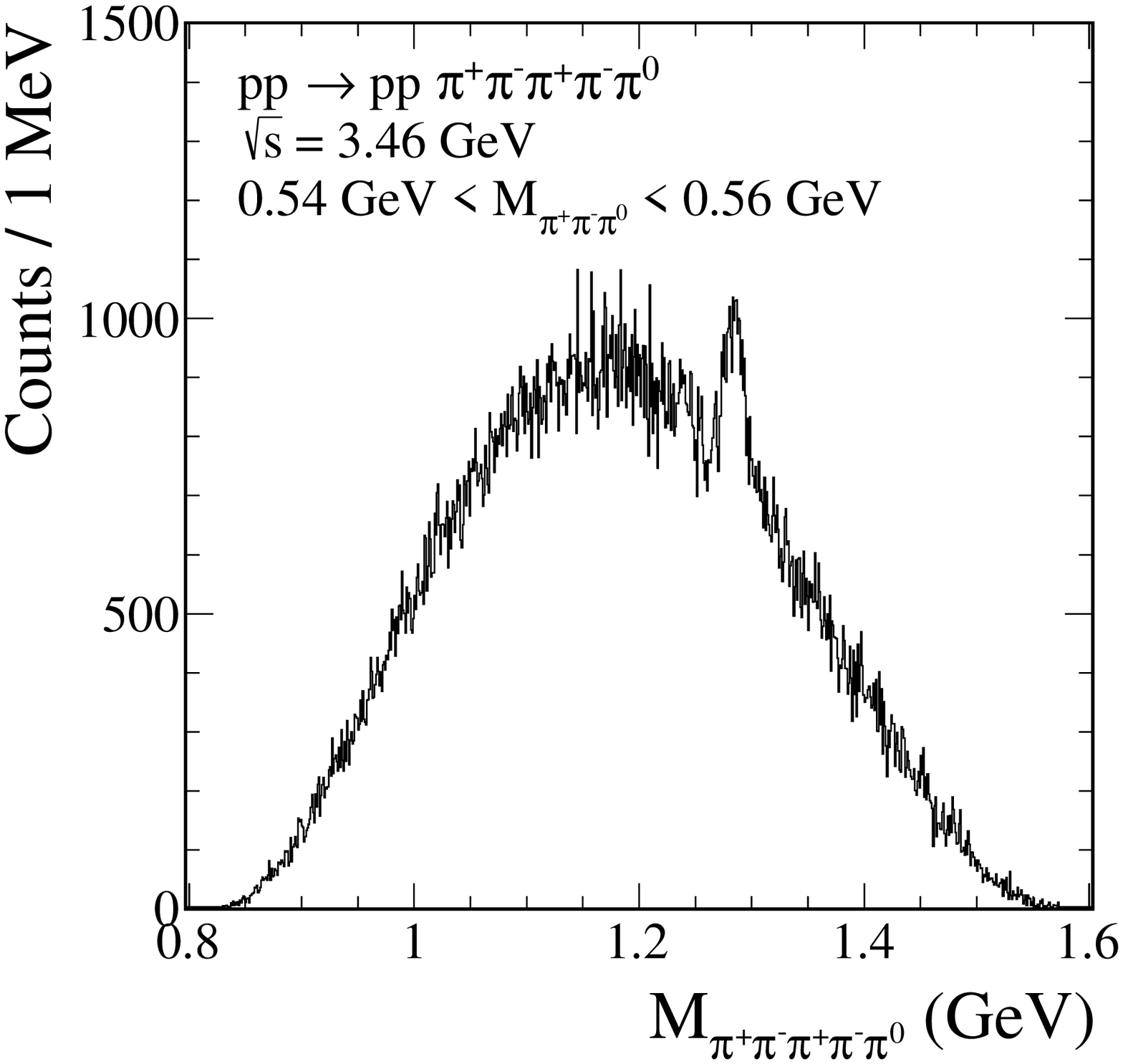}
\caption{Invariant mass distributions
of (a) $\pi^+\pi^-\pi^0$
and (b) $\pi^+ \pi^- \pi^+ \pi^- \pi^0$
of the $pp \pi^+ \pi^- \pi^+ \pi^- \pi^0$ final state
corresponding to the measurement
with the $p+p$ reactions at $E_{\rm kin} = 4.5$~GeV 
($\sqrt{s} = 3.46$~GeV) with the HADES apparatus.
All components listed in Table~\ref{tab:xsection}
were included in the simulation.
The result in panel (b) includes 
the cut on the $\eta$ meson mass
$0.54 \;{\rm GeV} < M_{\pi^+\pi^-\pi^0} < 0.56 \;{\rm GeV}$.}
\label{fig:f1_5pi}
\end{figure}

Figure~\ref{fig:f1_5pi}~(a) shows the reconstructed 
invariant mass of $\pi^+\pi^-\pi^0$ 
with a clear signal of $\eta$ meson on top of a large background. 
The shape of this background was also studied by multi-pion production 
with uniform phase space distribution. 
No significant difference was found.
A cut on the $\eta$ meson mass 
$0.54 \;{\rm GeV} < M_{\pi^+\pi^-\pi^0} < 0.56 \;{\rm GeV}$ allows 
for efficient background subtraction
and observation of $f_1(1285)$ meson peak
[see Fig.~\ref{fig:f1_5pi}~(b)].
The expected signal (about 4000 counts) 
and background distributions in Fig.~\ref{fig:f1_5pi}~(b)
display projections for about 30 days of measurement.
%-----------------------------------------------------
%-----------------------------------------------------

\begin{figure}[!ht]
\includegraphics[width=7.7cm]{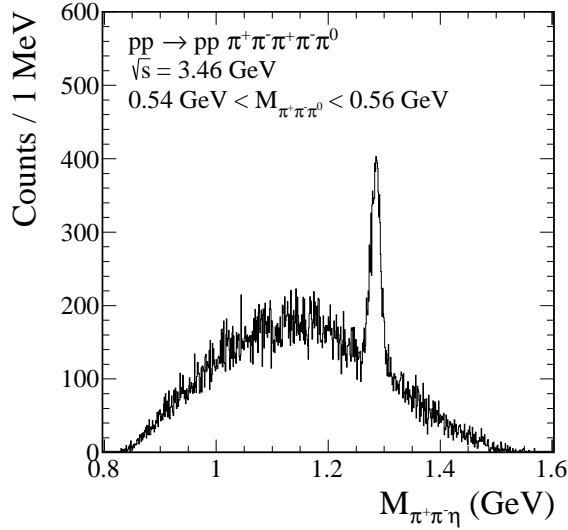}
\caption{Invariant mass distribution of $\pi^+\pi^-\eta$
observed in the $p p \pi^+ \pi^- \pi^+ \pi^- \pi^0$ final state
corresponding to the measurement
with the $p+p$ reactions at $E_{\rm kin} = 4.5$~GeV 
($\sqrt{s} = 3.46$~GeV) with the HADES apparatus.
Here, the two contributions (2) and (4) of Table~\ref{tab:xsection}
were included.
The result includes the cut on the $\eta$ meson mass
$0.54 \;{\rm GeV} < M_{\pi^+\pi^-\pi^0} < 0.56 \;{\rm GeV}$.}
\label{fig:eta_pipieta}
\end{figure}

So far we have considered the 5-pion background
with all components (1, 2, 3) listed in Table~\ref{tab:xsection}.
The contribution (1) can be, in principle, eliminated
by using side-band subtraction method.
%Therefore we concentrate on ...... 
We wish to discuss now separately the contribution (2),
in the $\pi^+ \pi^- \eta$ mesonic state,
to proof feasibility of the $f_1(1285)$ measurement.
In Fig.~\ref{fig:eta_pipieta} we make such a comparison.
The nonreducible background contribution
from double excitation of $N^{*}$ resonances
has a broader distribution than the $VV \to f_{1}$ signal.
With our estimate of the cross section
for the $p p \to p p f_1(1285)$ reaction
(see Table~\ref{tab:xsection})
we expect that the $f_1(1285)$ could be observed in
the $\pi^+ \pi^- \eta$ ($\to \pi^+ \pi^- \pi^0$) channel.

%-----------------------------------------------------
%-----------------------------------------------------

%--------------------------
\section{Conclusions}
\label{sec:conclusions}
%--------------------------

In the present paper we have discussed the possibility to observe 
the $f_{1}(1285)$ in the $p p \to p p f_{1}(1285)$ reaction at energies 
close to the threshold where the pomeron-pomeron fusion, known to be the
dominant mechanism at high energies,
is expected to give only a very small contribution.
Two different mechanisms have been considered:
(a) $\omega \omega \to f_{1}(1285)$ fusion and
(b) $\rho^0 \rho^0 \to f_{1}(1285)$ fusion.
We have estimated the cross section for $\sqrt{s} = 3.46$~GeV
for which a measurement will soon be possible for HADES@GSI.

We have presented our method for the derivation of the $V V \to f_{1}(1285)$
vertex for $V = \rho^{0}, \omega$.
The coupling constant $g_{\rho \rho f_1}$ has been extracted
from the decay rate of $f_1 \to \rho^0 \gamma$ 
using the VMD ansatz.
From naive quark model and VMD relations 
we have obtained equality of the $g_{\rho \rho f_{1}}$ and
the $g_{\omega \omega f_1}$ coupling constants;
see Appendix~\ref{sec:appendixA}.
In reality this relation can be expected to hold
at the $20\,\%$ level.
Then, we have fixed the cutoff parameters in the form factors and
the corresponding coupling constants
by fits to the CLAS experimental data 
for the process $\gamma p \to f_{1}(1285) p$.
There, the $\rho$- and $\omega$-exchange contributions
play a crucial role in reproducing
the forward-peaked angular distributions,
especially at higher energies, $W_{\gamma p} > 2.55$~GeV.

The corresponding $\rho \rho$ and $\omega \omega$
fusion amplitudes have been written out explicitly.
The two amplitudes have been used to estimate 
the total and differential cross sections
for c.m. energy $\sqrt{s} = 3.46$~GeV. 
The energy dependence close to the threshold has been discussed.
The distributions in $t$
(see Fig.~\ref{fig:HADES_t})
and the distributions in $\cos\theta_{M}$
(see Fig.~\ref{fig:z3})
seem particularly interesting.
The shape of these distributions gives information 
on the role of the individual fusion processes.

We have discussed the possibility of a measurement of
the $pp \to pp f_{1}(1285)$ reaction 
by the HADES collaboration at GSI. 
For this, the $\pi^+ \pi^- \pi^+ \pi^-$,
$\rho^0 \gamma$, and $\pi^+ \pi^- \eta$
channels, have been considered. 
For the four-pion channel we have estimated the background 
using the cross section from an old bubble chamber experiment 
\cite{Alexander:1967zz}. 
We have found that
the double excitation of the $N(1440)$ resonances 
via the $\sigma$-meson exchange
is probably the dominant mechanism
in the $pp \to pp 2\pi^+ 2\pi^-$ reaction.
The mechanisms considered by us: 
$\pi^0$-$\omega$-$\pi^0$ and $\omega$-$\pi^0$-$\omega$ exchanges
give much smaller background cross sections.
We conclude that it may be difficult to identify 
the $f_{1}(1285)$ meson in this channel. 
The $\rho^0 \gamma$ channel should be much
better suited as far as signal-to-background ratio is considered.
There, however, dominant background channel 
$p p \pi^{+} \pi^{-} \pi^{0}$ is of the order of $2$~mb \cite{Alexander:1967zz} 
and $\rho^{0}$ is so broad that it will not provide sufficient
reductions (as it is the case in $\eta$ decay channel).
In our opinion the $\pi^+ \pi^- \eta (\to \pi^+ \pi^- \pi^0)$ channel
is especially promising. 
We have performed feasibility studies
and estimated that a 30-days measurement with HADES
should allow to identify the $f_1(1285)$ meson 
in the $pp\pi^+ \pi^- \eta$ final state. No simulation 
of the $\pi^+ \pi^- \eta (\to \pi^+ \pi^- \pi^0)$
channel has been done for PANDA energies.

In \cite{Aceti:2015zva} the $f_{1}(1285)$ decays into 
$a_{0}(980) \pi^{0}$, $f_{0}(980) \pi^{0}$ 
and isospin breaking were studied.
An interesting proposal was also discussed 
in \cite{Achasov:2018mzt,Achasov:2019vcs}:
to study the anomalous isospin breaking decay
$f_{1}(1285) \to \pi^{+}\pi^{-}\pi^{0}$ 
in central exclusive production of the $f_{1}$.
There is another important decay channel, $K \bar{K} \pi$,
with branching fraction 9\% 
\cite{Zyla:2020zbs}
which can be used for $f_{1}$ meson studies in CEP.
See also \cite{Aceti:2015pma} for a discussion
of the $K \bar{K} \pi$ decay 
and the nature of the $f_{1}(1285)$ meson.

Predictions for the PANDA experiment at FAIR,
for the $p \bar{p} \to p \bar{p} f_{1}(1285)$ reaction, 
have also been presented.
The possibility to study the underlying reaction mechanisms 
have been discussed.
For the $VV \to f_{1}(1285)$ fusion processes
for $\sqrt{s} = 5.0$~GeV 
we have obtained about 10 times larger cross sections 
than for $\sqrt{s} = 3.46$~GeV.
Thus we predict a large cross section 
for the exclusive axial-vector $f_{1}(1285)$ production, 
compared to the background continuum processes 
via $VV$ and $\pi \pi$ fusion, in the $\rho^{0} \gamma$ channel.
The $\rho^0 \gamma$ channel seems, 
therefore, also promising for identifying the $f_1(1285)$ meson.

To conclude: we have shown that the study of $f_1(1285)$ production
at HADES and PANDA should be feasible.
From such experiments we will learn more on the nature of the $f_{1}$.
For instance, is it a normal $q \bar{q}$ state or $\bar{K} K^{*}$ molecule
\cite{Aceti:2015pma, Xie:2019iwz}? Can it be described in holographic QCD \cite{Domokos:2009cq}?
In particular, we shall learn from $f_{1}$ CEP at low energies
about the $\rho \rho f_{1}$ and $\omega \omega f_{1}$ coupling strengths.
These in turn are very interesting parameters for the calculations
of light-by-light contributions to the anomalous magnetic moment
of the muon \cite{Dorokhov:2019tjc,Roig:2019reh,Leutgeb:2019gbz,Cappiello:2019hwh,1852275,Szczurek:2020hpc}.
The final aim for studies of $f_{1}$ CEP in proton-proton collisions
should be to have a good understanding of this reaction,
both from theory and from experiment,
in the near threshold region,
in the intermediate energy region 
$8\;{\rm GeV} \lesssim \sqrt{s} \lesssim 30\;{\rm GeV}$, 
and up to high energies available at the LHC as discussed in \cite{Lebiedowicz:2020yre}.

%--------------------
\appendix
%--------------------

%-------------------------------------------------------------------
\section{On the relation of $g_{\rho \rho f_{1}}$ and $g_{\omega \omega f_{1}}$}
\label{sec:appendixA}
%-------------------------------------------------------------------

Here we discuss some simple theoretical ideas
on the relation of these two coupling parameters.
First we note that isospin symmetry alone provides \underline{no} relation.
We shall use the most naive quark model and VMD to get
a handle on the ratio $g_{\rho \rho f_{1}} / g_{\omega \omega f_{1}}$.

In the simple naive quark model the $f_{1}$, $\rho^{0}$ and $\omega$
mesons are represented with the following quark content
\begin{eqnarray}
f_{1}(1285) &\sim & \frac{1}{\sqrt{2}} \left( u \bar{u} + d \bar{d} \right)\,, \nonumber \\
\rho^{0} &\sim & \frac{1}{\sqrt{2}} \left( u \bar{u} - d \bar{d} \right)\,, \nonumber \\
\omega &\sim & \frac{1}{\sqrt{2}} \left( u \bar{u} + d \bar{d} \right)\,.
\label{A1}
\end{eqnarray}

Consider now a radiative decay of the $f_{1}$.
After the emission of the photon by the $f_{1}$ the quark state
should have the structure, with the quark charges
$e_{u} = 2/3$, $e_{d} = -1/3$, $e_{s} = -1/3$:
\begin{eqnarray}
\gamma f_{1} &\sim & \frac{1}{\sqrt{2}} \left( e_{u} \,u \bar{u} + e_{d} \,d \bar{d} \right) \nonumber \\
&& = \frac{1}{2} \left(e_{u} + e_{d}\right) 
     \frac{1}{\sqrt{2}}\left( u \bar{u} + d \bar{d} \right)
    +\frac{1}{2} \left(e_{u} - e_{d}\right) 
     \frac{1}{\sqrt{2}}\left( u \bar{u} - d \bar{d} \right) \nonumber \\
&& \sim \frac{1}{6} \,\omega + \frac{1}{2} \,\rho^{0} \,.
\label{A2}
\end{eqnarray}
Therefore, this simple argument suggests for the $f_{1} V \gamma$
coupling constants the relation
\begin{eqnarray}
\frac{g_{f_{1} \rho \gamma}}{g_{f_{1} \omega \gamma}} = 3 \,.
\label{A3}
\end{eqnarray}
This is the relation suggested, e.g., 
in \cite{Kochelev:2009xz}.

Now we can combine this with VMD which allows to relate 
the $g_{V V f_{1}}$ and $g_{f_{1} V \gamma}$ by the standard $V \gamma$
transition vertices; 
see e.g. (3.23)--(3.25) of \cite{Ewerz:2013kda}.
This gives, with $e = \sqrt{4 \pi \alpha_{\rm em}}$,
\begin{eqnarray}
g_{f_{1} V \gamma} = \frac{e}{\gamma_{V}} g_{V V f_{1}} \,,
\label{A4}
\end{eqnarray}
%
%Here $g_{f_{1} V \gamma}$ includes $e$,
%because we start from (\ref{VVA}). 
%But often in a radiative decays $e$ occurs separately, 
%i.e. $e g_{f_{1} V \gamma}$.
%The same remark is for Eq. (\ref{A9a}) below.
where $\gamma_{\rho} > 0$, $\gamma_{\omega} > 0$, and
\begin{eqnarray}
\frac{4 \pi}{\gamma_{\rho}^{2}} = 0.496 \pm 0.023\,, \quad
\frac{4 \pi}{\gamma_{\omega}^{2}} = 0.042 \pm 0.0015 \,.
\label{A5}
\end{eqnarray}

In the naive quark model plus VMD the hadronic light-quark-electromagnetic current
is written as follows:
\begin{eqnarray}
J_{\mu}^{\rm em} &=& 
e \left[ e_{u} \,\bar{u} \gamma_{\mu} u + e_{d} \,\bar{d} \gamma_{\mu} d
+ e_{s} \,\bar{s} \gamma_{\mu} s \right] \nonumber \\
& = &e \Big\lbrace 
 \frac{1}{\sqrt{2}} \left(e_{u} - e_{d}\right) 
 \frac{1}{\sqrt{2}} \left( \bar{u} \gamma_{\mu} u - \bar{d} \gamma_{\mu} d \right)
+\frac{1}{\sqrt{2}} \left(e_{u} + e_{d}\right) 
 \frac{1}{\sqrt{2}} \left( \bar{u} \gamma_{\mu} u + \bar{d} \gamma_{\mu} d \right)
+ e_{s} \, \bar{s} \gamma_{\mu} s 
\Big\rbrace \nonumber \\
& = &e \Big\lbrace 
\frac{m_{\rho}^{2}}{\gamma_{\rho}^{(\rm id)}} \, \rho_{\mu}^{(0)} +
\frac{m_{\omega}^{2}}{\gamma_{\omega}^{(\rm id)}} \, \omega_{\mu} +
\frac{m_{\phi}^{2}}{\gamma_{\phi}^{(\rm id)}} \, \phi_{\mu}
\Big\rbrace \,.
\label{A6}
\end{eqnarray}
Assuming $m_{\rho}^{2} = m_{\omega}^{2}$
(which is quite good) and $m_{\rho}^{2} = m_{\phi}^{2}$
(which is less good) we find from (\ref{A6})
the following ``ideal mixing'' coupling ratios:
\begin{eqnarray}
\frac{\gamma_{\omega}^{(\rm id)}}{\gamma_{\rho}^{(\rm id)}}
= \frac{e_{u} - e_{d}}{e_{u} + e_{d}} = 3\,, \quad
\frac{\gamma_{\phi}^{(\rm id)}}{\gamma_{\rho}^{(\rm id)}}
= \frac{e_{u} - e_{d}}{\sqrt{2} e_{s}} = -\frac{3}{\sqrt{2}}\,.
\label{A7}
\end{eqnarray}
From (\ref{A4}) and (\ref{A7}) we obtain with the ``ideal'' $\gamma V$ couplings
\begin{eqnarray}
\frac{g_{f_{1} \rho \gamma}}{g_{f_{1} \omega \gamma}} = 
\frac{\gamma_{\omega}^{(\rm id)}}{\gamma_{\rho}^{(\rm id)}}
\frac{g_{\rho \rho f_{1} }}{g_{\omega \omega f_{1} }} =
3 \,\frac{g_{\rho \rho f_{1} }}{g_{\omega \omega f_{1} }} \,.
\label{A8}
\end{eqnarray}
With (\ref{A3}) plus (\ref{A8}) we obtain, thus,
the simple estimate
\begin{eqnarray}
\frac{g_{\rho \rho f_{1} }}{g_{\omega \omega f_{1} }} = 1 \,,
\label{A9}
\end{eqnarray}
based on naive quark-model relations plus the simplest VMD ansatz.

If we include form factors in our considerations as in
(\ref{VVA_vertex})--(\ref{ff_pow}) we will get instead of (\ref{A4})
\begin{eqnarray}
g_{f_{1} V \gamma} = \frac{e}{\gamma_{V}} g_{V V f_{1}} \tilde{F}_{V}(0)\,;
\label{A9a}
\end{eqnarray}
see also Appendix~B below. 
Assuming $\tilde{F}_{\rho}(0) = \tilde{F}_{\omega}(0)$
we get again the relation (\ref{A9}).

In reality we get from (\ref{A5}), using the central values there,
\begin{eqnarray}
\frac{\gamma_{\omega}}{\gamma_{\rho}} 
%= \frac{\sqrt{4 \pi / \gamma_{\rho}^{2}}}
%       {\sqrt{4 \pi / \gamma_{\omega}^{2}}}
= 3.44 \,.
\label{A10}
\end{eqnarray}
That is, ideal mixing (\ref{A7}) gives only an approximation,
valid to within $15\,\%$, compared to the experimental value (\ref{A10}).
We can, therefore, expect that also the relation (\ref{A9}) may be violated
in the real world by 15~to~20~$\%$.

We emphasize that the arguments presented in this Appendix
depend crucially on the assumption made in (\ref{A1}) that
the $f_{1}(1285)$ is a normal $q \bar{q}$ state.
The relation (\ref{A3}) in particular could be quite different
if this assumption is violated and the $f_{1}(1285)$
has another structure.
In \cite{Aceti:2015pma,Xie:2019iwz}, for instance, the $f_{1}(1285)$ 
is described as a $K^{*}\bar{K}$ molecule, 
not as a $q \bar{q}$ state.

%-------------------------------------------------------------------
\section{The radiative decays of the $f_{1}(1285)$ meson and the $\rho \rho f_{1}$ coupling constant}
\label{sec:appendixB}
%-------------------------------------------------------------------

In this appendix we shall discuss the radiative decays of the $f_{1}(1285)$ meson
using VMD and the $VVf_1$ coupling vertex (\ref{VVA_vertex})
for $V = \rho^{0}$.
We consider two theoretical treatments.
In the first method we consider the decay $f_{1} \to \rho^{0} \gamma$
with a fixed mass for the $\rho^{0}$ meson:
\begin{eqnarray}
f_{1}\,(k,\epsilon^{(f_{1})}) \to \rho^{0}\,(k_{\rho},\epsilon^{(\rho)}) + \gamma\,(k_{\gamma},\epsilon^{(\gamma)}) \,.
\label{1to2}
\end{eqnarray}
In the second method we consider the decay $f_{1} \to \pi^{+}\pi^{-} \gamma$
via an intermediate $\rho^{0}$ meson
taking its mass distribution into account:
\begin{eqnarray}
f_{1}\,(k,\epsilon^{(f_{1})}) \to 
[\rho^{0}\,(k_{\rho}) \to \pi^{+}(k_{1}) + \pi^{-}(k_{2})] + \gamma\,(k_{\gamma},\epsilon^{(\gamma)}) \,.
\label{1to3}
\end{eqnarray}
From (\ref{1to2}) and (\ref{1to3}) we will estimate the $g_{\rho \rho f_{1}}$ coupling constant and the cutoff parameter $\Lambda_{V}$
in the form factor $F_{\rho \rho f_{1}}$ from experiment.

The amplitude for the reaction (\ref{1to2}) is given by
%has the form
%
\begin{eqnarray}
%\braket{\rho^{0}\,(k_{\rho}), \gamma\,(k_{\gamma}) |{\cal T}|f_{1}(k,\epsilon)} 
{\cal M}_{\lambda_{f_{1}} \to \lambda_{\rho} \lambda_{\gamma}}
&=& (-i) 
(\epsilon^{(\rho)\,\mu}(\lambda_{\rho}))^*
(\epsilon^{(\gamma)\,\nu}(\lambda_{\gamma}))^*\,
i\Gamma_{\mu \nu'' \alpha}^{(\rho \rho f_{1})}(-k_{\rho},-k_{\gamma})
%F_{\rho \rho f_{1}}(k_{\rho}^{2}, k_{\gamma}^{2}, k^{2})
\nonumber \\
&&\times i\Delta^{(\rho)\,\nu'' \nu'}(k_{\gamma}) \,
i\Gamma^{(\rho \to \gamma)}_{\nu' \nu}(k_{\gamma})\,  
\epsilon^{(f_{1})\,\alpha}(\lambda_{f_{1}})
\nonumber \\
&=& \frac{e}{\gamma_{\rho}}
(\epsilon^{(\rho)\,\mu}(\lambda_{\rho}))^*
(\epsilon^{(\gamma)\,\nu}(\lambda_{\gamma}))^*\,
\Gamma_{\mu \nu \alpha}^{(\rho \rho f_{1})}(-k_{\rho},-k_{\gamma})
%\mid_{\rm bare} \tilde{F}_{\rho}(0)\,
\epsilon^{(f_{1})\,\alpha}(\lambda_{f_{1}})\,, \quad
\label{T_1to2}
\end{eqnarray}
where $\epsilon^{(\rho)}$, $\epsilon^{(\gamma)}$ 
and $\epsilon^{(f_{1})}$ are the polarisation vectors 
for $\rho^{0}$, photon and $f_{1}(1285)$ meson 
with the four-momenta and helicities
$k_{\rho}$, $\lambda_{\rho} = \pm 1, 0$,
$k_{\gamma}$, $\lambda_{\gamma} = \pm 1$ and 
$k$, $\lambda_{f_{1}} = \pm 1, 0$, respectively.
We use the VMD ansatz
for the coupling of the $\rho^{0}$ meson to the photon; 
see e.g. (3.23)--(3.25) of \cite{Ewerz:2013kda}.
We assume in the $\Gamma^{(\rho \rho f_{1})}$ vertex (\ref{VVA_vertex})
the form factor according to (\ref{F_VVA})
\begin{eqnarray}
F_{\rho \rho f_{1}}(k_{\rho}^{2}, k_{\gamma}^{2}, k^{2}) 
= F_{\rho \rho f_{1}}(m_{\rho}^{2}, 0, m_{f_{1}}^{2}) 
= \tilde{F}_{\rho}(m_{\rho}^{2}) \tilde{F}_{\rho}(0) F_{f_{1}}(m_{f_{1}}^{2}) 
= \tilde{F}_{\rho}(0)\,,
\end{eqnarray}
with $\tilde{F}_{\rho}(0)$ given in (\ref{ff_pow}).

The amplitude for the reaction (\ref{1to3}),
${\cal M}_{\lambda_{f_{1}} \to \pi^{+} \pi^{-} \lambda_{\gamma}}$,
is obtained from (\ref{T_1to2})
by making the replacement
\begin{eqnarray}
(\epsilon^{(\rho)\,\mu}(\lambda_{\rho}))^* \to
i\Delta^{(\rho)\,\mu \mu'}(k_{\rho})\, 
i\Gamma^{(\rho \pi \pi)}_{\mu'}(k_{1},k_{2}) 
 = -\frac{g_{\rho \pi \pi}}{2} (k_{1}-k_{2})^{\mu}
     \Delta^{(\rho)}_{T}(k_{\rho}^{2}) 
     \,,
\label{1to3_replacement}
\end{eqnarray}
and taking 
%for the $\Gamma_{\mu \nu \alpha}^{(\rho \rho f_{1})}$
%vertex the form factor 
%
\begin{eqnarray}
F_{\rho \rho f_{1}}(k_{\rho}^{2}, k_{\gamma}^{2}, k^{2})
= F_{\rho \rho f_{1}}(k_{\rho}^{2}, 0, m_{f_{1}}^{2})
= \tilde{F}_{\rho}(k_{\rho}^{2}) \tilde{F}_{\rho}(0)
\end{eqnarray}
with \mbox{$k_{\rho}^{2} = (k_{1}+k_{2})^{2}$}.
The $\rho^{0}$~propagator function 
and the $\rho^{0} \pi^{+} \pi^{-}$
coupling in (\ref{1to3_replacement})
are taken from (4.1)--(4.6)
and (3.35), (3.36) of \cite{Ewerz:2013kda}, respectively.

Then, the coupling constant $g_{\rho \rho f_{1}}$,
occurring in $\Gamma^{(\rho \rho f_{1})}$
in the amplitudes above,
can be adjusted to the experimental decay width
$\Gamma(f_{1}(1285) \to \gamma \rho^{0})$.
For the $1 \to 2$ decay process (\ref{1to2})
this is straightforward.
For the $1 \to 3$ decay process (\ref{1to3})
this will be done with the help of a new Monte Carlo generator \textsc{Decay} \cite{Kycia:2020mgf}
designed for a general decay of the $1 \to n$ type.

Unfortunately the partial decay width 
$\Gamma(f_{1}(1285) \to \gamma \rho^{0})$ 
appears to be not well known in the literature,
see also the discussion in Sec.~VII~C and Table~IV 
in \cite{Dickson:2016gwc},
\begin{eqnarray}
&&{\rm from \;PDG}\; [44]: \quad\;\,
\Gamma(f_{1}(1285) \to \gamma \rho^{0}) 
= 1384.7^{+305.1}_{-283.1} \;{\rm keV}\,,
\label{Gamma_rhogamma_PDG}\\
&&{\rm from\; CLAS}\; [16]: \quad 
\Gamma(f_{1}(1285) \to \gamma \rho^{0}) = (453 \pm 177) \;{\rm keV}\,.
\label{Gamma_rhogamma_CLAS}
\end{eqnarray}
Using the values of total widths accordingly 
from PDG (\ref{f1_PDG}) and the CLAS experiment 
(\ref{f1_CLAS}) we get
\begin{eqnarray}
&&{\rm from \;PDG}\; [44]: \quad\;\,
{\cal BR}(f_{1}(1285) \to \gamma \rho^{0}) = (6.1 \pm 1.0) \, \%\,,
\label{BR_f1_1285_rhogam_PDG}\\
&&{\rm from\; CLAS}\; [16]: \quad 
{\cal BR}(f_{1}(1285) \to \gamma \rho^{0}) = (2.5^{+0.7}_{-0.8}) \, \%\,.
\label{BR_f1_1285_rhogam_CLAS}
\end{eqnarray}
We note that the CLAS result is in agreement 
with that found in \cite{Amelin:1994ii},
\begin{eqnarray}
{\cal BR}(f_{1}(1285) \to \gamma \rho^{0}) =
(2.8 \pm 0.7 \,{\rm (stat)} \pm 0.6 \,{\rm (syst)}) \, \%\,,
\label{BR_f1_1285_rhogam_VES}
\end{eqnarray}
where the decay $f_{1}(1285) \to \rho^{0} \gamma$
was studied in the reaction $\pi^{-} N \to \pi^{-} f_{1} N$.
Theoretical estimates based on 
the QCD inspired models
such as the covariant oscillator quark model \cite{Ishida:1988uw} 
and the Nambu--Jona-Lasinio model \cite{Osipov:2017ray}, 
which assume that the $f_{1}(1285)$ has a quark-antiquark nature,
suggest (\ref{Gamma_rhogamma_CLAS})
rather than (\ref{Gamma_rhogamma_PDG}).
We hope that the future experimental measurements can clarify this issue.
In the following we shall use both values,
(\ref{Gamma_rhogamma_PDG}) and (\ref{Gamma_rhogamma_CLAS}),
to highlight the problem.

In Table~\ref{tab:table1}
we collect our results for the two processes
(\ref{1to2}) and (\ref{1to3}) obtained
from (\ref{Gamma_rhogamma_PDG}) and (\ref{Gamma_rhogamma_CLAS}).
In the calculations we take $m_{\rho} = 775$~MeV.
We show results for 
the cutoff parameter 
from $\Lambda_{\rho} = 0.65$~GeV to 2~GeV in (\ref{ff_pow}).
We expect the upper limit of the 
$\rho \rho f_{1}$ coupling constant 
to be not much larger 
than $|g_{\rho \rho f_{1}}| \simeq 20$.
Otherwise one gets
a nonrealistically large cutoff parameter 
$\Lambda_{VNN}$ in the $VNN$ vertex 
(see the discussion in Appendix~\ref{sec:appendixC}).
%%%%%%%%%%%%%%%%%%%%%%%%%%%%%%%%%%%%%%%%
\begin{table}[!ht]
\centering
\caption{Coupling constant $|g_{\rho \rho f_{1}}|$
extracted from our model analysis of the radiative decays
of the $f_{1}(1285)$.
The results correspond to the two central values of
$\Gamma(f_{1}(1285) \to \gamma \rho^{0})$ from (\ref{Gamma_rhogamma_PDG}) 
(``PDG'') and (\ref{Gamma_rhogamma_CLAS}) (``CLAS''),
respectively,
and to various values of the cutoff parameter
$\Lambda_{\rho}$ in (\ref{ff_pow}).}
\label{tab:table1}
\begin{tabular}{|l|l|r|r|}
\hline
Process & Cutoff parameter
& PDG, $|g_{\rho \rho f_{1}}|$ 
%& $|g_{f_{1} \rho \gamma}|$, Eq.~(\ref{A4})
& CLAS, $|g_{\rho \rho f_{1}}|$ 
%& $|g_{f_{1} \rho \gamma}|$, Eq.~(\ref{A4})
\\  
\hline
$f_{1} \to \rho^{0} \gamma$, Eq.~(\ref{1to2})&  
$\Lambda_{\rho} = 0.65$~GeV 
& 27.37
& 15.66
\\
&  
$\Lambda_{\rho} = 0.7$~GeV 
& 22.68
& 12.97
\\
&
$\Lambda_{\rho} = 1.0$~GeV 
& 12.33
& 7.05
\\
&
$\Lambda_{\rho} = 2.0$~GeV 
& 9.27
& 5.30
\\
\hline 
$f_{1} \to \pi^{+}\pi^{-} \gamma$, Eq.~(\ref{1to3})& 
%$\Lambda_{\rho} = 0.6$~GeV 
%& 44.80
%& 25.62
%\\
$\Lambda_{\rho} = 0.65$~GeV 
& 35.02
& 20.03
\\ 
& $\Lambda_{\rho} = 0.7$~GeV 
& 28.54
& 16.33
\\ 
& $\Lambda_{\rho} = 0.8$~GeV 
& 20.98
& 12.00
\\ 
& $\Lambda_{\rho} = 0.9$~GeV 
& 17.05
& 9.75
\\ 
& $\Lambda_{\rho} = 1.0$~GeV 
& 14.85
& 8.49
\\ 
& $\Lambda_{\rho} = 1.5$~GeV 
& 11.52
& 6.59
\\
& $\Lambda_{\rho} = 2.0$~GeV 
& 10.97
& 6.27
\\
\hline 
\end{tabular}
\end{table}
%%%%%%%%%%%%%%%%%%%%%%%%%%%%%%%%%%%%%%%%

It is also interesting to compare 
our results with those of \cite{Xie:2019iwz}.
In \cite{Xie:2019iwz} the radiative decays 
$f_{1}(1285) \to \gamma V$
were evaluated with the assumption that the $f_{1}(1285)$ is
dynamically generated from the $K^{*}\bar{K}$ interaction.
In this model the partial decay widths strongly depend on the cutoff parameter $\Lambda$,
for instance, $\Gamma(f_{1}(1285) \to \gamma \rho^{0}) = 560$~keV,
or 1360~keV,
for $\Lambda = 1.0$~GeV, or 2.5~GeV, respectively; 
see Table~I of \cite{Xie:2019iwz}.
Moreover, there were also determined the ratios
\begin{eqnarray}
&& R_{1} = \frac{\Gamma(f_{1} \to \gamma \rho^{0})}{\Gamma(f_{1} \to \gamma \phi)} \simeq 62\,, 
\label{ratio_R1}\\
&& R_{2} = \frac{\Gamma(f_{1} \to \gamma \rho^{0})}{\Gamma(f_{1} \to \gamma \omega)} \simeq 28\,.
\label{ratio_R2}
\end{eqnarray}
The dependence of both ratios on the cutoff parameter 
is rather weak.
In the model of \cite{Xie:2019iwz} the partial decay width 
of $\Gamma(f_{1} \to \gamma \rho^{0})$
is much larger than the ones of the $\gamma \omega$ 
and $\gamma \phi$ channels due to
constructive (destructive) interference 
of the triangle loop diagrams 
for the $\rho^{0}$ ($\omega$ and $\phi$) production.
%This is in contradiction to the quark model predictions
%\cite{Ishida:1988uw,Osipov:2017ray}.

Now we consider the decay $f_{1} \to \omega \gamma$ in our approach.
We use the formula of (\ref{T_1to2})
with the replacements $\rho \to \omega$
[$\gamma_{\rho} \to \gamma_{\omega}$ (\ref{A5}),
$g_{f_{1} \rho \rho} \to g_{f_{1} \omega \omega}$,
$m_{\rho} \to m_{\omega}$].
In the calculation we take $m_{\omega} = 783$~MeV.
We assume $g_{\omega \omega f_1} = g_{\rho \rho f_1}$ (\ref{A9})
and take $g_{\rho \rho f_1}$ corresponding to
$\Lambda_{\rho} = 0.65$~GeV and 2.0~GeV
from Table~\ref{tab:table1} ($f_{1} \to \rho^{0} \gamma$).

With $\Lambda_{\rho} = 0.65$~GeV
(first line in Table~\ref{tab:table1}), 
we obtain $\Gamma(f_{1} \to \gamma \omega) = 106.61$~keV
for $|g_{f_{1} \omega \omega}| = 27.37$ and 
$\Gamma(f_{1} \to \gamma \omega) = 34.90$~keV
for $|g_{f_{1} \omega \omega}| = 15.66$.
Using the central values of (\ref{Gamma_rhogamma_PDG}) and (\ref{Gamma_rhogamma_CLAS})
these correspond to the ratios of 
$R_{2} = 12.98$ and $R_{2} = 12.99$, respectively.
With $\Lambda_{\rho} = 2.0$~GeV 
(fourth line in Table~\ref{tab:table1}), we obtain
$\Gamma(f_{1} \to \gamma \omega) = 112.61$~keV
for $|g_{f_{1} \omega \omega}| = 9.27$ 
and 
$\Gamma(f_{1} \to \gamma \omega) = 36.81$~keV
for $|g_{f_{1} \omega \omega}| = 5.30$.
With the central values of (\ref{Gamma_rhogamma_PDG}) and (\ref{Gamma_rhogamma_CLAS}) we obtain
the ratios $R_{2} = 12.31$ and $R_{2} = 12.30$, respectively.
These values for $R_{2}$ 
are about 2 times smaller than (\ref{ratio_R2}) 
estimated in \cite{Xie:2019iwz}.

The recent average for $R_{1}$ given by PDG \cite{Zyla:2020zbs} 
is $R_{1} = 82.4^{+11.4}_{-23.8}$.
This is about 1 s.d. away from the theoretical result
(\ref{ratio_R1}) of \cite{Xie:2019iwz}.
But we have to keep in mind the differences 
in the width of $f_{1} \to \gamma \rho^{0}$ given by PDG and CLAS; 
see (\ref{Gamma_rhogamma_PDG}) and (\ref{Gamma_rhogamma_CLAS}).
There are currently no experimental data available for 
$f_{1}(1285) \to \gamma \omega$ decay.
Further experiments will hopefully clarify the situation.

%-------------------------------------------------------------------
\section{Photoproduction of the $f_{1}(1285)$ meson and comparison with the CLAS experimental data}
\label{sec:appendixC}
%-------------------------------------------------------------------

\begin{figure}[!ht]
\includegraphics[width=0.35\textwidth]{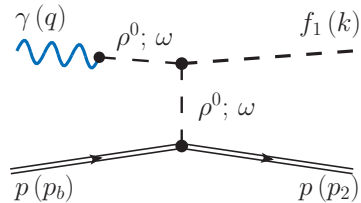}
  \caption{\label{fig:diagrams_f1_photoproduction}
  \small
Photoproduction of an $f_{1}$ meson via vector-meson exchanges.}
\end{figure}

Here we discuss the photoproduction of the $f_{1}(1285)$ meson.
Using VMD and the $g_{VV f_1}$ coupling constants
introduced in (\ref{VVA_vertex})
we have to calculate the diagram shown in
Fig.~\ref{fig:diagrams_f1_photoproduction}.
The differential cross section 
for the reaction $\gamma p \to f_{1}(1285) p$ will be 
compared with the CLAS data \cite{Dickson:2016gwc}.
From this we will estimate the form factor 
and cutoff parameters of the model.

%--------------------------------------------------------
%--------------------------------------------------------

The unpolarized differential cross section 
for the reaction $\gamma p \to f_{1}(1285) p$ is given by
\begin{eqnarray}
&& \frac{d\sigma}{d\Omega} = \frac{1}{64 \pi^{2} s} \frac{| \bk |}{| \bq |} \frac{1}{4} \sum_{{\rm spins}}|{\cal M}_{\gamma p \to f_{1}(1285) p}|^{2}\,, 
\nonumber \\
&& d\Omega = \sin\theta \,d\theta \,d\phi \,.
\label{dsig_dz}
\end{eqnarray}
Here we work in the center-of-mass (c.m.) frame,
$s$ is the invariant mass squared of the $\gamma p$ system,
and $\bq$ and $\bk$ are the c.m. three-momenta of the initial photon and final $f_{1}(1285)$, respectively.
Taking the direction of $\bq$ as a $z$ axis
we denote the polar and azimuthal angles of $\bk$
by $\theta$ and $\phi$.

We use standard kinematic variables
\begin{eqnarray}
&&s = W_{\gamma p}^{2} = (p_{b} + q)^{2} = (p_{2} + k)^{2}\,, \nonumber \\
&&q_{t} = p_{b} - p_{2} = k - q  \,, \quad t = q_{t}^{2}\,.
\label{2to2_kinematic}
\end{eqnarray}

The amplitude for the $\gamma p \to f_{1}(1284) p$ reaction
via the vector-meson exchange includes two terms
\begin{eqnarray}
{\cal M}_{\gamma p \to f_{1}(1285) p} = 
{\cal M}^{(\rho \, {\rm exchange})} + {\cal M}^{(\omega \, {\rm exchange})}\,.
\label{amp_sum_photo}
\end{eqnarray}
The generic amplitude with $V = \rho^{0}, \omega$,
for the diagram in Fig.~\ref{fig:diagrams_f1_photoproduction},
can be written as 
\begin{eqnarray}
&&\Braket{f_{1}(k,\lambda_{f_{1}}),
p(p_{2},\lambda_{2})
|{\cal T}|
\gamma(q,\lambda_{\gamma}),
p(p_{b},\lambda_{b})} 
\equiv  
\nonumber \\  
&& {\cal M}^{(V \, {\rm exchange})}_{\lambda_{\gamma} \lambda_{b} \to \lambda_{f_{1}} \lambda_{2}} =
(-i) \, (\epsilon^{(f_{1})\,\alpha}(\lambda_{f_{1}}))^* \,
i\Gamma_{\mu'' \nu' \alpha}^{(VV f_{1})}(q, q_{t}) \,  
i\Delta^{(V)\,\mu'' \mu'}(q) \,
i\Gamma^{(\gamma \to V)}_{\mu' \mu}(q)\,
\epsilon^{(\gamma)\,\mu}(\lambda_{\gamma}) 
\nonumber \\ 
&& \qquad \qquad \qquad \quad \times 
i\tilde{\Delta}^{(V)\,\nu' \nu}(s,t)\, 
\bar{u}(p_{2}, \lambda_{2}) 
i\Gamma_{\nu}^{(V pp)}(p_{2},p_{b}) 
u(p_{b}, \lambda_{b}) \,,
\label{gamp_f1p}
\end{eqnarray}
where $p_{b}$, $p_{2}$ and $\lambda_{b}$, $\lambda_{2} = \pm \frac{1}{2}$ 
denote the four-momenta and helicities of the incoming and outgoing protons.

We use the relations for the $\gamma$-$V$ couplings 
($V = \rho^{0}, \omega$)
from (\ref{A4}) and (\ref{A5}).
For the other building blocks of the amplitude (\ref{gamp_f1p}) 
see (\ref{VVA_vertex})--(\ref{trajectory}) 
in Sec.~\ref{sec:theoretical_fr}.
We can then write
\begin{eqnarray}
&&{\cal M}^{(V  \, {\rm exchange})}_{\lambda_{\gamma} \lambda_{b} \to \lambda_{f_{1}} \lambda_{2}} =
- \dfrac{e}{\gamma_{V}}\,
(\epsilon^{(f_{1})\, \alpha}(\lambda_{f_{1}}))^*\,
i\Gamma_{\mu \nu \alpha}^{(VV f_{1})}(q, q_{t}) \,  
\epsilon^{(\gamma)\, \mu}(\lambda_{\gamma})  
\nonumber \\ 
&& \qquad \qquad \qquad \quad \times 
\tilde{\Delta}_{T}^{(V)}(s,t) \, \bar{u}(p_{2}, \lambda_{2}) 
i\Gamma^{(V pp)\,\nu}(p_{2},p_{b}) 
u(p_{b}, \lambda_{b}) \,.
\label{gamp_f1p_aux}
\end{eqnarray}

We perform the calculation of the total 
and differential cross sections
with the cutoff parameter $\Lambda_{\rho}$ 
and corresponding
$VV f_{1}$ coupling constant $g_{VV f_{1}}$
from Table~\ref{tab:table1}.
We choose the values from the last column (CLAS).
For instance, $|g_{\rho \rho f_{1}}| = 8.49$
corresponds to $\Lambda_{\rho} = 1.0$~GeV and 
$|g_{\rho \rho f_{1}}| = 20.03$
corresponds to $\Lambda_{\rho} = 0.65$~GeV.
We assume $g_{\omega \omega f_{1}} = g_{\rho \rho f_{1}}
\equiv g_{VV f_{1}}$; see (\ref{A9}).
For the $Vpp$ coupling constants we take (\ref{Vpp_couplings}).
For the $V$-proton form factor $F_{VNN}(t)$
we take the monopole form as in (\ref{F_V_M}) 
with the parameter $\Lambda_{VNN}$ 
to be extracted from the CLAS data.

\begin{figure}[!ht]  
\center
\includegraphics[width = 0.45\textwidth]{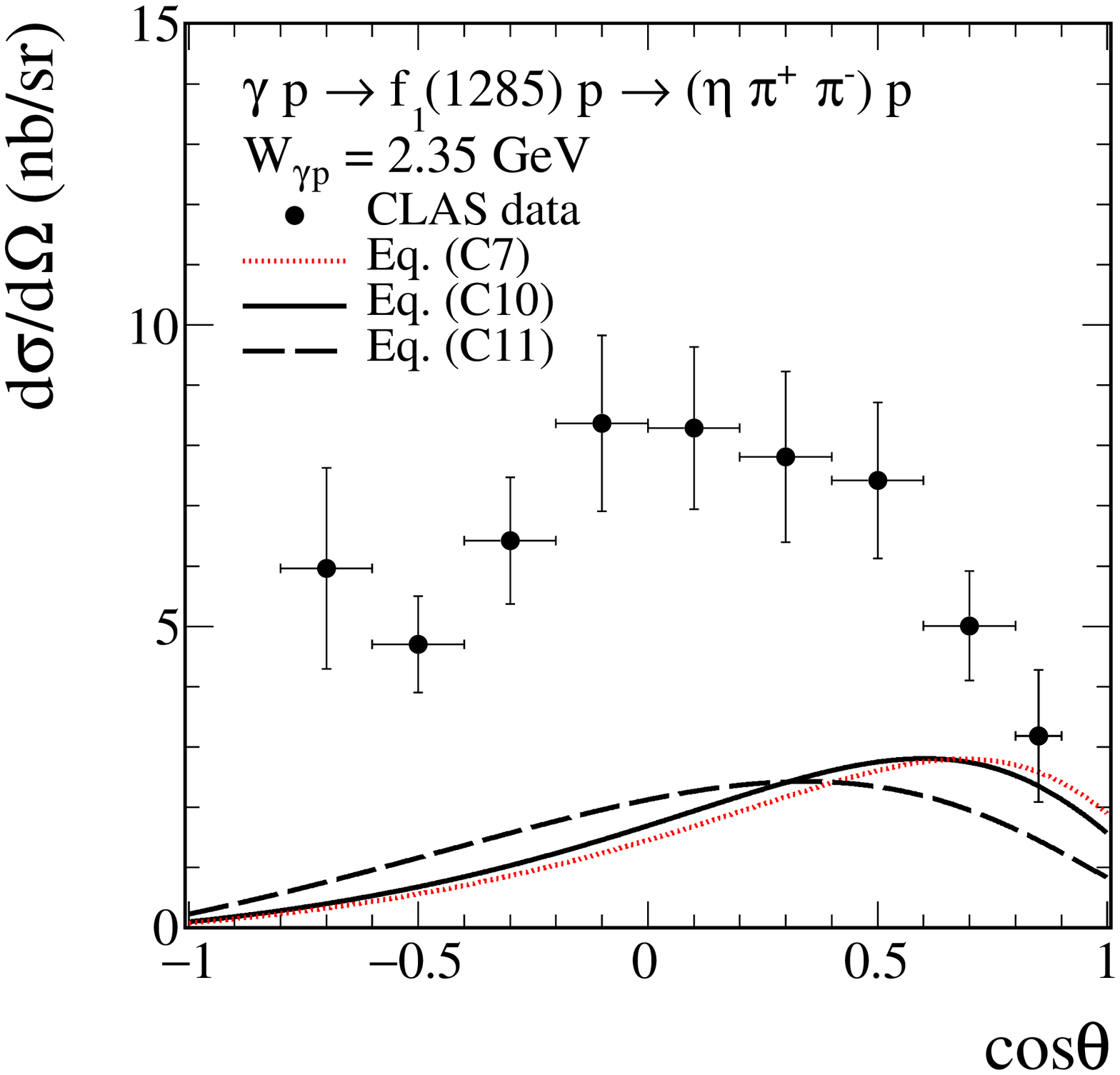}
\includegraphics[width = 0.45\textwidth]{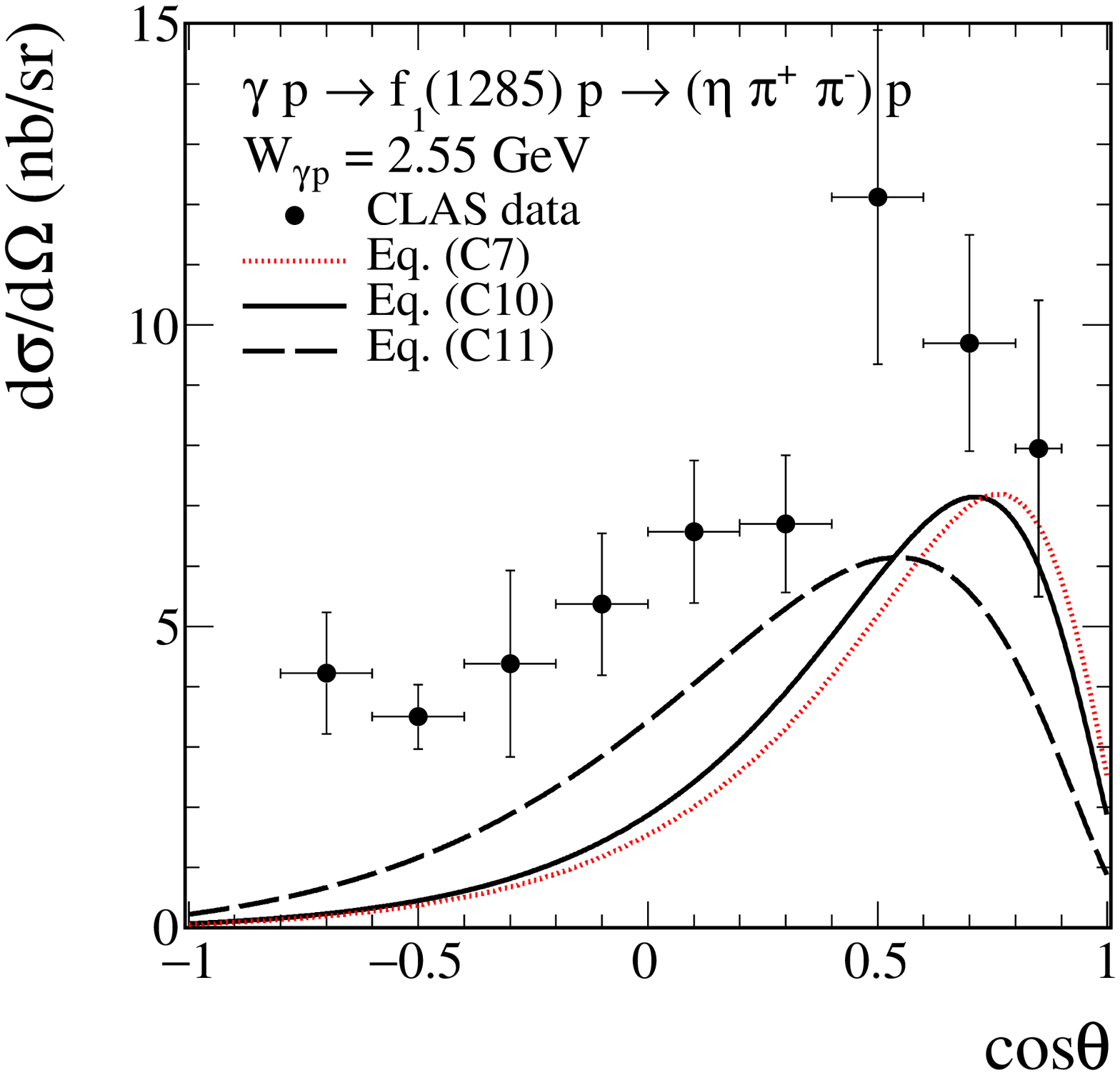}
\includegraphics[width = 0.45\textwidth]{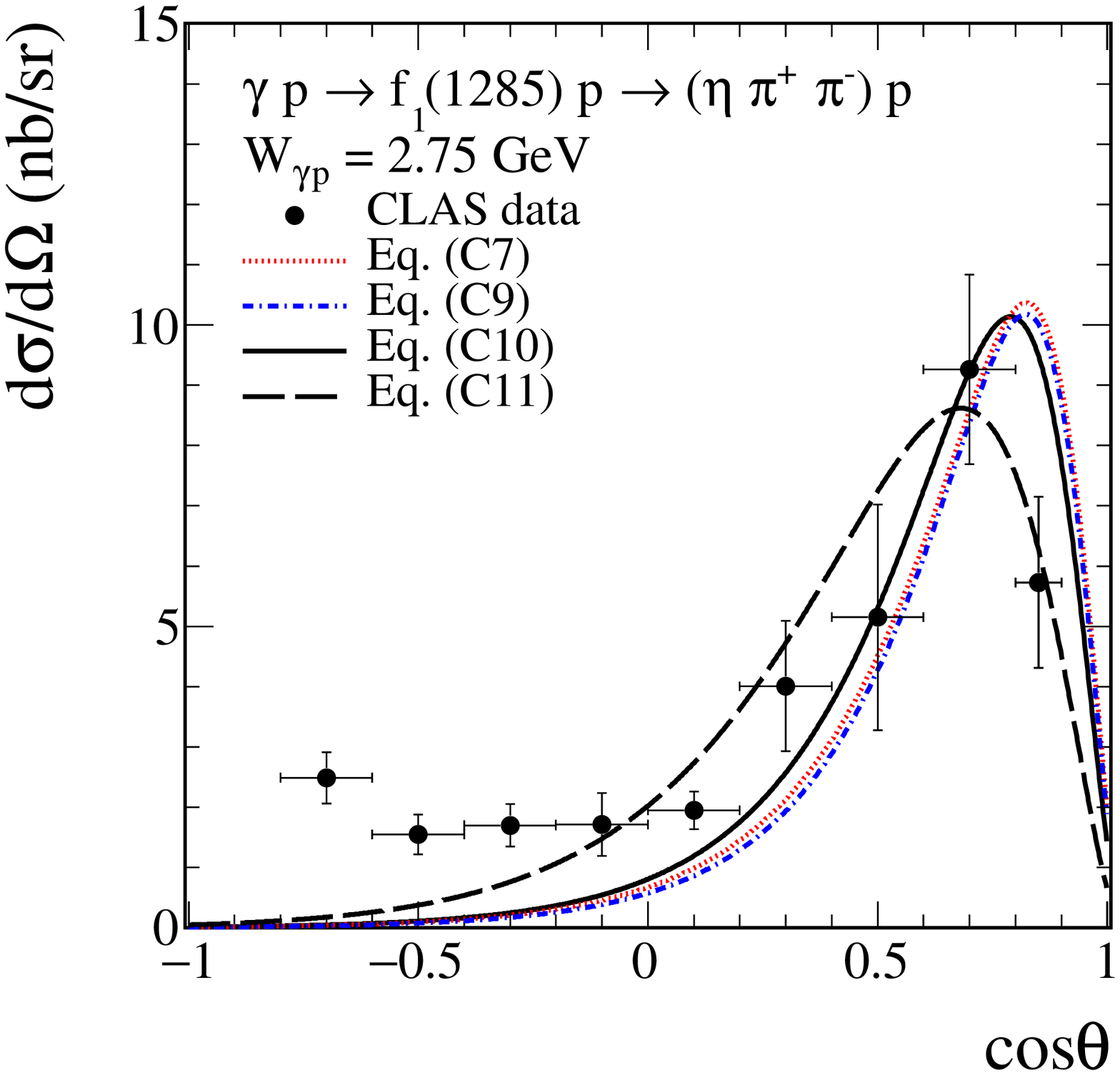}
\includegraphics[width = 0.45\textwidth]{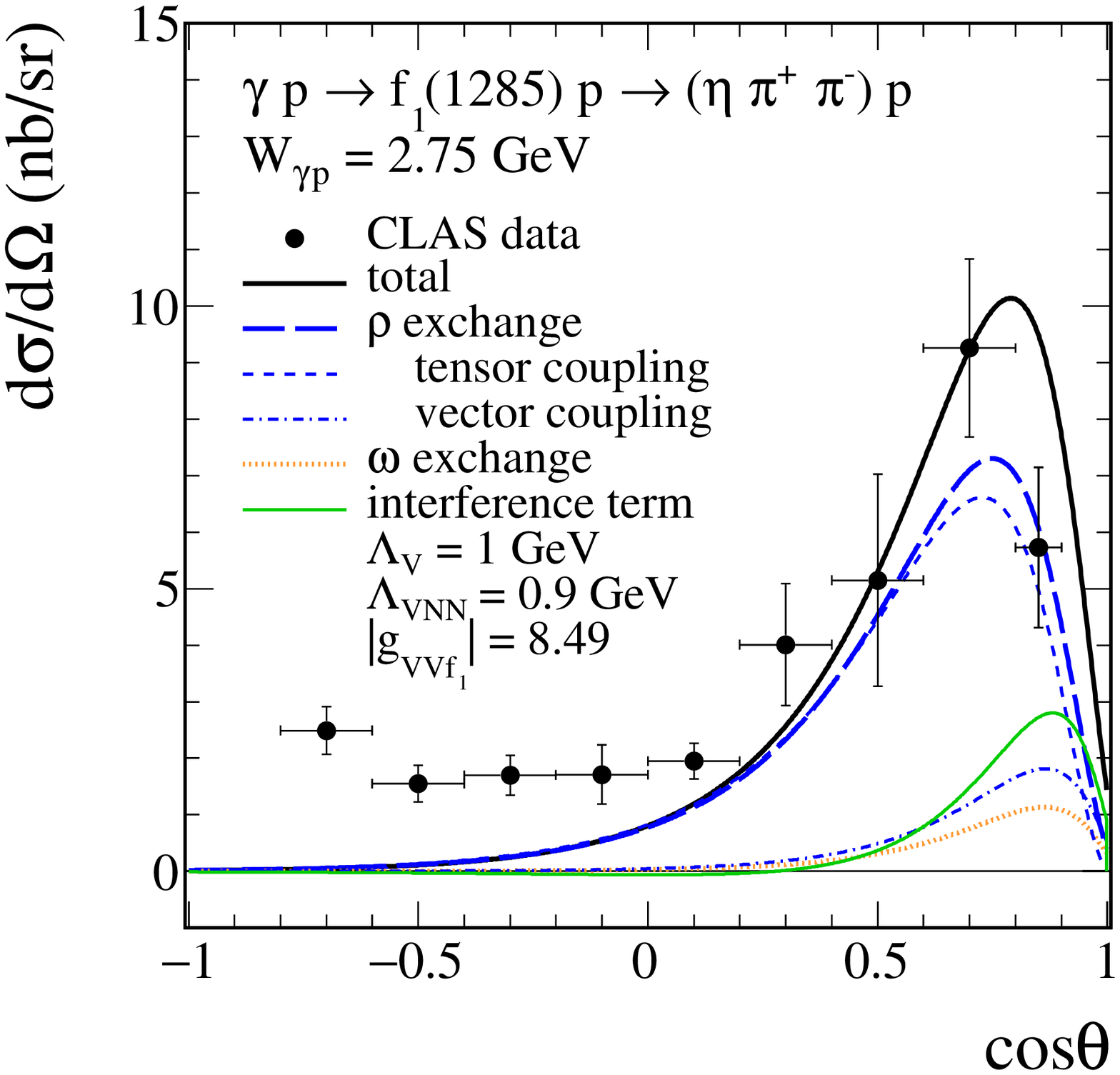}
  \caption{\label{fig:B1}
  \small
The differential cross sections 
for the reaction $\gamma p \to f_{1}(1285) p \to \eta \pi^{+}\pi^{-} p$.
Data are taken from Table~V of \cite{Dickson:2016gwc}.
The vertical error bars are the statistical and systematic uncertainties.
Our results are scaled by a factor of 0.35 to account 
for the branching fraction from $f_{1}(1285) \to \eta \pi^{+}\pi^{-}$ (\ref{BR_etapippim}).
We take the $Vpp$ coupling constants 
from (\ref{Vpp_couplings})
and the different values of $g_{VV f_{1}}$
corresponding to $\Lambda_{V}$ from the column ``CLAS''
of Table~\ref{tab:table1}.
In the bottom right panel we show the individual contributions of
$\rho$ and $\omega$ exchanges and 
their coherent sum (total) at $W_{\gamma p} = 2.75$~GeV.
For the $\rho$-exchange contribution 
also the results for only one type of coupling,
tensor or vector, in the $\rho$-proton vertex (\ref{vertex_VNN}) are shown.}
\end{figure}

In Fig.~\ref{fig:B1} we compare our results
for the photoproduction of $f_{1}(1285)$ meson
with the CLAS data \cite{Dickson:2016gwc}.
Since the CLAS Collaboration presents 
the differential cross sections 
for $\gamma p \to f_{1}(1285) p \to \eta \pi^{+}\pi^{-} p$,
our theoretical curves in Fig.~\ref{fig:B1}
have been scaled by the branching fraction 
${\cal BR}(f_{1}(1285) \to \eta \pi^{+}\pi^{-}) = 0.35$ from
\cite{Zyla:2020zbs} 
\begin{equation}
{\cal BR}(f_{1}(1285) \to \eta \pi^{+}\pi^{-}) = 0.35 \pm 0.15\,.
\label{BR_etapippim}
\end{equation}
From Fig.~\ref{fig:B1} we see that 
with the reggeized $\rho$ and $\omega$ exchanges we describe
the CLAS experimental data \cite{Dickson:2016gwc}
at $W_{\gamma p} = 2.75$~GeV
in the forward scattering region.
$d\sigma / d\cos\theta$ decreases
rapidly at forward angles, $\cos\theta = 1$.
We cannot expect our exchange model to also describe the data
at the lower CLAS energies since there we must expect
$s$-channel resonances 
and $u$-channel nucleon exchanges
to become important; see e.g.~\cite{Wang:2017hug}.
The model results are very sensitive on the 
form-factor cutoff parameters $\Lambda_{V}$ and $\Lambda_{VNN}$ 
at the $VVf_{1}$ and $VNN$ vertices ($V = \rho, \omega$).
Fitting the CLAS experimental data at $W_{\gamma p} = 2.75$~GeV we find
\begin{eqnarray}
%&& \Lambda_{VNN} = 1.7 \; {\rm GeV}\; {\rm for}\;
%\Lambda_{V} = 0.6 \; {\rm GeV}\,, 
%|g_{VV f_{1}}| = 25.62\,;
%\label{aLam0.6}\\
&& \Lambda_{VNN} = 1.35 \; {\rm GeV}\; {\rm for}\; 
\Lambda_{V} = 0.65 \; {\rm GeV}\,, 
|g_{VV f_{1}}| = 20.03\,;
\label{aLam0.65}\\
&& \Lambda_{VNN} = 1.17 \; {\rm GeV}\; {\rm for}\;
\Lambda_{V} = 0.7 \; {\rm GeV}\,, 
|g_{VV f_{1}}| = 16.33\,;
\label{aLam0.7}\\
&& \Lambda_{VNN} = 1.01 \; {\rm GeV}\; {\rm for}\;
\Lambda_{V} = 0.8 \; {\rm GeV}\,, 
|g_{VV f_{1}}| = 12.0\,;
\label{aLam0.8}\\
&& \Lambda_{VNN} = 0.9 \; {\rm GeV}\; {\rm for}\;
\Lambda_{V} = 1.0 \; {\rm GeV}\,, 
|g_{VV f_{1}}| = 8.49\,;
\label{aLam1.0}\\
&& \Lambda_{VNN} = 0.834 \; {\rm GeV}\; {\rm for}\;
\Lambda_{V} = 1.5 \; {\rm GeV}\,, 
|g_{VV f_{1}}| = 6.59\,;
\label{aLam1.5}\\
&& \Lambda_{VNN} = 0.82 \; {\rm GeV}\; {\rm for}\;
\Lambda_{V} = 2.0 \; {\rm GeV}\,, 
|g_{VV f_{1}}| = 6.27\,.
\label{aLam2.0}
\end{eqnarray}
There is not much difference between
the angular distributions resulting from 
the four parameter sets (\ref{aLam0.65})--(\ref{aLam1.0}).
The results for $\Lambda_{VNN} < 0.9$~GeV
in the monopole form factor (\ref{F_V_M}) 
are rather unrealistic,
$\Lambda_{VNN}$ being too close to the vector meson masses 
$m_{\rho}$, $m_{\omega}$.
Note that the value of $\Lambda_{VNN} = 1.35$~GeV 
is close to the values used in the Bonn potential model:
$\Lambda_{\rho NN} = 1.4$~GeV and $\Lambda_{\omega NN} = 1.5$~GeV;
see Table~4 of \cite{Machleidt:1987hj}.
Therefore, we are left with the sets of parameters
(\ref{aLam0.65})--(\ref{aLam1.0}) 
for our considerations in the $pp \to pp f_{1}$ reaction.

In the bottom right panel of Fig.~\ref{fig:B1}
the individual $\rho$- and $\omega$-exchange
contributions at $W_{\gamma p} = 2.75$~GeV are shown.
Here we use the parameters given in (\ref{aLam1.0}).
The $\rho$-exchange term is larger than 
the $\omega$-exchange term due to larger coupling constants
both in the $\gamma \to V$ transition vertex
(\ref{A5}), (\ref{A10}) and
for the tensor coupling in the $V$-proton vertex
(\ref{vertex_VNN}), (\ref{Vpp_couplings}).
The differential distribution at $W_{\gamma p} = 2.75$~GeV 
peaks for $\cos\theta = 0.7$ corresponding to
$-t = 0.66$~GeV$^{2}$.
The tensor coupling in the $\rho$-proton vertex
with parameters
$\kappa_{\rho} F_{\rho NN}(t)$
plays the most important role there.
One can observe also an interference effect
between the $\rho$ and $\omega$ 
exchange terms.\footnote{In \cite{Yu:2019wly} a different relation 
for the coupling constants was used, namely 
$g_{f_{1} \omega \gamma} = -g_{f_{1} \rho \gamma}/3$
instead of (\ref{A3}).
The relative sign of these couplings has physical significance
in the process $\gamma p \to f_{1}(1285) p$
as the $\rho$- and $\omega$-exchange terms interfere.
Therefore, the corresponding cross sections, 
see Figs.~5 and 6 of \cite{Yu:2019wly}, are different from ours.}
%--------------------------------------------------------
%-------------------------------------------------------

\begin{figure}[!ht]
\includegraphics[width=0.45\textwidth]{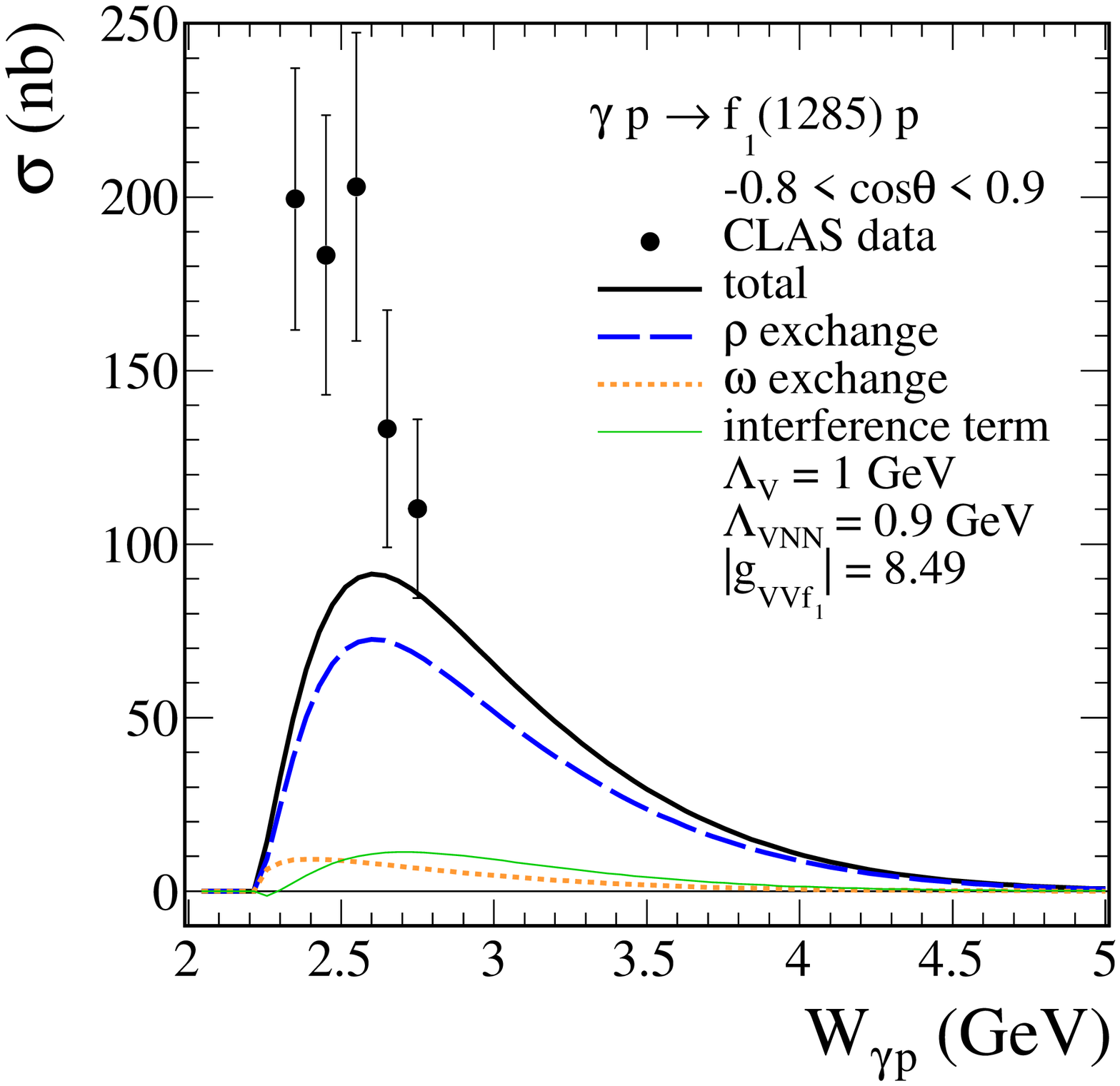}
\includegraphics[width=0.45\textwidth]{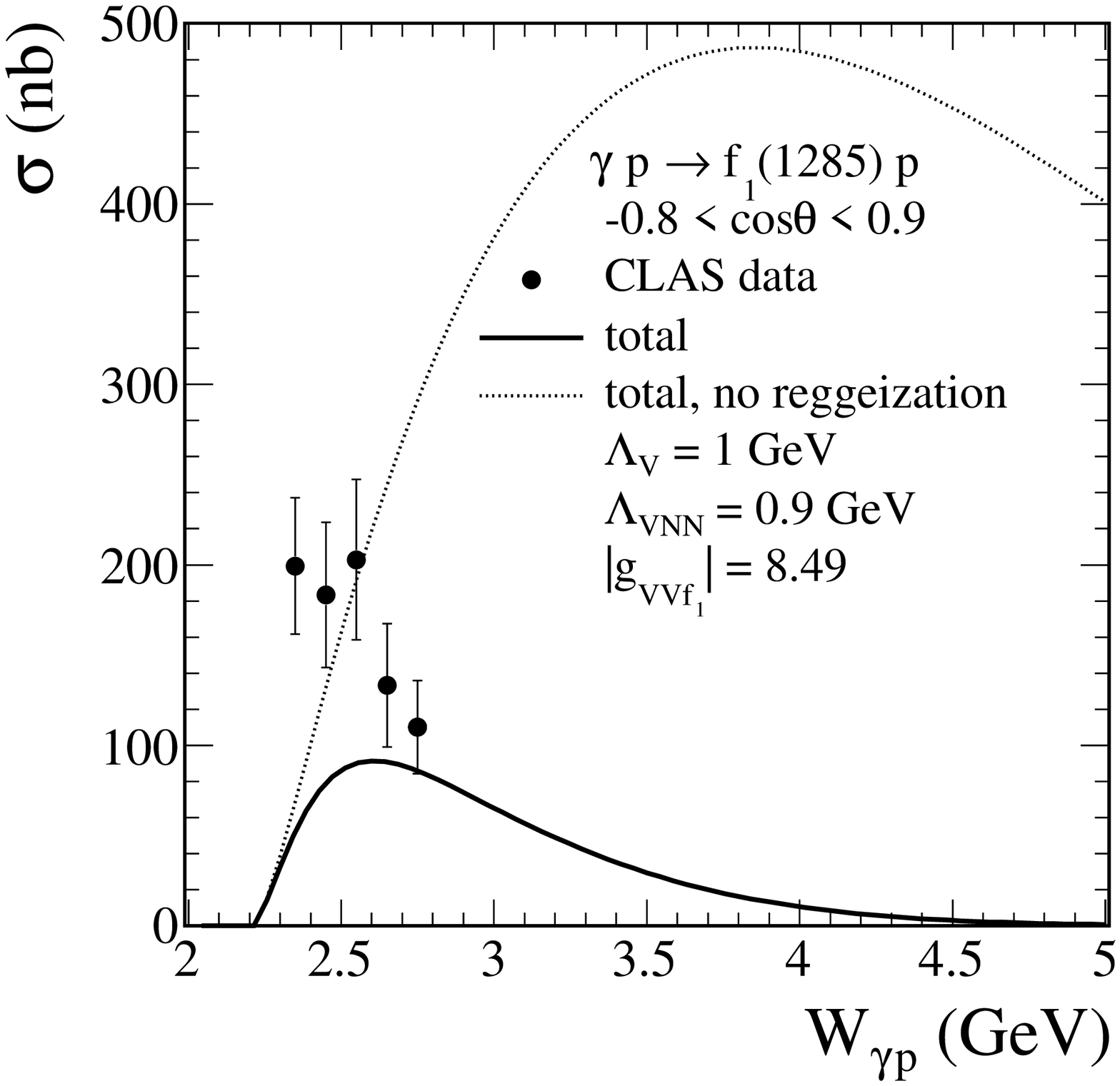}
\caption{\label{fig:B2}
\small
The elastic $f_{1}(1285)$ photoproduction cross section
as a function of the center-of-mass energy $W_{\gamma p}$.
Five data points are obtained by integrating out
the differential cross sections
given in Table~V of \cite{Dickson:2016gwc}.
The experimental results have been scaled by 
the branching fraction 
${\cal BR}(f_{1}(1285) \to \eta \pi^{+}\pi^{-}) = 0.35$; 
see (\ref{BR_etapippim}).
We take the coupling parameters
the same as in the bottom right panel of
Fig.~\ref{fig:B1}.
We integrate for $-0.8 < \cos\theta < 0.9$.
In the left panel
the reggeized contributions of $\rho$
and $\omega$ exchanges, their coherent sum (total),
and the interference term are shown.
In the right panel the solid line is the result 
from the reggeized model, the dotted line
indicates the result without the reggeization.}
\end{figure}

In Fig.~\ref{fig:B2} we show the integrated cross sections for
the reaction $\gamma p \to f_{1}(1285) p$
together with the CLAS data.
Results for $-0.8 < \cos\theta < 0.9$ are presented.
In the calculation we take (\ref{aLam1.0}).
In the left panel we show
the respective contributions of $\rho$
and $\omega$ exchanges and their coherent sum 
with the same $V$-proton coupling parameters 
as in the bottom right panel of Fig.~\ref{fig:B1}.
There, for $W_{\gamma p} \simeq 2.7$~GeV,
a large interference
between the $\rho$ exchange term
and the $\omega$ exchange term can be observed.
In the right panel we compare our reggeized model 
results with those of the model without this effect.
We note that the form of reggeization used in our model,
calculated according to (\ref{reggeization_2})--(\ref{trajectory}),
affects both, the $t$-dependence of the $V$ exchanges,
and the size of the cross section.
%-------------------------------------------------------
%--------------------------------------------------------

%--------------------
\acknowledgments
%--------------------
We are indebted to Kanzo Nakayama for a discussion of FSI effect 
in the $pp \to pp M$ reactions and Johann Haidenbauer 
for exchange of information on excitations of nucleon resonances.
This work was partially supported by
the Polish National Science Centre under Grant No. 2018/31/B/ST2/03537
and by the Center for Innovation and Transfer of Natural Sciences 
and Engineering Knowledge in Rzesz\'ow (Poland).

%------------------------------------------------------------------
\bibliography{refs}
%------------------------------------------------------------------

\end{document}